\documentclass[12pt,a4paper]{article}
\usepackage{graphicx}
\usepackage{axodraw}
\newcommand{\be}{\begin{equation}}
\newcommand{\ee}{\end{equation}}
\newcommand{\ba}{\begin{eqnarray}}
\newcommand{\ea}{\end{eqnarray}}
\newcommand{\la}{\langle}
\newcommand{\ra}{\rangle}
\newcommand{\lla}{\langle\langle}
\newcommand{\rra}{\rangle\rangle}
\newcommand{\mphys}{M^2_{\mathrm{phys}}}
\newcommand{\fphys}{F_{\mathrm{phys}}}
\newcommand{\trf}[1]{\langle#1\rangle}

\begin{document}

\begin{titlepage}
\begin{flushright}
LU TP 11-07\\
February 2011\\
\end{flushright}
\vfill
\begin{center}
{\Large\bf Meson-meson Scattering in QCD-like Theories}
\vfill
{\bf Johan Bijnens and Jie Lu}\\[0.3cm]
{Department of Astronomy and Theoretical Physics, Lund University,\\
S\"olvegatan 14A, SE 223-62 Lund, Sweden}
\end{center}
\vfill
\begin{abstract}
We discuss meson-meson scattering at next-to-next-to-leading order
in the chiral expansion for QCD-like theories with general
$n$ degenerate flavours for the cases with a complex, real and pseudo-real
representation. I.e. with global symmetry and breaking pattern
$SU(n)_L\times SU(n)_R\to SU(n)_V$, $SU(2n)\to SO(2n)$
and $SU(2n)\to Sp(2n)$. We obtain fully analytical expressions for all
these cases. We discuss the general structure of the amplitude and the
structure of the possible intermediate channels for all three cases.
We derive the expressions for the lowest partial wave scattering length
in each channel and present some representative numerical results.
We also show various relations between the different cases in the limit of
large $n$.
\end{abstract}
\vfill
{\bf PACS:}
\vfill
\end{titlepage}

\section{Introduction}

In an earlier paper \cite{paper1} we started the phenomenology of
QCD-like theories at next-to-next-to-leading (NNLO) order in the light mass
expansion in their respective low-energy effective theories.
The motivation for this work is that these theories are interesting as
variations on QCD and could play some role as models for a nonperturbative
Higgs sector. Early work in this context are the technicolor
variations of \cite{Peskin,Preskill,Dimopoulos}.
Recent reviews of more modern developments are \cite{Techni1,Techni2}.
Lattice calculations
have started to explore these type of theories as well,
some references are \cite{lattice}.
The main interest in these theories is in the massless limit but lattice
simulations are necessarily performed at a finite fermion mass.
In \cite{paper1} we worked out a number of simple observables,
the mass, decay constant and vacuum-expectation-value to NNLO
in these theories. Here we work out the amplitude for meson-meson
scattering to the same order. In lattice calculations the amplitude
for meson-meson scattering is not directly accessible but the
scattering lengths can be derived from the dependence on the volume of
the lattice \cite{Luscher}. We therefore also provide
explicit expressions for the scattering lengths.

The EFT relevant for dynamical electroweak symmetry breaking
can have different patterns of spontaneous breaking
of the global symmetry than QCD. The resulting Goldstone Bosons,
or pseudo-Goldstone bosons in the presence of mass terms,
are thus in different manifolds and the low-energy EFT is also different.

In this paper we only discuss the same cases as in \cite{paper1} where the
underlying strong interaction is vectorlike and all
fermions have the same mass. Three main patterns of global symmetry
show up. A thorough discussion tree level
or lowest order (LO) is \cite{Kogut}.
With $n$ fermions\footnote{We use $n$ rather than $N_F$ for
the number of flavours since it makes the formulas shorter.}
in a complex representation
the global symmetry group is $SU(n)_L\times SU(n)_R$
and it is expected to be spontaneously broken to the diagonal subgroup
$SU(n)_V$. This is the direct extension of the QCD case.
For $n$ fermions in a real representation the global symmetry group is
$SU(2n)$ and it is expected to be spontaneously broken to $SO(2n)$.
In the case of two colours and $n$ fermions in the fundamental (pseudo-real)
representation the global symmetry group is again $SU(2n)$ but here
it is expected to be spontaneously broken to an $Sp(2n)$ subgroup.
Earlier references are \cite{Kogan,Leutwyler,SV}.
Some earlier work for the complex case and the pseudo-real case at NLO
can be found in  \cite{GL2,GL4,Splittorff}.

In the remainder of this paper we refer to the complex representation case
as complex or QCD, the real representation case as adjoint or real and the
pseudo-real representation case as two-colour or pseudo-real.
In \cite{paper1} we extended the construction of the general Lagrangian
to NLO\footnote{References to some related work
can be found in \cite{Techni1}.} including the divergence structure.
The NNLO for the QCD case is in \cite{BCE1} and the divergence structure
in \cite{BCE2}. The Lagrangian constructed in \cite{BCE1} is
with the changes discussed in \cite{paper1} and in Sect.~\ref{EFT}
also a complete Lagrangian for the other two cases but we have not
shown it to be minimal nor calculated the divergence structure.

We do not repeat the discussion of the three different cases
at the underlying fermion (quark) level. This can be found in \cite{Kogut}
and \cite{paper1}, Sect.~2.
In Sect.~\ref{EFT} we quote the structure of the effective field
theories for the three cases but we again refer to \cite{paper1} for
more details. Sect.~\ref{groups} discusses in detail the
general structure of the amplitude. The amplitude can be expressed
in terms of two functions $B(s,t,u)$ and $C(s,t,u)$ which are
generalizations of the amplitude $A(s,t,u)$ in $\pi\pi$-scattering
\cite{Weinbergpipi}. We work out the possible intermediate
states using the relevant group theory
and using a projection operator formalism obtain the
amplitudes in the different channels. The results for the amplitude
are discussed in Sect.~\ref{results} and for the scattering lengths
in Sect.~\ref{scatteringlengths}. Here we present some representative numerical
results for the scattering lengths as well as some large $n$ relations
between the different cases.
The lengthier formulas at two-loop
order are given in an appendix.
This work needed a few more integrals at intermediate stages than
\cite{BCEGS1,BCEGS2}, these are given in \ App.~\ref{loopintegrals}.
In Sect.~\ref{conclusions} we summarize our results.

\section{Effective Field Theory}
\label{EFT}

\subsection{Generators}

The notation for the three cases can be brought in a very similar
form. More details can be found in \cite{paper1}.
The Goldstone boson live on a manifold $G/H$ where $G$ is the full global
symmetry group and $H$ is the part that remains unbroken after spontaneous
symmetry-breaking. We label the unbroken
generators as $T^a$ and the broken ones as $X^a$.

The space $SU(n)\times SU(n)/SU(n)$ is isomorphic to $SU(n)$
so we use the $X^a$ as the generators of $SU(n)$ for the QCD case.
They are traceless, hermitian $n\times n$ matrices.

The adjoint or real case has the generators in $SU(2n)/SO(2n)$
where the broken generators satisfy
\be
\label{brokenSON}
J_S X^a = \left(X^a\right)^T J_S\,,
\quad\mathrm{with}\quad
J_S=\left(\begin{array}{cc}0 &I\\I & 0\end{array}\right)\,.
\ee
$I$ is the $n\times n$ unit matrix and the superscript $T$ indicates the
transpose.
The $X^a$ are traceless, hermitian $2n\times 2n$ matrices in this case.
Multiplying (\ref{brokenSON}) with $J_S$ from left and right
leads immediately to
\be
X^a J_S = J_S \left(X^a\right)^T\,.
\ee

The two-colour or pseudo-real case has the generators in $SU(2n)/Sp(2n)$
where the broken generators satisfy
\be
\label{brokenSPN}
J_A X^a = \left(X^a\right)^T J_A\,,
\quad\mathrm{with}\quad
J_A=\left(\begin{array}{cc}0 &-I\\I & 0\end{array}\right)\,.
\ee
The $X^a$ are traceless, hermitian $2n\times 2n$ matrices
also in this case.
Multiplying (\ref{brokenSPN}) with $J_A$ similar to above gives
\be
X^a J_A = J_A \left(X^a\right)^T\,.
\ee

The unbroken generators satisfy
\ba
SO(2n): && T^a J_S + J_S T^{aT}=0\,,
\nonumber\\
Sp(2n): && T^a J_S + J_S T^{aT}=0\,.
\ea
This allows in both cases to derive using $J=J_S$ or $J=J_A$ respectively:
\be
\label{propertyh}
h^\dagger J = J h^T\quad\mathrm{with}\quad h=\exp{ih^a T^2}\,.
\ee

We always use
generators normalized to one:
\be
\langle T^a T^b\rangle
=\langle X^a X^b\rangle
=\delta^{ab}.
\ee
$\langle A\rangle =\mathrm{tr}_F(A)$, is the trace over the flavour indices.
This is over $n$ for the QCD case and $2n$ for the real and pseudo-real case.

During the course of the calculation, we often have to sum over the
Goldstone Bosons. These sums can be easily performed using
\ba
\mathrm{complex:}\qquad&&\nonumber\\
\trf{X^a A X^a B}&=& \trf{A}\trf{B}
 -\frac{1}{n}\trf{AB}\,,
\nonumber\\
\trf{ X^a A}\trf{X^a B}&=&
 \trf{AB}
 -\frac{1}{n}\trf{A}\trf{B}\,.
\nonumber\\
\mathrm{Real:}\qquad&&\nonumber\\
\trf{X^a A X^a B}&=& \frac{1}{2}\trf{A}\trf{B}
 +\frac{1}{2}\trf{AJ_S B^T J_S}-\frac{1}{2n}\trf{AB}\,,
\nonumber\\
\trf{ X^a A}\trf{X^a B}&=&
 \frac{1}{2}\trf{AB}+\frac{1}{2}\trf{AJ_S B^T J_S}
 -\frac{1}{2n}\trf{A}\trf{B}\,.
\nonumber\\
\mathrm{Pseudoreal:}\qquad&&\nonumber\\
\trf{X^a A X^a B}&=& \frac{1}{2}\trf{A}\trf{B}
+\frac{1}{2}\trf{AJ_A B^T J_A} -\frac{1}{2n}\trf{AB}\,,
\nonumber\\
\trf{ X^a A}\trf{X^a B}&=&
 \frac{1}{2}\trf{AB}-\frac{1}{2}\trf{AJ_A B^T J_A}
 -\frac{1}{2n}\trf{A}\trf{B}\,.
\label{tracesum}
\ea

There is a relation that the broken generators
satisfy for the real and pseudo-real case.
\ba
\label{tracerelation}
\trf{X^{a}X^{b}\ldots X^{k}X^{l}} = \trf{X^{l}X^{k}\ldots X^{b}X^{a}}\,.
\ea
The proof for the real case is
\ba
\trf{X^{a}X^{b}\ldots X^{k}X^{l}}&=&\trf{X^{a}X^{b}\ldots X^{k}X^{l}J_S^2}
\nonumber\\
&=&\trf{X^{a}X^{b}\ldots X^{k}J_S X^{lT}J_S}
\nonumber\\
&=&\trf{X^{a}X^{b}\ldots J_SX^{kT} X^{lT}J_S}
\nonumber\\
&=&\trf{J_SX^{aT}X^{bT}\ldots X^{kT} X^{lT}J_S}
\nonumber\\
&=&\trf{X^{aT}X^{bT}\ldots X^{kT} X^{lT}}
\nonumber\\
&=&\trf{\left(X^{l}X^{k}\ldots X^{b} X^{a}\right)^T}
\nonumber\\
&=&\trf{X^{l}X^{k}\ldots X^{b} X^{a}}
\ea
The pseudo-real case is proven by replacing $J_S^2$ by $-J_A^2$ and following
the same steps. (\ref{tracerelation}) is also the reason why the Lagrangian
in \cite{BCE1} is not minimal for the real and pseudo-real case.

In the group theory references there is a conjecture mentioned
that to get from $SO(2n)$ to $Sp(2n)$ it is sufficient to take
$n\to-n$. This feature is indeed visible in most of our formulas.

\subsection{Lagrangians}
\label{lagrangian}

As described in more detail in \cite{paper1} we can write the
Lagrangians in the three cases in a very similar way.
The Goldstone Boson manifold $G/H$ is parametrized by
\be
u = \exp\left(\frac{i}{\sqrt{2}F}\phi\right)\,,
\quad \phi = \phi^a X^a\,.
\ee

These transform under the symmetry transformation in the QCD case
for $g_L\times g_R \in SU(n)_L\times SU(n)_R$ as
\be
\label{defh}
u\to g_R u h(g_L,g_R,\phi)^\dagger =  h(g_L,g_R,\phi) u g_L^\dagger\,.
\ee
$h$ is the socalled compensator field and is defined by (\ref{defh})
and is also an $SU(n)$ matrix. This can be derived from
the standard general formulation \cite{CCWZ} as done in \cite{paper1}.
For a transformation in the conserved part of the group we have that
$g_L=g_R=g_V$ and $h=g_V$.

The notation for the other two cases is directly that of \cite{CCWZ}.
A symmetry transformation $g\in G=SU(2n)$ transforms
$u$ as
\be
u \to g\,u\,h(g,\phi)^\dagger\,,
\quad\mathrm{with}\quad
h = \exp\left(i h^a T^a\right)\,.
\ee
I.e. $h$ is in the unbroken part $H$ of the group.
In case the transformation $g$ is in the conserved part of the group, $g\in H$,
we have that $h=g$.

We can now define the quantities
\ba
u_\mu &=&i \left(u^\dagger \partial_\mu u-u\partial_\mu u^\dagger\right)\,,
\nonumber\\
\Gamma_\mu &=&
\frac{1}{2} \left(u^\dagger \partial_\mu u-u\partial_\mu u^\dagger\right)\,.
\ea
Under the group transformation in all cases we have $u_\mu\to h u_\mu h^\dagger$
and $\Gamma_\mu$ can be used to define a covariant derivative.
\be
\nabla_\mu u_\nu\equiv \partial_\mu u_\nu+\Gamma_\mu u_\nu-u_\nu \Gamma_\mu
\to h\nabla_\mu u_\nu h^\dagger\,.
\ee
In \cite{paper1} we also showed how the external fields can be included
in a similar way as for the QCD case in \cite{GL1,GL2}.
In particular the quark masses can be put in a quantity $\chi_\pm$
that transforms as $\chi_\pm\to h\chi_\pm h^\dagger$.

The lowest order Lagrangian takes on the standard form
\be
\label{LOlag}
\mathcal{L}_{LO} = \frac{F^2}{4}\trf{u_\mu u^\mu+\chi_+}
\ee
for all three cases and the same is true for the NLO Lagrangian.
\ba
\label{NLOlag}
\mathcal{L}_{NLO} &=&
L_0 \langle u^\mu u^\nu u_\mu u_\nu \rangle
+L_1 \langle u^\mu u_\mu\rangle\langle u^\nu u_\nu \rangle
+L_2 \langle u^\mu u^\nu\rangle\langle u_\mu u_\nu \rangle
\nonumber\\&&
+L_3 \langle u^\mu u_\mu u^\nu u_\nu \rangle
+L_4  \langle u^\mu u_\mu\rangle\langle\chi_+\rangle
+L_5  \langle u^\mu u_\mu\chi_+\rangle
+L_6 \langle\chi_+\rangle^2
\nonumber\\&&
+L_7 \langle\chi_-\rangle^2
+\frac{1}{2} L_8 \langle\chi_+^2+\chi_-^2\rangle\,.
\ea
We have kept only the terms contributing to meson-meson scattering in
(\ref{NLOlag}).

The NNLO Lagrangian is known for the complex or QCD case \cite{BCE1}
as well as its divergence structure \cite{BCE2}. The same Lagrangian with the
changes mentioned above is complete for the other two cases but probably
not minimal. We have nonetheless chosen to leave the contributions from
those terms in the results quoted here.

\subsection{Renormalization}
\label{renormalization}

We use the standard renormalization procedure in ChPT \cite{GL1,GL2}
with the extension to NNLO described in great detail in \cite{BCEGS2,BCE2}.
The divergences at NLO are canceled by the subtractions as calculated
in \cite{paper1}. At NNLO the divergences for the QCD case are
canceled by the subtractions calculated in \cite{BCE2}. The other two
cases satisfy all the expected constraints.
Nonlocal divergences fully cancel, the $\epsilon$ parts of the loop
integrals as defined in App.~\ref{loopintegrals}
always cancel and the double divergences satisfy the
Weinberg relations \cite{BCE2}.

As usual in ChPT we apply the $\overline{\mathrm{M}S}$
scheme of dimensional regularization,
in which the bare LECs $L_i$ are defined as
\be
\label{defLir}
L_i = \left(c \mu\right)^{d-4}
\left[\Gamma_i\Lambda+L_i^r(\mu)\right]\,
\ee
Where the dimension $d=4-2\epsilon$, and
\ba
\Lambda &=& {1\over16\pi^2(d-4)}\ ,\\
\ln c &=& -{1\over2}[\ln 4\pi+\Gamma^\prime(1)+1]\ .
\ea
The coefficients $\Gamma_i$ can be found in \cite{GL1,paper1} for
the complex and in \cite{paper1} for the real and pseudo-real$Sp(2n)$ case.

The NNLO terms can be made finite with the subtractions
\ba
K_i = \left(c\mu\right)^{2(d-4)}
\left[K^r_i-\Gamma_i^{(2)}\Lambda^2
-\left(\frac{1}{16\pi^2}\Gamma_i^{(1)}+\Gamma_i^{(L)}
\right)\Lambda
\right]\,.
\ea
The coefficients $\Gamma_i^{(2)}$, $\Gamma_i^{(1)}$ and
$ \Gamma_i^{(L)}$ for the complex case have been derived in \cite{BCE2}.
For the real and pseudo-real case, the results do not exist.
We have checked that all remaining divergences are local and can thus
be subtracted.

\section{General results for the amplitudes}
\label{groups}

\subsection{$\pi\pi$ case}

The $\pi\pi$ scattering amplitude, which correspond to the QCD case with $n=2$
is well known.
Due to crossing and the possible $SU(2)$ (isospin) invariants the
amplitude can be written as \cite{Weinbergpipi,ChewMandelstampipi}
\be
\label{defAstu}
M_{\pi\pi}(s,t,u)= \delta^{ab}\delta^{cd} A(s,t,u)
+\delta^{ac}\delta^{bd} A(t,u,s) +\delta^{ad}\delta^{bc} A(u,s,t)\,.
\ee
The function $A(s,t,u)$ is symmetric under the interchange of $t$ and $u$.

The possible states of two pions are isospin 0, 1 or 2.
The amplitude for the three channels are given by
\cite{ChewMandelstampipi}
\ba
T^0(s,t,u) &=& 3A(s,t,u) + A(t,s,u) + A(u,t,s),
\nonumber\\
T^1(s,t,u) &=& A(t,s,u) - A(u,s,t),
\nonumber\\
T^2(s,t,u) &=& A(t,s,u) + A(u,s,t).
\ea
Where $I$ is isospin, and  $P_I$ is the projection operator on isospin $I$
They satisfy the relation
\be
\label{decomposition2}
M_{\pi\pi}(s,t,u)=\sum_{I=0,2} T^I(s,t,u) P_I.
\ee
and
\be
\label{useprojection}
T^I(s,t,u) P_I~(\mathrm{no~sum}) = P_I M_{\pi\pi}(s,t,u)\,.
\ee

In the remainder of this section we will generalize these results.
(\ref{defAstu}) is generalized in terms of two functions in
Sect.~\ref{general}. The possible intermediate states and the corresponding
amplitudes are derived for the three cases separately in the last three
subsections of this section.

\subsection{General amplitude}
\label{general}

The amplitude for meson-meson scattering is given by
\be
\langle \phi^c(p_c)\phi^d(p_d)|\phi^a(p_a)\phi^b(p_d)\rangle
= M(s,t,u)\,.
\ee
The Mandelstam variables $s,t,u$ are defined by
\be
s = (p_a+p_b)^2/\mphys\,,\qquad  t = (p_a-p_c)^2/\mphys\,,
\qquad  u = (p_a-p_d)^2/\mphys\,.
\ee
These satisfy
\be
\label{relstu}
 s+ t+ u = 4\,.
\ee
We have chosen here to use the dimensionless
versions in order to simplify later formulas.

The flavour structure of the amplitude for meson-meson scattering can be
described by constructing all possible invariants from the four
corresponding generator matrices $X^e$, $e=a,b,c,d$.
Taking into account that $\trf{X^e}=0$ there are
9 invariants possible
\ba
\label{invariants}
\trf{X^a X^b X^c X^d},&&
\trf{X^a X^c X^d X^b},\quad
\trf{X^a X^d X^b X^c},
\nonumber\\
\trf{X^a X^d X^c X^b},&&
\trf{X^a X^b X^d X^c},\quad
\trf{X^a X^c X^b X^d},
\nonumber\\
\trf{X^aX^b}\trf{X^cX^d},&&
\trf{X^aX^c}\trf{X^bX^d},\quad
\trf{X^aX^d}\trf{X^bX^c}\,.
\ea
Under charge conjugation $X^a\longrightarrow X^{aT}$.
This means that the amplitudes
multiplying the first row in (\ref{invariants}) must be the same as
those multiplying the second row.\footnote{Alternatively use
(\ref{tracerelation}) for the real and pseudo-real case.}
As a result the full amplitude can
be written in terms of two invariant amplitudes $B(s,t,u)$ and $C(s,t,u)$.
\ba
\label{defBC}
M(s,t,u)&=& \left[\trf{X^a X^b X^c
X^d}+\trf{X^a X^d X^c X^b}\right] B(s,t,u)
\nonumber\\&&
+\left[\trf{X^a X^c X^d X^b}+\trf{X^a X^b X^d X^c}\right] B(t,u,s)
\nonumber\\&&
+\left[\trf{X^a X^d X^b X^c}+\trf{X^a X^c X^b X^d}\right] B(u,s,t)
\nonumber\\&&
+\delta^{ab}\delta^{cd} C(s,t,u)+\delta^{ac}\delta^{bd} C(t,u,s)
+\delta^{ad}\delta^{bc} C(u,s,t)\,.
\ea
The flavour structure also implies that
\be
\label{BCsymmetry}
B(s,t,u) = B(u,t,s)\,\qquad C(s,t,u) = C(s,u,t)\,.
\ee
For $n=3$ there is the Cayley-Hamilton relation
\be
\sum_{\rm 6~permutations}\trf{X^a X^b X^c X^d} = \sum_{\rm 3~permutations}
\trf{X^a X^b}\trf{X^c X^d}\,,
\ee
which allows for an ambiguity in the split of $B$ and $C$.
For $n=2$ we can perform all the traces with four matrices $X^a$ in terms of
Kronecker deltas.

The relation between the general amplitudes and the $\pi\pi$
scattering case (\ref{defAstu}) is
\be
\label{relABC} A(s,t,u) = C(s,t,u) +
B(s,t,u)+B(t,u,s)-B(u,s,t)\,.
\ee Note that the property
(\ref{BCsymmetry}) insures that $A(s,t,u)$ is symmetric under the
interchange of $t$ and $u$ as it should be. The form (\ref{defAstu})
holds also for any set of pions. I.e., taking any $SU(2)$ subgroup
of the unbroken group and any three of the pseudo Goldstone bosons
that form a triplet under such a group, one can rewrite the general
amplitude (\ref{defBC}) into (\ref{defAstu}) using (\ref{relABC}).

\subsection{QCD case: channels and amplitudes}

The Goldstone boson transform under the conserved part of the group, $SU(n)$,
as
\be
\phi\to h \phi h^\dagger\,.
\ee
This means that they are in the adjoint representation of $SU(n)$.
For the $n=2,3$ case we have an isospin triplet under $SU(2)$ and an octet
under $SU(3)$. The intermediate states for these are well known:
\ba
SU(2):~~3\otimes3&=&
 1\oplus3\oplus5~(\mathrm{or}~I=0,1,2)\,,
\nonumber\\
SU(3):~~8\otimes8&=&
 1\oplus 8_S\oplus 8_A\oplus 10\oplus\overline{10}\oplus27\,.
\ea
The group theory for $SU(n)$ can be done in many ways. One is via
Young diagrams and the second using tensor methods. We will do both.
The $SU(n)$ case derived here was in fact known \cite{Neville} and our
results are in agreement with his. Young diagrams for $SU(n)$ are explained
in \cite{PDG} page 370.
The Young diagrams for $SU(n)$ give
\be
\label{youngSUN}
\!\raisebox{-52pt}{
\setlength{\unitlength}{1pt}
\begin{picture}(20,60)
\SetScale{1.0}
\newcommand{\youngbox}[1]
{\SetScaledOffset(#1)\EBox(0.,0.)(10.,10.)\SetScaledOffset(0.,0.)}
\youngbox{0,0}
\youngbox{0,10}
\Text(5,32)[]{$\displaystyle\vdots$}
\youngbox{0,40}
\youngbox{0,50}
\youngbox{10,50}
\end{picture}}
~\otimes\!
\raisebox{-52pt}{
\setlength{\unitlength}{1pt}
\begin{picture}(20,60)
\SetScale{1.0}
\newcommand{\youngbox}[1]
{\SetScaledOffset(#1)\EBox(0.,0.)(10.,10.)\SetScaledOffset(0.,0.)}
\youngbox{0,0}
\youngbox{0,10}
\Text(5,32)[]{$\displaystyle\vdots$}
\youngbox{0,40}
\youngbox{0,50}
\youngbox{10,50}
\end{picture}}
\,\,=
\,\,\cdot
~\oplus\!
\raisebox{-52pt}{
\setlength{\unitlength}{1pt}
\begin{picture}(20,60)
\SetScale{1.0}
\newcommand{\youngbox}[1]
{\SetScaledOffset(#1)\EBox(0.,0.)(10.,10.)\SetScaledOffset(0.,0.)}
\youngbox{0,0}
\youngbox{0,10}
\Text(5,32)[]{$\displaystyle\vdots$}
\youngbox{0,40}
\youngbox{0,50}
\youngbox{10,50}
\end{picture}}
~\oplus\!
\raisebox{-52pt}{
\setlength{\unitlength}{1pt}
\begin{picture}(20,60)
\SetScale{1.0}
\newcommand{\youngbox}[1]
{\SetScaledOffset(#1)\EBox(0.,0.)(10.,10.)\SetScaledOffset(0.,0.)}
\youngbox{0,0}
\youngbox{0,10}
\Text(5,32)[]{$\displaystyle\vdots$}
\youngbox{0,40}
\youngbox{0,50}
\youngbox{10,50}
\end{picture}}
~\oplus\!
\raisebox{-52pt}{
\setlength{\unitlength}{1pt}
\begin{picture}(30,60)
\SetScale{1.0}
\newcommand{\youngbox}[1]
{\SetScaledOffset(#1)\EBox(0.,0.)(10.,10.)\SetScaledOffset(0.,0.)}
\youngbox{0,10}
\Text(5,32)[]{$\displaystyle\vdots$}
\youngbox{0,40}
\youngbox{0,50}
\youngbox{10,50}
\youngbox{20,50}
\end{picture}}
~\oplus\!
\raisebox{-52pt}{
\setlength{\unitlength}{1pt}
\begin{picture}(30,60)
\SetScale{1.0}
\newcommand{\youngbox}[1]
{\SetScaledOffset(#1)\EBox(0.,0.)(10.,10.)\SetScaledOffset(0.,0.)}
\youngbox{0,0}
\youngbox{0,10}
\Text(5,32)[]{$\displaystyle\vdots$}
\youngbox{0,40}
\youngbox{0,50}
\youngbox{10,0}
\youngbox{10,10}
\Text(15,32)[]{$\displaystyle\vdots$}
\youngbox{10,40}
\youngbox{10,50}
\youngbox{20,50}
\youngbox{20,40}
\end{picture}}
~\oplus\!
\raisebox{-52pt}{
\setlength{\unitlength}{1pt}
\begin{picture}(20,60)
\SetScale{1.0}
\newcommand{\youngbox}[1]
{\SetScaledOffset(#1)\EBox(0.,0.)(10.,10.)\SetScaledOffset(0.,0.)}
\youngbox{0,10}
\Text(5,32)[]{$\displaystyle\vdots$}
\youngbox{0,40}
\youngbox{0,50}
\youngbox{10,40}
\youngbox{10,50}
\end{picture}}
~\oplus\!
\raisebox{-52pt}{
\setlength{\unitlength}{1pt}
\begin{picture}(40,60)
\SetScale{1.0}
\newcommand{\youngbox}[1]
{\SetScaledOffset(#1)\EBox(0.,0.)(10.,10.)\SetScaledOffset(0.,0.)}
\youngbox{0,0}
\youngbox{0,10}
\Text(5,32)[]{$\displaystyle\vdots$}
\youngbox{0,40}
\youngbox{0,50}
\youngbox{10,0}
\youngbox{10,10}
\Text(15,32)[]{$\displaystyle\vdots$}
\youngbox{10,40}
\youngbox{10,50}
\youngbox{20,50}
\youngbox{30,50}
\end{picture}}
\ee
Note that the $\vdots$ stand for $n-5$ boxes.
In terms of free indices the right hand side of (\ref{youngSUN}) is
no indices (singlet), twice one upper and one lower index (adjoint).
The remaining four have all two lower and two upper indices, where the upper
indices are produced from the columns with length $n-1$ and $n-2$ boxes using
the Levi-Civita tensor $\epsilon^{i_1\ldots i_n}$. For these
we use the notation $R_X^Y$ where $X=S,A$ indicate whether the lower
indices are symmetric or antisymmetric and $Y=S,A$ the same for the upper
indices.
The decomposition (\ref{youngSUN}) can thus be written as
\be
\label{decompositionSUN}
{Adj.}\otimes {Adj.} =
R_I \oplus R_S \oplus R_A \oplus+ R^{\ A}_S \oplus R^{\ S}_A
\oplus R^{\ A}_A \oplus R^{\ S}_S
\ee
The order of the irreducible representation on the right-hand side is
the same in (\ref{youngSUN}) and (\ref{decompositionSUN}).

If we have a particular two-meson state $\phi^a(p_1)\phi^b(p_2)$
we have to write it in terms of states that belong to the irreducible
multiplets to obtain the amplitudes for the different channels.
We use a simplified notation below with $A=X^a, B=X^b, C=X^c$
and $D=X^d$ for many of the terms. All traces connecting a lower with an
upper index must vanish.

\begin{itemize}
\item
$R_I$: singlet representation. All indices should be contracted,
so this must be proportional to
$A_i^j B_j^i=\trf{AB}$. Summing over the $n^2-1$
states that are present tells us that the correct normalized state is
\be
R_I = {1\over \sqrt{n^2-1}}\sum_{a,b}\trf{X^aX^b}\phi^a\phi^b\,.
\ee
A projection operator $P_I^{abcd}$ that projects on the singlet component
is
\be
\label{PISUN}
P_I^{abcd} = \frac{1}{n^2-1}\trf{AB}\trf{CD}\,.
\ee
It can be checked that this is a projection operator
\be
\label{PI2}
P_I^{abcd}P_I^{cdef} = P_I^{abef}\,,
\ee
using (\ref{tracesum}) and $\sum_{a,b}\delta_{ab} = n^2-1$.

\item
$R_S$: adjoint symmetric representation. This needs in the end an upper
and a lower index and must be traceless. We choose here to split
up the two possible contractions of the indices of $A$ and $B$ in way that
is symmetric under the interchange of $A$ and $B$.
\be
\label{RSSUN}
\left(R_S\right)_j^{i} =
\sqrt{n\over2(n^2-4)}\, \Bigg[ A^{m}_j B^i_m +
B^{m}_j A^i_m -\frac{2}{n} \delta^{i}_j \trf{AB}\!\Bigg]\,.
\ee
The last term is needed to make $R_S$ traceless.
The normalization can be worked out by checking the normalization
of a particular state or by checking that the projection operator\footnote{We
suppress the superscript $abcd$ from now on.}
\be
P_S = R_S(A,B)_j^i\,R_S(C,D)_i^j
\ee
has the correct normalization, its square is equal to itself.
This leads finally to the projection operator
\be
\label{PSSUN}
P_S = \frac{n}{2(n^2-4)}\left[\trf{(AB+CD)(CD+DC)}
-\frac{4}{n}\trf{AB}\trf{CD}\right]\,.
\ee
Suppressing the indices we get similar to (\ref{PI2}) $P_S^2=P_S$
and $P_S P_I = P_I P_S=0$.

\item
$R_A$: adjoint anti-symmetric representation
\be
\left(R_A\right)_j^i = \frac{1}{\sqrt{2n}}
\left( A^{m}_jB^i_{\ m}
       - B^{m}_j A^i_{m}\right)\,.
\ee
$R_A(A,B)$ is traceless and antisymmetric in $A$ and $B$.
The projection operator corresponding to this is
\be
\label{PASUN}
P_A = \frac{-1}{2n}\trf{(AB-BA)(CD-DC)}\,.
\ee

\item
$ R^{A}_S$: symmetric for lower indices
and antisymmetric for upper.
A state here corresponds to
\ba
\left(R^{A}_S\right)_{kl}^{ij} &=&
\frac{1}{2}\left[
 A^{i}_k B^{j}_l
  + {1\over n}\delta^i_k \left(A^j_m B^m_l-A^m_l B^j_m\right)  \right]
\nonumber\\&&
   - (i\leftrightarrow j) + (k\leftrightarrow l)
   - (i\leftrightarrow j,k\leftrightarrow l)\,.
\ea
The projection operator on this type of states is
\ba
\label{PSASUN}
P_{SA} &=& \left(R_S^A(A,B)\right)_{kl}^{ij}\left(R_S^A(C,D)\right)_{ij}^{kl}
\nonumber\\
&=&\frac{1}{4n}\Big[\trf{(AB-BA)(CD-DC)}+n\Big(\trf{ACBD}
\nonumber\\&&
-\trf{ADBC}\Big)+n\Big(\trf{AC}\trf{BD}-\trf{AD}\trf{CD}\Big)\Big]\,.
\ea

\item
$ R^{S}_A$: antisymmetric for lower indices
and symmetric for upper.
A state here corresponds to
\ba
\left(R_{A}^S\right)_{kl}^{ij} &=&
\frac{1}{2}\left[
 A^{i}_k B^{j}_l
  - {1\over n}\delta^i_k \left(A^j_m B^m_l-A^m_l B^j_m\right)  \right]
\nonumber\\&&
   + (i\leftrightarrow j) - (k\leftrightarrow l)
   - (i\leftrightarrow j,k\leftrightarrow l)\,.
\ea
The projection operator on this type of states is
\ba
\label{PASSUN}
P_{AS} &=&
\frac{1}{4n}\Big[\trf{(AB-BA)(CD-DC)}-n\Big(\trf{ACBD}
\nonumber\\&&
-\trf{ADBC}\Big)+n\Big(\trf{AC}\trf{BD}-\trf{AD}\trf{CD}\Big)\Big]\,.
\ea

\item
$ R^{S}_S$: symmetric for both upper index and lower index.
The states are
\ba
(R^{S}_S)_{kl}^{ij} &=&
\frac{1}{2}\Bigg[ A^{i}_k B^{j}_l
- \frac{1}{n+2}\delta^{i}_k
\Big( A^{m}_l B^j_m + B^{ m}_l A^j_m \Big)
\nonumber\\&&
+ \frac{1}{(n+1)(n+2)}\delta^{i}_k \delta^{j}_l
\trf{AB} \Bigg]
\nonumber\\&&
 + (i\leftrightarrow j) + (k\leftrightarrow l) + (i\leftrightarrow j,k\leftrightarrow l)
\ea
and the projection operator is
\ba
\label{PSSSUN}
P_{SS}&=&
\frac{-1}{4(n+2)}\trf{(AB+BA)(CD+DC)}
+\frac{1}{4}\Big(\trf{ACBD}
\nonumber\\&&
+\trf{ADBC}\Big)
+\frac{1}{4}\Big(\trf{AC}\trf{BD}+\trf{AD}\trf{CD}\Big)
\nonumber\\&&
+\frac{1}{2(n+1)(n+2)}\trf{AB}\trf{CD}\,.
\ea

\item
$ R^{A}_A$: antisymmetric for both upper index and lower index.
The states are
\ba
(R^{A}_A)_{kl}^{ij} &=&
\frac{1}{2}\Bigg[ A^{i}_k B^{j}_l
+ \frac{1}{n-2}\delta^{i}_k
\Big(A^{m}_l B^j_m + B^{ m}_l A^j_m \Big)
\nonumber\\&&
- \frac{1}{(n-1)(n-2)}\delta^{i}_k \delta^{j}_l
\trf{AB} \Bigg]
\nonumber\\&&
 - (i\leftrightarrow j) - (k\leftrightarrow l) + (i\leftrightarrow j,k\leftrightarrow l)
\ea
and the projection operator is
\ba
\label{PAASUN}
P_{AA}&=&
\frac{-1}{4(n-2)}\trf{(AB+BA)(CD+DC)}
-\frac{1}{4}\Big(\trf{ACBD}
\nonumber\\&&
+\trf{ADBC}\Big)
+\frac{1}{4}\Big(\trf{AC}\trf{BD}+\trf{AD}\trf{CD}\Big)
\nonumber\\&&
+\frac{1}{2(n-1)(n-2)}\trf{AB}\trf{CD}\,.
\ea
\end{itemize}
The projection operators agree with those of \cite{Neville} and using
(\ref{tracerelation}) can be shown to satisfy $P_r P_{r'} = P_r\delta_{rr'}$
for $r$ the various representations. One last check is that
\be
\sum_r P_r = \trf{AC}\trf{BD}\,.
\ee
The right-hand side is the unit operator when acting on the product of two states
in the adjoint representation. It can also be seen that $R_I, R_S,R_A^A$
and $R_S^S$ are symmetric under interchanging $A$ and $B$ while $R_A,R_A^S$
and $R_S^A$ are antisymmetric.

The amplitude in the different intermediate states can now be extracted from
the general amplitude in two equivalent ways.
We can pick a state $R_r$ in a representation $r$ and get it via
\be
\label{stateamplitude}
T_r = \langle R_r|M(s,t,u)|R_r\rangle\,,
\ee
or apply the projection operators on the full amplitude
with
\be
\label{projectamplitude}
P_r T_r = P_r M(s,t,u)\,.
\ee
Both methods give as expected the same result but the second one is
much easier to apply. For the first method it is best to choose
a state where the terms with $\delta$ functions are not present.
E.g. for the four index representations
take a state $R_r$ with $i=1, j=2, k=3, l=4$.
For evaluating (\ref{projectamplitude}) one can use (\ref{tracesum}).

\ba
\label{TISUN}
T_I &=&  2\left(n-{1\over n}\right) [B(s,t,u) + B(t,u,s)]- {2\over n}B(u,s,t)
\nonumber\\
&& + (n^2-1) C(s,t,u) + C(t,u,s) +  C(u,s,t)\ ,
\nonumber\\
T_S &=&  \left(n - {4\over n}\right)[B(s,t,u) + B(t,u,s)]- {4\over n}B(u,s,t)
\nonumber\\&&
+ C(t,u,s) + C(u,s,t)\ ,
\nonumber\\
T_A &=& n[-B(s,t,u)+B(t,u,s)]+C(t,u,s)-C(u,s,t) \ ,
\nonumber\\
T_{SA} &=& C(t,u,s)- C(u,s,t) \ ,
\nonumber\\
T_{AS} &=& C(t,u,s) - C(u,s,t)\ ,
\nonumber\\
T_{SS}&=& 2B(u,s,t) + C(t,u,s) + C(u,s,t)\ ,
\nonumber\\
T_{AA}&=& -2B(u,s,t) + C(t,u,s) + C(u,s,t)\ .
\ea
They satisfy the relation similar to (\ref{decomposition2}),
\be
\label{decompositionN}
M(s,t,u)=\sum_r T_r(s,t,u) P_r\ .
\ee

One also notices that $T_{SA}=T_{AS}$ in general from (\ref{TISUN}).

\subsection{Real case: channels and amplitudes}

In this subsection we work out the possible two meson intermediate states for
the case of $SU(2n)/SO(2n)$. One problem is that the mesons transform under
$SO(2n)$ as $\phi\to h\phi h^\dagger$.
The matrices $h\in SO(2n)$ in the embedding introduced here do not simply
satisfy $h h^T=1$ either. In our case the $SO(2n)$ is instead defined as
$h J_S h^T = J_S$. The easiest way to obtain objects that appear in the usual
way is to note that using (\ref{propertyh})
\be
\phi J_S \to h\phi h^\dagger J_S = h \phi J_S h^T
\ee
and that an invariant trace on these object needs an extra factor of $J_S$.
E.g.
\be
(\phi^a J_S) J_S (\phi^b J_S) \to
h(\phi^a J_S)h^T J_S h(\phi^b J_S)h^T = h(\phi^a J_S)J_S(\phi^b J_S)h^T\,.
\ee
Keeping that in mind we can use the standard way of dealing with
$SO(2n)$. Note that $(\phi J_S)^T = J_S \phi^T = \phi J_S$ so the
Goldstone bosons live in the symmetric representation of $SO(2n)$.

The method of Young tableaux has been generalized to $SO(2n)$ \cite{youngSON}.
Putting together two symmetric representations gives
\be
\label{youngSON}
\!\raisebox{-12pt}{
\setlength{\unitlength}{1pt}
\begin{picture}(20,20)
\SetScale{1.0}
\newcommand{\youngbox}[1]
{\SetScaledOffset(#1)\EBox(0.,0.)(10.,10.)\SetScaledOffset(0.,0.)}
\youngbox{0,10}
\youngbox{10,10}
\end{picture}}
~\otimes\!
\raisebox{-12pt}{
\setlength{\unitlength}{1pt}
\begin{picture}(20,20)
\SetScale{1.0}
\newcommand{\youngbox}[1]
{\SetScaledOffset(#1)\EBox(0.,0.)(10.,10.)\SetScaledOffset(0.,0.)}
\youngbox{0,10}
\youngbox{10,10}
\end{picture}}
\,\,=
\,\,\cdot
~\oplus\!
\raisebox{-12pt}{
\setlength{\unitlength}{1pt}
\begin{picture}(10,20)
\SetScale{1.0}
\newcommand{\youngbox}[1]
{\SetScaledOffset(#1)\EBox(0.,0.)(10.,10.)\SetScaledOffset(0.,0.)}
\youngbox{0,0}
\youngbox{0,10}
\end{picture}}
~\oplus\!
\raisebox{-12pt}{
\setlength{\unitlength}{1pt}
\begin{picture}(20,20)
\SetScale{1.0}
\newcommand{\youngbox}[1]
{\SetScaledOffset(#1)\EBox(0.,0.)(10.,10.)\SetScaledOffset(0.,0.)}
\youngbox{0,10}
\youngbox{10,10}
\end{picture}}
~\oplus\!
\raisebox{-12pt}{
\setlength{\unitlength}{1pt}
\begin{picture}(40,20)
\SetScale{1.0}
\newcommand{\youngbox}[1]
{\SetScaledOffset(#1)\EBox(0.,0.)(10.,10.)\SetScaledOffset(0.,0.)}
\youngbox{0,10}
\youngbox{10,10}
\youngbox{20,10}
\youngbox{30,10}
\end{picture}}
~\oplus\!
\raisebox{-12pt}{
\setlength{\unitlength}{1pt}
\begin{picture}(30,20)
\SetScale{1.0}
\newcommand{\youngbox}[1]
{\SetScaledOffset(#1)\EBox(0.,0.)(10.,10.)\SetScaledOffset(0.,0.)}
\youngbox{0,0}
\youngbox{0,10}
\youngbox{10,10}
\youngbox{20,10}
\end{picture}}
~\oplus\!
\raisebox{-12pt}{
\setlength{\unitlength}{1pt}
\begin{picture}(20,20)
\SetScale{1.0}
\newcommand{\youngbox}[1]
{\SetScaledOffset(#1)\EBox(0.,0.)(10.,10.)\SetScaledOffset(0.,0.)}
\youngbox{0,0}
\youngbox{10,0}
\youngbox{0,10}
\youngbox{10,10}
\end{picture}}
\ee
We can write this in the form
\be
Sym.\otimes\!Sym. =
R_I \oplus R_A \oplus R_S \oplus R_{FS} \oplus R_{MA} \oplus R_{MS}\,.
\ee
The states are given
in Tab.~\ref{tabSONstates}.
They are made traceless but remember that indices of $\phi^a J_S$
and $\phi^b J_S$ are always contracted with $J_S$.

As an example the singlet representation $R_I$ is proportional to
$$
(\phi^a J_S)_{ij} J_{Sik} J_{Sjl} (\phi^b J_S)kl = \trf{\phi^a\phi^b}\,.
$$
The two-index
antisymmetric representation $R_A$ is, now using $A,B$ for $\phi^a,\phi^b$,
\be
(AJ_S)_{ik}(BJ_S)_{jl} J_{Skl}-(AJ_S)_{jk}(BJ_S)_{il} J_{Skl}=
(ABJ-BAJ)_{ij}
\ee
where we heavily used $J_S^2=1$ and the fact that $AJ_S$ and $BJ_S$ are
symmetric matrices. One can also easily check that the trace
$(ABJ_S-BAJ_S)_{ij} J_{Sij}$ vanishes.

Then $R_A$ and $R_S$ are antisymmetric respectively symmetric both under
$i\leftrightarrow j$ and $A\leftrightarrow B$.
The remaining ones are the four-index representations.
$R_{FS}$ is fully symmetric in all indices and under $A\leftrightarrow B$.
The two remaining representations have a mixed symmetry in the indices
but are antisymmetric respectively symmetric under $A\leftrightarrow B$.
\begin{table}
\begin{center}
\begin{tabular}{|c|c|}
\hline
\rule{0cm}{1.3em}$R_I$       & $\frac{1}{\sqrt{(2n-1)(n+1)}} \trf{AB}$\\[3mm]
\hline
\rule{0cm}{1.3em}$R_A$       & $(ABJ -BAJ)_{ij}$ \\[1mm]
\hline
\rule{0cm}{1.3em}$R_S$       & $(ABJ +BAJ)_{ij}- {1\over n} J_{ij} \la AB\ra$
\\[1mm]
\hline
\rule{0cm}{1.3em}$R_{FS}$    & $ (AJ)_{ij}\ (BJ)_{kl}
 -{1\over n+2} \Big[J_{ij}(ABJ + BAJ)_{kl}\Big]+{1\over2 (n+2)(n+1)} J_{ij} J_{kl} \la AB \ra$\\
 & $+(ijkl\leftrightarrow ikjl)
   +(ijkl\leftrightarrow iljk)
  +(ijkl\leftrightarrow klij)
$\\&$
  +(ijkl\leftrightarrow jlik)
  +(ijkl\leftrightarrow jkil)$
\\[2mm]
\hline
\rule{0cm}{1.3em}$R_{MA}$       & $(AJ)_{ij}\ (BJ)_{kl}-(AJ)_{kl}\ (BJ)_{ij}
 - {1\over 2n+2} \Big[J_{ik}(ABJ - BAJ)_{jl}$
\\&$+  J_{jk}(ABJ - BAJ)_{il}
+J_{il}(ABJ - BAJ)_{jk}+  J_{jl}(ABJ - BAJ)_{ik}\Big]$
\\[1mm]
\hline
\rule{0cm}{1.3em}$R_{MS}$       &
 $(AJ)_{ij}\ (BJ)_{kl}+ (AJ)_{kl}\ (BJ)_{ij}- (AJ)_{ik}\ (BJ)_{jl} - (AJ)_{jl}\ (BJ)_{ik} $   \\
            & $+ {1\over 2(n-1)}\Big[J_{ij}(ABJ+BAJ)_{kl} + J_{kl} (ABJ+BAJ)_{ij}$\\
            & \qquad\qquad$- J_{ik}(ABJ+BAJ)_{jl} - J_{jl} (ABJ+BAJ)_{ik} \Big]$   \\
            & $-{1\over (n-1)(2n-1)} \Big( J_{ij} J_{kl} - J_{ik} J_{jl} \Big) \la AB\ra$
\\[2mm]
\hline
\end{tabular}
\end{center}
\caption{The intermediate states for the real or adjoint case, $SU(2n)/SO(2n)$.
The notations $A,B$ stands for $\phi^a$ and $\phi^b$ and $J$ is
$J_S$ everywhere.}
\label{tabSONstates}
\end{table}

The corresponding projection operators are given in Tab.~\ref{tabSONproj}.
These can be obtained by contracting the indices with $J_S$
of the states once with $A,B$ and once with $C,D$.

\begin{table}
\label{projection operator}
\begin{center}
\begin{tabular}{|c|c|}
\hline
\rule{0cm}{1.3em}$P_I$       & ${1\over (2n-1)(n+1)} \la AB\ra \la CD\ra$
\\[3mm]
\hline
\rule{0cm}{1.3em}$P_A$       & $ -{1\over 2(n+1)} \la (AB-BA)(CD-DC)\ra$
\\[3mm]
\hline
\rule{0cm}{1.3em}$P_S$       & ${n\over  2(n-1)(n+2)} \Big(\la(AB+BA)(CD+DC)\ra - {2\over n}\la AB\ra\la CD\ra\Big)$
\\[3mm]
\hline
$P_{FS}$       & ${1\over6} \Big[{2\over (n+1)(n+2)}\la AB\ra\la CD\ra + \la AC\ra\la BD\ra + \la AD\ra \la BC\ra$\\
            & \qquad\qquad $ +2 \la ACBD+ADBC \ra - {2\over n+2}\la(AB+BA)(CD+DC)\ra \Big]$
\\[3mm]
\hline
\rule{0cm}{1.3em}$P_{MA}$    & $
    {1\over 2(n+1)}\la (AB-BA)(CD-DC)\ra + {1\over 2} (\la AC\ra\la BD\ra-\la AD\ra\la BC\ra )$
\\[3mm]
\hline
\rule{0cm}{1.3em}$P_{MS}$    & ${1\over6}\Big[{2\over(n-1)(2n-1)}\la AB\ra\la CD\ra + 2\Big(\la AC \ra \la BD\ra +\la AD \ra \la BC\ra\Big) $ \\
            & $-2\la ADBC+ACBD\ra-{1\over n-1}\la(AB+BA)(CD+DC)\ra\Big]$\\[3mm]
\hline
\end{tabular}
\end{center}
\caption{The projection operator for the different intermediate states
for the real or adjoint case, $SU(2n)/SO(2n)$.}
\label{tabSONproj}
\end{table}

We can now use Tabs.~\ref{tabSONstates} and \ref{tabSONproj}
to project out the amplitudes in the various channels
using (\ref{stateamplitude}) or (\ref{projectamplitude}).
The results are
\ba
\label{TISON}
T_I &=& {1\over n}(2n-1)(n+1)[B(s,t,u) + B(t,u,s)] + {1\over n}(n-1)B(u,s,t)
\nonumber\\&&
 +(2n-1)(n+1)C(s,t,u) + C(t,u,s) + C(u,s,t)\,,
\nonumber\\
T_A &=& - (1+n)[B(s,t,u) - B(t,u,s)]+ C(t,u,s) - C(u,s,t)\,,
\nonumber\\
T_S &=&{1\over n}(n-1)(n+2)[B(s,t,u)+B(t,u,s)] + {1\over n}(n-2)B(u,s,t)
\nonumber\\&&
             +C(t,u,s) + C(u,s,t)\,,
\nonumber\\
T_{FS} &=& 2B(u,s,t)+ C(t,u,s) + C(u,s,t)\,,
\nonumber\\
T_{MA} &=& C(t,u,s) - C(u,s,t)\,,
\nonumber\\
T_{MS} &=& -B(u,s,t)+C(t,u,s) + C(u,s,t)\,.
\ea

\subsection{Pseudo-real case: channels and amplitudes}

In this subsection we work out the possible two meson intermediate states for
the case of $SU(2n)/Sp(2n)$. One problem is that the mesons transform under
$Sp(2n)$ as in $\phi\to h\phi h^\dagger$.
The easiest way to obtain objects that appear in the usual
way is to note that using (\ref{propertyh})
\be
\phi J_A \to h\phi h^\dagger J_A = h \phi J_A h^T
\ee
and that an invariant trace on these object needs an extra factor of $J_A$.
Keeping that in mind we can use the standard way of dealing with
$Sp(2n)$. Note that $(\phi J_A)^T = J_A^T \phi^T = - J_A \phi^T =\phi J_S$
so the Goldstone bosons live in the antisymmetric representation of $Sp(2n)$.

The method of Young tableaux also is developed for $Sp(2n)$ \cite{youngSPN}.
Putting together two antisymmetric representations gives
\be
\label{youngSPN}
\!\raisebox{-32pt}{
\setlength{\unitlength}{1pt}
\begin{picture}(10,40)
\SetScale{1.0}
\newcommand{\youngbox}[1]
{\SetScaledOffset(#1)\EBox(0.,0.)(10.,10.)\SetScaledOffset(0.,0.)}
\youngbox{0,20}
\youngbox{0,30}
\end{picture}}
~\otimes\!
\raisebox{-32pt}{
\setlength{\unitlength}{1pt}
\begin{picture}(10,40)
\SetScale{1.0}
\newcommand{\youngbox}[1]
{\SetScaledOffset(#1)\EBox(0.,0.)(10.,10.)\SetScaledOffset(0.,0.)}
\youngbox{0,20}
\youngbox{0,30}
\end{picture}}
\,\,=
\,\,\cdot
~\oplus\!
\raisebox{-32pt}{
\setlength{\unitlength}{1pt}
\begin{picture}(10,40)
\SetScale{1.0}
\newcommand{\youngbox}[1]
{\SetScaledOffset(#1)\EBox(0.,0.)(10.,10.)\SetScaledOffset(0.,0.)}
\youngbox{0,20}
\youngbox{0,30}
\end{picture}}
~\oplus\!
\raisebox{-32pt}{
\setlength{\unitlength}{1pt}
\begin{picture}(20,40)
\SetScale{1.0}
\newcommand{\youngbox}[1]
{\SetScaledOffset(#1)\EBox(0.,0.)(10.,10.)\SetScaledOffset(0.,0.)}
\youngbox{0,30}
\youngbox{10,30}
\end{picture}}
~\oplus\!
\raisebox{-32pt}{
\setlength{\unitlength}{1pt}
\begin{picture}(10,40)
\SetScale{1.0}
\newcommand{\youngbox}[1]
{\SetScaledOffset(#1)\EBox(0.,0.)(10.,10.)\SetScaledOffset(0.,0.)}
\youngbox{0,0}
\youngbox{0,10}
\youngbox{0,20}
\youngbox{0,30}
\end{picture}}
~\oplus\!
\raisebox{-32pt}{
\setlength{\unitlength}{1pt}
\begin{picture}(20,40)
\SetScale{1.0}
\newcommand{\youngbox}[1]
{\SetScaledOffset(#1)\EBox(0.,0.)(10.,10.)\SetScaledOffset(0.,0.)}
\youngbox{0,10}
\youngbox{0,20}
\youngbox{0,30}
\youngbox{10,30}
\end{picture}}
~\oplus\!
\raisebox{-32pt}{
\setlength{\unitlength}{1pt}
\begin{picture}(20,30)
\SetScale{1.0}
\newcommand{\youngbox}[1]
{\SetScaledOffset(#1)\EBox(0.,0.)(10.,10.)\SetScaledOffset(0.,0.)}
\youngbox{0,20}
\youngbox{10,20}
\youngbox{0,30}
\youngbox{10,30}
\end{picture}}
\ee

We can write this in the form
\be
Asym.\otimes\!Asym. =
R_I \oplus R_A \oplus R_S \oplus R_{FA} \oplus R_{MA} \oplus R_{MS}\,.
\ee
The states are given
in Tab.~\ref{tabSONstates}.
They are made traceless but remember that indices of $\phi^a J_A$
and $\phi^b J_A$ are always contracted with $J_A$.

The representations are the singlet representation, symmetric under
$A\leftrightarrow B$, $R_A$ which is antisymmetric under the interchange
$i\leftrightarrow j$ but symmetric under $A\leftrightarrow B$
and $R_S$ which is symmetric under the interchange
$i\leftrightarrow j$ but antisymmetric under $A\leftrightarrow B$
Let us show the latter on $R_A$
The two-index
antisymmetric representation $R_A$ is, now using $A,B$ for $\phi^a,\phi^b$,
\ba
\lefteqn{(AJ_A)_{ik}(BJ_A)_{jl} J_{Akl}-(AJ_A)_{jk}(BJ_A)_{il} J_{Akl}}&&
\nonumber\\
&=& -(AJ_A)_{ik} J_{Akl}(BJ_A)_{lj} J_{Akl}-(BJ_A)_{il} J_{Alk}(AJ_A)_{ki}
\nonumber\\
&=& (ABJ+BAJ)_{ij}\,,
\ea
where we used $J_A^2=-1$ and the fact that $AJ_A$, $BJ_A$ and $J_A$ are
antisymmetric matrices.

The remaining ones are the four-index representations.
$R_{FA}$ is fully antisymmetric in all indices and symmetric
under $A\leftrightarrow B$.
The two remaining representations have a mixed symmetry in the indices
but are antisymmetric respectively symmetric under $A\leftrightarrow B$.
The states are give in Tab.~\ref{tabstatesSPN}.

\begin{table}
\begin{center}
\begin{tabular}{|c|c|}
\hline
\rule{0cm}{1.3em}$R_I$       & ${1\over\sqrt{ (2n+1)(n-1)}} \la AB\ra$
\\[2mm]
\hline
\rule{0cm}{1.3em}$R_A$       & $(ABJ+BAJ )_{ij}- {1\over n} J_{ij} \la AB\ra$
\\[2mm]
\hline
\rule{0cm}{1.3em}$R_S$       & $(ABJ -BAJ)_{ij}$
\\[1mm]
\hline
\rule{0cm}{1.3em}$R_{FA}$       &
 $(AJ)_{ij}\ (BJ)_{kl}+ {1\over n-2}  J_{ij}(ABJ + BAJ)_{kl}
   -{1\over 2 (n-1)(n-2)} J_{ij} J_{kl} \la AB \ra$
\\
 & $-(ijkl\leftrightarrow ikjl)
   +(ijkl\leftrightarrow iljk)
   +(ijkl\leftrightarrow klij)
$\\&$
  +(ijkl\leftrightarrow jlik)
  +(ijkl\leftrightarrow jkil)$
\\[1mm]
\hline
\rule{0cm}{1.3em}$R_{MA}$
  & $(AJ)_{ij}\ (BJ)_0{kl}-(AJ)_{kl}\ (BJ)_0{ij}
 - {1\over 2n-2} \Big[J_{ik}(ABJ - BAJ)_{jl}$\\
 &$ + J_{jl}(ABJ - BAJ)_{ik} -  J_{il}(ABJ - BAJ)_{jk}
  - J_{jk}(ABJ - BAJ)_{il}\Big]$\\[1mm]
\hline
\rule{0cm}{1.3em}$R_{MS}$       &
 $\Big\{(AJ)_{ij}\ (BJ)_{kl}+ (AJ)_{ik}\ (BJ)_{jl}
 -{1\over 2(n+1)}\Big[J_{ij}(ABJ+BAJ)_{kl}
$\\&$
 - J_{ik}(ABJ+BAJ)_{jl} \Big]+(ij\leftrightarrow kl)\Big\}$\\
&$+{1\over 2(n+1)(2n-1)} \Big( J_{ij} J_{kl} - J_{ik} J_{jl} \Big) \la AB\ra$
\\[3mm]
\hline
\end{tabular}
\end{center}
\caption{The intermediate states for the pseudo-real or two-colour case,
$SU(2n)/Sp(2n)$.
$J$ means $J_A$.}
\label{tabstatesSPN}
\end{table}

The projection operators can be constructed by contracting all indices with
$J_A$ of the states once with $A,B$ and once with $C,D$. The results are given
in Tab.~\ref{tabprojSPN}.
\begin{table}
\begin{center}
\begin{tabular}{|c|c|}
\hline
\rule{0cm}{1.3em}$P_I$       & ${1\over (2n+1)(n-1)} \la AB\ra \la CD\ra$
\\[3mm]
\hline
\rule{0cm}{1.3em}$P_A$       & ${n\over 2(n+1)(n-2)} \Big(\la(AB+BA)(CD+DC)\ra
 - {2\over n}\la AB\ra\la CD\ra\Big)$
\\[3mm]
\hline
\rule{0cm}{1.3em}$P_S$       & $ -{1\over 2(n-1)} \la (AB-BA)(CD-DC)\ra$
\\[3mm]
\hline
\rule{0cm}{1.3em}$P_{FA}$     &
$ {1\over6}\Big[ {2\over (n-1)(n-2)}\la AB\ra\la CD\ra
 + \la AC\ra \la BD\ra +  \la AD\ra \la BC\ra
$\\  & $
 -2\la ACBD+ADBC\ra
 -{2\over n-2}\la(AB+BA)(CD+DC)\ra\Big]$
\\[3mm]
\hline
\rule{0cm}{1.3em}$P_{MA}$       &
$ {1\over2(n-1)}\la (AB-BA)(CD-DC)\ra+{1\over 2} (\la AC \ra \la BD\ra
 -\la AD \ra \la BC\ra)  $
\\[3mm]
\hline
\rule{0cm}{1.3em}$P_{MS}$       &
 ${1\over3}\Big[{1\over (n+1)(2n+1)}\la AB\ra\la CD\ra + \la AC\ra\la BD\ra
+\la AD\ra\la BC\ra$
\\
  & $+ \la ACBD+ADBC \ra-{1\over 2(n+1)}\la (AB+BA)(CD+DC)\ra \Big]$
 \\[3mm]
\hline
\end{tabular}
\end{center}
\caption{The projection operator for the different channels
for the pseudo-real or two-colour case, $SU(2n)/Sp(2n)$.}
\label{tabprojSPN}
\end{table}

We can now use again (\ref{stateamplitude}) or (\ref{projectamplitude})
to obtain the amplitudes in the different channels. The results are very
similar to the real case and read
\ba
\label{TISPN}
T_I &=& {1\over n}(2n+1)(n-1)[B(s,t,u) + B(t,u,s)] - {1\over n}(n+1)B(u,s,t)
\nonumber\\&&
 +(2n+1)(n-1)C(s,t,u) + C(t,u,s) + C(u,s,t)\,,
\nonumber\\
T_A &=& {1\over n}(n+1)(n-2)[B(s,t,u)+B(t,u,s)] - {1\over n}(n+2)B(u,s,t)
\nonumber\\&&
 +C(t,u,s) + C(u,s,t)\,,
\nonumber\\
T_S &=& (1-n)[B(s,t,u) - B(t,u,s)]+ C(t,u,s) - C(u,s,t)\,,
\nonumber\\
T_{FA} &=& - 2B(u,s,t)+ C(t,u,s) + C(u,s,t)\,,
\nonumber\\
T_{MA} &=& C(t,u,s) - C(u,s,t)\,.
\nonumber\\
T_{MS} &=& B(u,s,t)+C(t,u,s) + C(u,s,t)\,,
\ea

\section{Results for the amplitude $M(s,t,u)$}
\label{results}

We have rewritten the amplitudes here in terms of the
physical decay constant $\fphys$ and mass $\mphys$.
The relation of these to the lowest order results can be found
in \cite{paper1}.
We also use the abbreviations
\be
x_2 = \frac{\mphys}{\fphys^2}\,,\quad
L = \frac{1}{16\pi^2}\ln\frac{\mphys}{\mu^2}\,,
\quad\pi_{16}=\frac{1}{16\pi^2}.
\ee
In addition we have often used (\ref{relstu}) to simplify the expressions.

\subsection{Lowest order}

\begin{figure}
\centering
\includegraphics[width=1.0\textwidth]{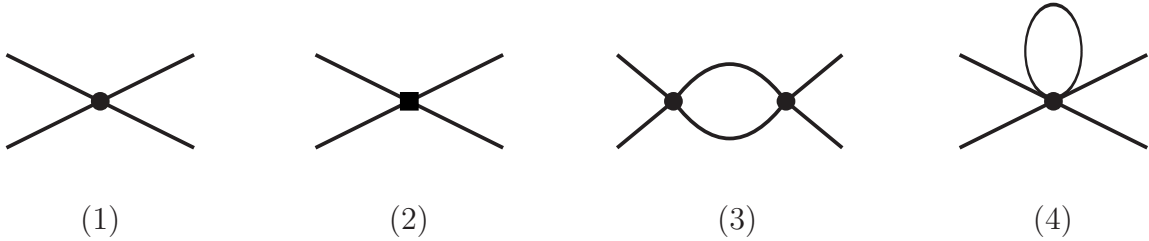}
\caption{The leading order and next-to leading order for meson-meson scattering
$\phi\phi\to\phi\phi$.
The filled circle is a vertex from ${\cal L}_2$,
and the filled square is a vertex from ${\cal L}_{4}$.}
\label{pipi1loop}
\end{figure}

The lowest order result comes from the simple tree-level diagram (1)
of Fig.~\ref{pipi1loop}.

\be
B_{LO}(s,t,u) = x_2\left(-\frac{1}{2}t+1\right)\,,\qquad C_{LO}(s,t,u) = 0\,,
\ee
for all cases. This reproduces using the relation (\ref{relABC}) Weinberg's
result \cite{Weinbergpipi} for $\pi\pi$ scattering
\be
A_{LO} = x_2\left(s-1\right)\,.
\ee

\subsection{Next-to-leading order}

The next-to-leading order contains the 3 diagrams (2-4) in
Fig.~\ref{pipi1loop} in addition to wave-function-renormalization.
The divergences from loop diagram (3) and (4) can be canceled by the bare
low energy constants (LECs) of $\mathcal{L}_4$ in the diagram (2).

The functions $B(s,t,u)$ and $C(s,t,u)$ can be calculated from the
one-loop graphs shown in Fig.~\ref{pipi1loop}(2-4) and wave-function
renormalization. It also contains terms from rewriting the lowest-order result
in the physical mass and decay constant.

The functions $B(s,t,u)$ and $C(s,t,u)$ can be rewritten
in the form
\ba
B(s,t,u) &=& x_2^2\left[B_P(s,t,u) + B_S(s,t-u)+B_S(u,t-s)+ B_T(t)\right]\,,
\nonumber\\
C(s,t,u) &=& x_2^2\left[C_P(s,t,u) +C_S(s)+C_T(t)+C_T(u)\right]\,.
\ea
$B_P(s,t,u)$ and $C_P(s,t,u)$ are the polynomial part, the remaining pieces
are often called the unitarity correction.
This can be proven using an extension of the methods of \cite{Stern}.

Using (\ref{relstu}) we rewrite the polynomial part in its simplest
form satisfying the symmetry constraints (\ref{BCsymmetry}):
\ba
\label{defpol}
B_P(s,t,u) &=& \alpha_1+ \alpha_2 t+\alpha_3 t^2 +\alpha_4 (s-u)^2\,,
\nonumber\\
C_P (s,t,u) &=&\beta_1+ \beta_2 s+\beta_3 s^2 +\beta_4 (t-u)^2\,.
\ea
The polynomial part for the three cases is give in Tab.~\ref{NLOpoly}.
\begin{table}
\begin{center}
\begin{tabular}{|c|c|c|}
\hline
\multicolumn{3}{|c|}{QCD: $SU(n)\times SU(n)/SU(n)$} \\
\hline
\rule{0em}{1.2em}
$B_P(s,t,u)$&  $\alpha_1$ &
$ {2\over n} L + {2\over n} \pi_{16} + 16 L_8^r + 16 L_0^r-{2\over3} n L - {5\over9} n \pi_{16}$ \\
  &$\alpha_2 $
&  $ - 4 L_5^r - 16 L_0^r + {5\over12} n L + {11\over36} n \pi_{16}$\\
 &  $\alpha_3$
&  $L_3^r + 4 L_0^r - {1\over16} n L - {1\over24} n \pi_{16}$\\
  & $\alpha_4$
&  $ L_3^r - {1\over48} n L - {1\over36} n \pi_{16}$\\[1mm]
\hline
\rule{0em}{1.2em}
$C_P(s,t,u)$ & $\beta_1$
&  $ 32 (L_1^r - L_4^r + L_6^r) -  {2\over n^2}( L + \pi_{16}) $ \\
 & $\beta_2$           &  $ 16 L_4^r - 32\,L_1^r$\\
 & $\beta_3$ & $- {3\over8} L + 2 L_2^r + 8 L_1^r - {3\over8} \pi_{16} $\\
  & $\beta_4$ &  $2 L_2^r - {1\over8} L  - {1\over8} \pi_{16}$\\[1mm]
\hline
\multicolumn{3}{|c|}{Adjoint: $SU(2n)/SO(2n)$} \\
\hline
\rule{0em}{1.2em}
$B_P(s,t,u)$&$\alpha_1$& $  16 (L^r_0+L^r_8)  -\left(\frac{7 }{6}+\frac{2 n}{3}-\frac{1}{n}\right)L
                        -\left(\frac{19}{18}+\frac{5 n}{9}-\frac{1}{n}\right)\pi_{16}$  \\
            &$\alpha_2$& $-16L^r_0-4 L^r_5+\left(\frac{5 n}{12}+\frac{2 }{3}\right)L
                    +\left(\frac{5}{9}+\frac{11 n }{36}\right)\pi_{16} $ \\
            &$\alpha_3$& $4L^r_0+L^r_3-\left(\frac{1}{8}+\frac{n}{16}\right)L
                      -\left(\frac{5}{48}+\frac{n}{24}\right)\pi_{16}$  \\
            &$\alpha_4$& $L^r_3+\left(\frac{1}{24}-\frac{n }{48}\right)L
                   +\left(\frac{5}{144}-\frac{n}{36}\right)\pi_{16} $\\[1mm]
\hline
\rule{0em}{1.2em}
$C_P(s,t,u)$&$\beta_1$&$  32(L^r_1 - L^r_4 +L^r_6  ) - {1\over2n^2} (L+  \pi_{16})  $ \\
      &$\beta_2$& $ 16\left( L^r_4 - 2 L^r_1\right) $\\
      &$\beta_3$&  $ 8 L^r_1+ 2L^r_2  - {3\over16} \pi_{16} - {3\over16} L$\\
      &$\beta_4$& $  2L^r_2- {1\over16} L - {1\over16} \pi_{16} $ \\[1mm]
\hline
\multicolumn{3}{|c|}{Two-colour: $SU(2N)/Sp(2N)$}\\
\hline
\rule{0em}{1.2em}
$B_P(s,t,u)$&$\alpha_1$& $ 16 (L^r_0+L^r_8) +\left(\frac{7 }{6}+\frac{1}{n}-\frac{2  n}{3}\right)L
                         +\left(\frac{19 }{18}-\frac{5 n }{9}+\frac{1}{n}\right)\pi_{16}$\\
            &$\alpha_2$& $-16L^r_0-4 L^r_5+ \left(\frac{5 n}{12}-\frac{2 }{3}\right)L
                           +\left(\frac{5}{9}-\frac{11 n }{36}\right)\pi_{16}$\\
            &$\alpha_3$& $ 4L^r_0+L^r_3+\left(\frac{1}{8}-\frac{n}{16}\right)L
                     +\left(\frac{5}{48} - \frac{n}{24}\right)\pi_{16} $\\
            &$\alpha_4$& $L^r_3-\left(\frac{1}{24}+\frac{n }{48}\right)L
                       -\left(\frac{5}{144}+\frac{n}{36}\right)\pi_{16}$\\[1mm]
\hline
\rule{0em}{1.2em}
$C_P(s,t,u)$&$\beta_1$&  $32 (L^r_1- L^r_4+ L^r_6)-\frac{1}{2 n^2}(L+\pi_{16})$\\
            &$\beta_2$& $ 16(L^r_4-2 L^r_1)$\\
          &$\beta_3$& $8 L^r_1+2 L^r_2-\frac{3}{16}\pi_{16}-\frac{3}{16} L$\\
            &$\beta_4$& $2L^r_2-\frac{1}{16}L-\frac{1}{16}\pi_{16}$\\[1mm]
\hline
\end{tabular}
\end{center}
\caption{The next-to-leading results for all three cases for the polynomial
part. The coefficients are defined in (\ref{defpol}).}
\label{NLOpoly}
\end{table}

The unitarity correction is given in Tab.\ref{NLOunitarity}.
We noticed that the $C$ functions for the $SO(2n)$ and $Sp(2n)$ case are
the same.

\begin{table}
\begin{center}
\begin{tabular}{|c|c|}
\hline
\multicolumn{2}{|c|}{ QCD: $SU(n)\times SU(n)/SU(n)$ }\\
\hline
\rule{0em}{1.2em}
$B_S(s,t-u)$&  $\bar J(s) \left[ -\frac{1}{n} +\frac{n}{16} s^2
                    +\frac{n}{12}\left(1 -\frac{s}{4} \right)(t-u)\right]$ \\
$B_T(t)$ & 0 \\[1mm]
\hline
\rule{0em}{1.2em}
$C_S(s)$    &  $\bar J(s)\left(\frac{2}{n^2}+\frac{1}{4}s^2\right)$  \\
$C_T(t)$    &  $\frac{1}{4} \bar J(t)\left(t-2\right)^2$   \\[1mm]
\hline
\multicolumn{2}{|c|}{ Adjoint: $SU(2n)/SO(2n)$} \\
\hline
\rule{0em}{1.2em}
$B_S(s,t-u)$& $  \bar J(s) \left[-\frac{1}{2 n}+\frac{s}{4}+\frac{1}{16}\left( n - 1\right)s^2
                        + \frac{1}{12}(n+1)\left(1-\frac{s}{4}\right)(t-u) \right]$\\
$B_T(t)$    & $ {1\over8}\bar J(t)( t-2)^2$ \\[1mm]
\hline
\rule{0em}{1.2em}
$C_S(s)$    &  $\bar J(s) \left(  {1\over2n^2} + {1\over8}s^2 \right)$  \\
$C_T(t)$    &  ${1\over8}\bar J(t)( t-2)^2$   \\[1mm]
\hline
\multicolumn{2}{|c|}{Two-colour: $SU(2N)/Sp(2N)$}\\
\hline
\rule{0em}{1.2em}
$B_S(s,t-u)$ & $\bar J(s)\left[  - {1\over2n}   - {1\over4}s  +
                    {1\over16}( n+1 )s^2+
          \frac{1}{12}(n-1)\left(1-\frac{s}{4}\right)(t-u)\right]$ \\
$B_T(t)$    & $ -{1\over8}\bar J(t)( t-2)^2$ \\[1mm]
\hline
\rule{0em}{1.2em}
$C_S(s)$    &   $\bar J(s)\left( {1\over2n^2}+ {1\over8} s^2 \right)$ \\
$C_T(t)$    & ${1\over8}\bar J(t)( t-2)^2$ \\[1mm]
\hline
\end{tabular}
\end{center}
\caption{The next-to-leading results for all three cases for the unitarity
correction.}
\label{NLOunitarity}
\end{table}

\subsection{Next-to-next-to-leading order}
\label{NNLO}

\begin{figure}
\centering
\includegraphics[width=1.0\textwidth]{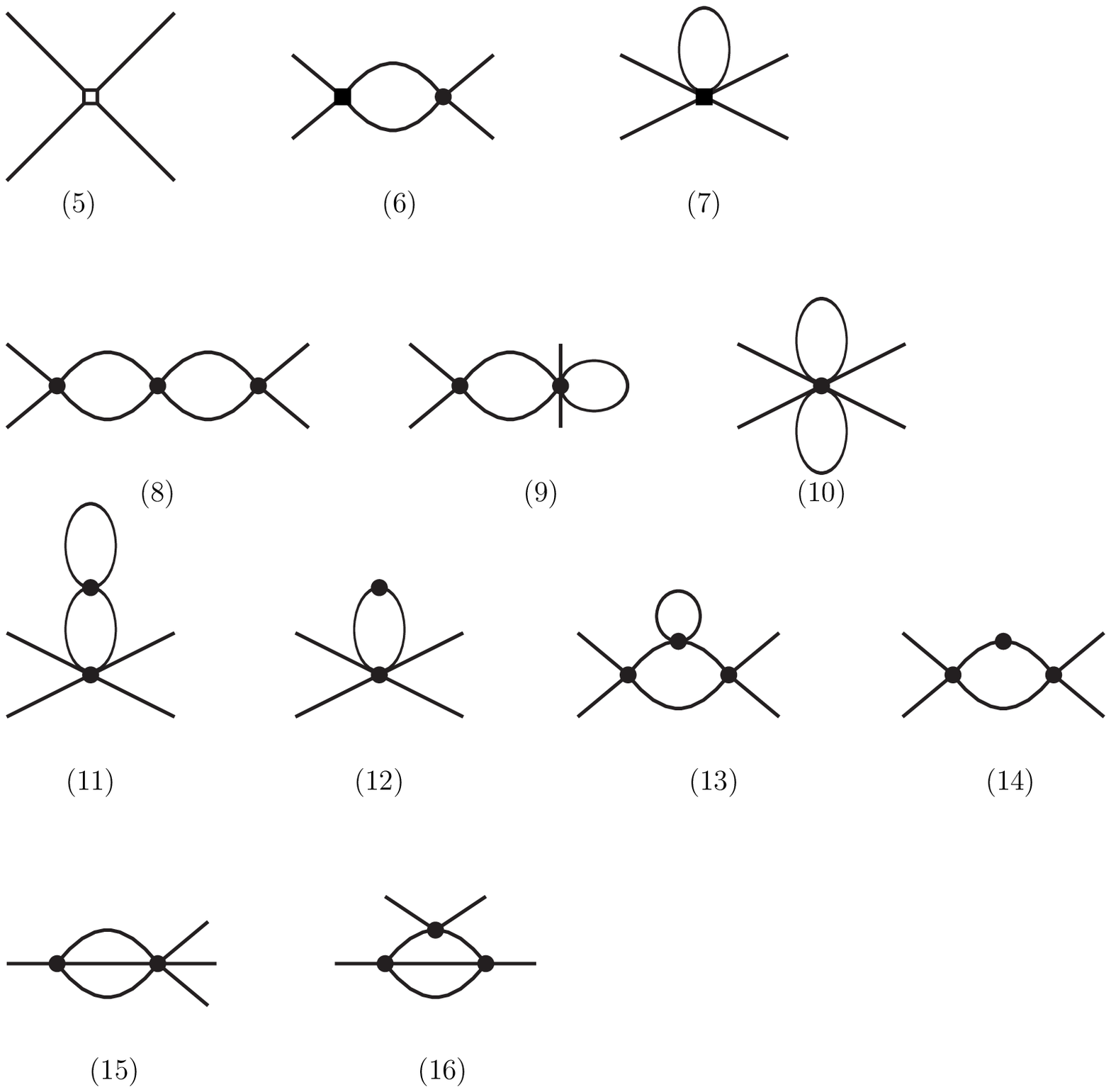}
\caption{The next-to-next-to leading order  diagrams for $\pi\pi\to\pi\pi$.
The filled circle a vertex from ${\cal L}_{2}$,
The filled square is a vertex from ${\cal L}_{4}$,
and the open square is a vertex from ${\cal L}_6$.}
\label{pipi2loop}
\end{figure}

There are 13 diagrams at next-to-next-to-leading order shown
in Fig.~\ref{pipi2loop}. We have checked that the nonlocal divergence
cancels for all three cases and that for the complex or QCD case
the result is fully finite with the subtractions calculated in \cite{BCE2}.

The diagrams (15) and (16) are often called the sunset and vertex or fish
diagram respectively. These require the most difficult integrals.
At intermediate stages we needed more integrals than those calculated
for \cite{BCEGS1,BCEGS2}. They were calculated with the methods of
\cite{GS} and are given in App.~\ref{loopintegrals}.

\ba
\label{NNLOresults}
B(s,t,u) &=& x_2^3\left[B_P(s,t,u) + B_S(s,t-u)+B_S(u,t-s)+B_T(t)\right]\,,
\nonumber\\
C(s,t,u) &=& x_2^3\left[C_P(s,t,u) +C_S(s)+C_T(t)+C_T(u)\right]\,.
\ea

The polynomial parts
we rewrite using (\ref{relstu}) in their simplest
form satisfying the symmetry constraints (\ref{BCsymmetry}):
\ba
\label{defpol2}
B_P(s,t,u) &=& \gamma_1+ \gamma_2 t+\gamma_3 t^2 +\gamma_4 (s-u)^2
+\gamma_5 t^3 +\gamma_6 t(s-u)^2\,,
\nonumber\\
C_P (s,t,u) &=&\delta_1+ \delta_2 s+\delta_3 s^2 +\delta_4 (t-u)^2
+\delta_5 s^3+\delta_6 s(t-u)^2\,.
\ea
The coefficients in these polynomials as well as the functions in
(\ref{NNLOresults}) are given in App.~\ref{appNNLO}.
The FORM expressions can be downloaded from \cite{website}.
We stress once more that the result is fully analytical
and expressed in terms of $L$ and $\overline J$.

\section{Scattering lengths}
\label{scatteringlengths}

The threshold parameters of general meson meson scattering
are defined similar to those of $\pi \pi$ scattering.
First we calculate the amplitudes for the different channels
using (\ref{TISUN}), (\ref{TISON}) and (\ref{TISPN}) for each of the possible
intermediate states of representation or channels $I$.

The scattering amplitude for each channel $I$
can be projected out using the partial wave expansion
\be
T^I_\ell(s) = {1\over64\pi} \int^1_{-1} d(cos\theta) P_\ell(cos\theta)
 T_I(s,t,u)\,.
\ee

Near the threshold $s=4$, we can expand the amplitude above the threshold
using $s=4(1+q^2/M^2_\pi)$ in the small three-momentum $q$.
\be
\mathrm{Re}\;T_\ell^I(s)=q^{2\ell}[a_\ell^I +q^2 b_\ell^I +O(q^4)]\,,
\label{eqthrex}
\ee
where $a_\ell^I$ is the scattering length, and $b_\ell^I$ is the slope.

In App.~\ref{scattlength},
we give the expressions of the lowest partial wave scattering length
for each channel in all three cases.
As mentioned in Sect.~\ref{groups},
some channels are symmetric under $A\leftrightarrow B$, hence the
lowest order partial wave is $\ell=0$. The other channels are antisymmetric
under $A\leftrightarrow B$, so that the lowest order partial wave is  $\ell=1$.
$A$ and $B$ are the incoming mesons here using the notation of
Sect.~\ref{groups}.

For the purpose of illustration, we plot the scattering length for the singlet
and the fully-symmetric (fully-antisymmetric) channels
as a function of the physical meson mass $\mphys$.
Since currently we do not have knowledge for
the values of the low energy constants for these,
we take the values of the $L_i^r$ of fit 10 of \cite{ABT} for
the complex or QCD case and half that for the other two
as suggested by the large $n$ relations discussed below.
The values of the NNLO LECs we simply put to zero. We also choose the
subtraction scale $\mu=0.77$~GeV and the physical decay constant
$\fphys = 0.0924$~GeV.

\begin{figure}
\begin{center}
\begin{minipage}{7.2cm}
\includegraphics[angle=0,width=0.99\textwidth]{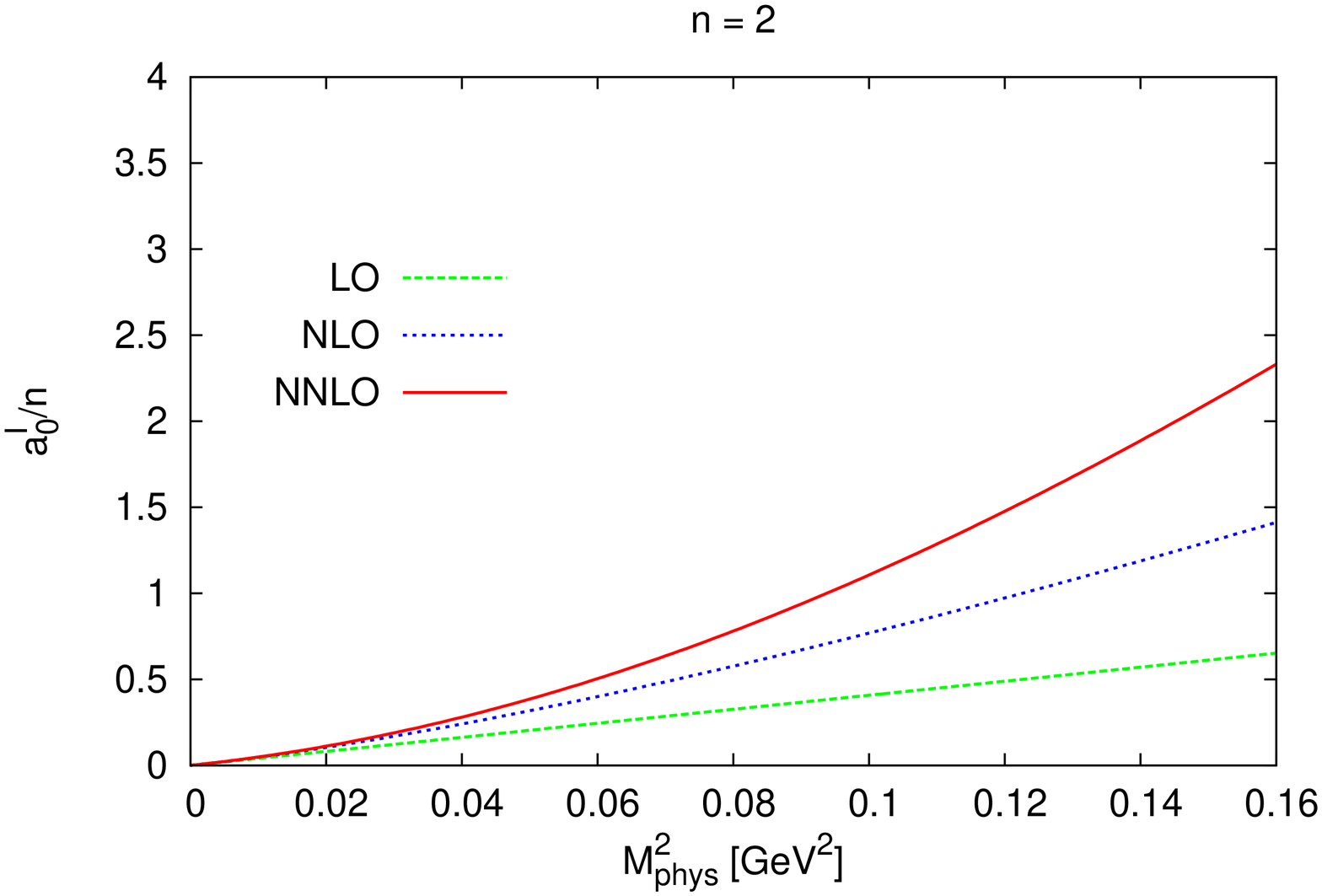}
\end{minipage}
\begin{minipage}{7.2cm}
\includegraphics[angle=0,width=0.99\textwidth]{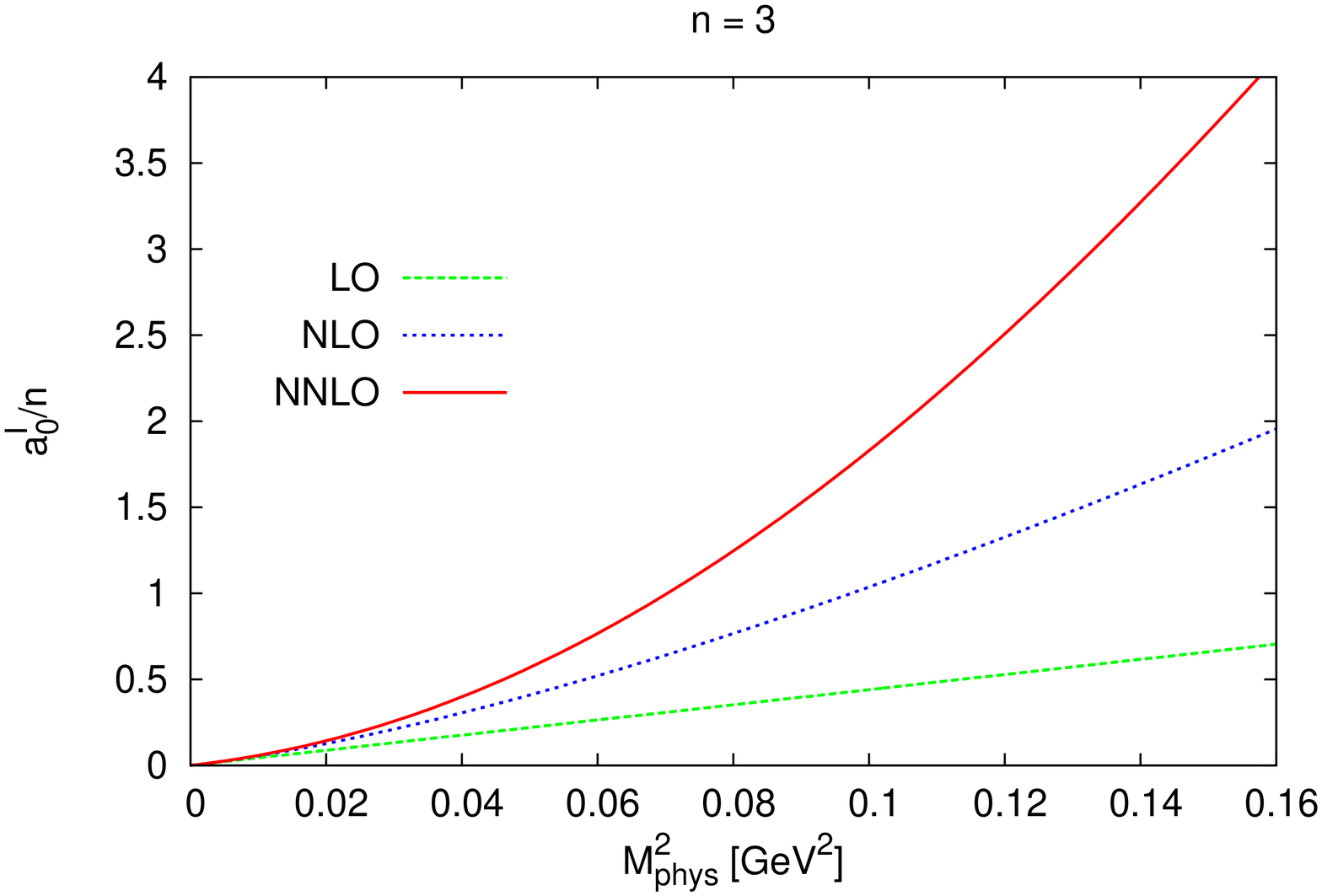}
\end{minipage}
\begin{minipage}{7.2cm}
\includegraphics[angle=0,width=0.99\textwidth]{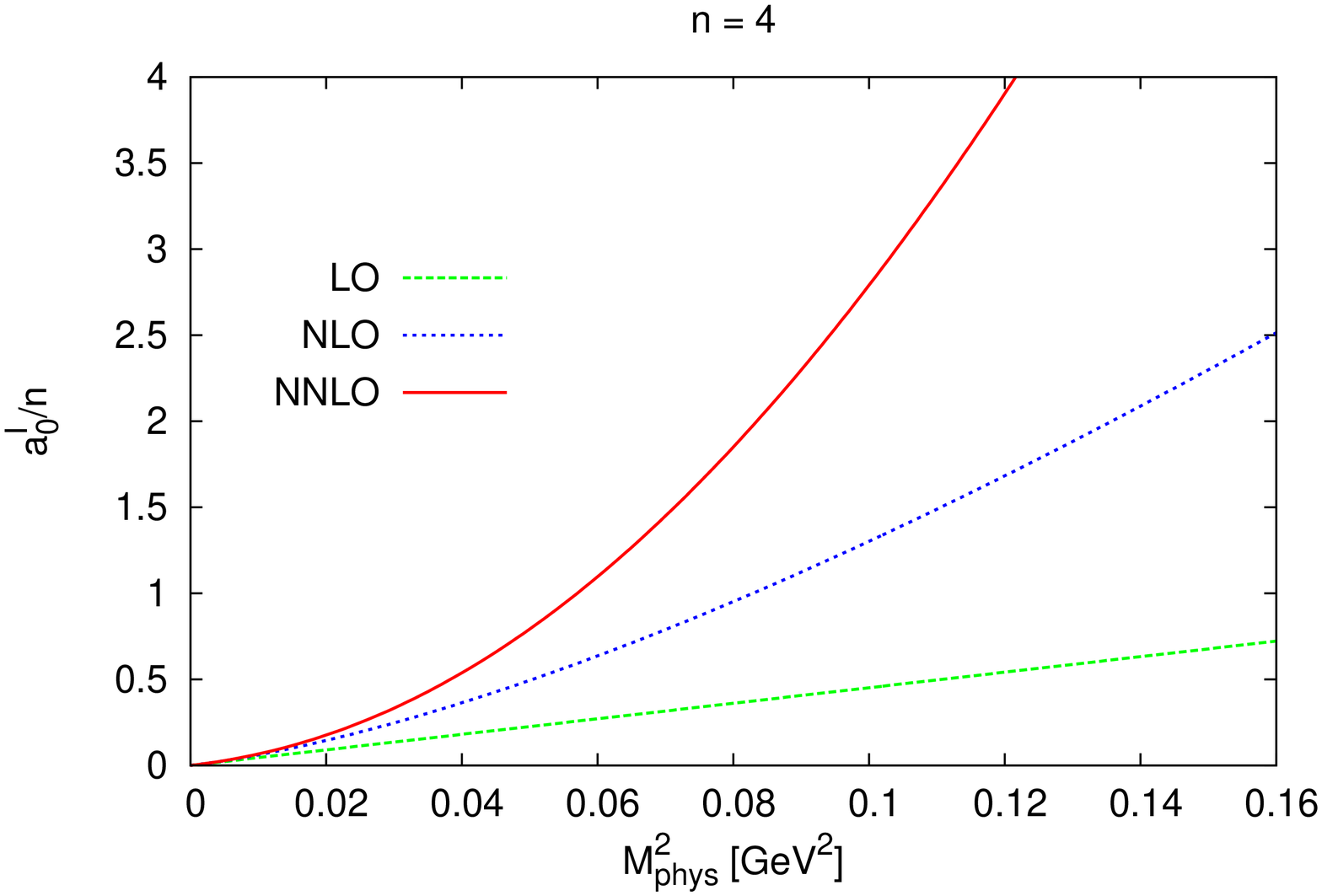}
\end{minipage}
\begin{minipage}{7.2cm}
\includegraphics[angle=0,width=0.99\textwidth]{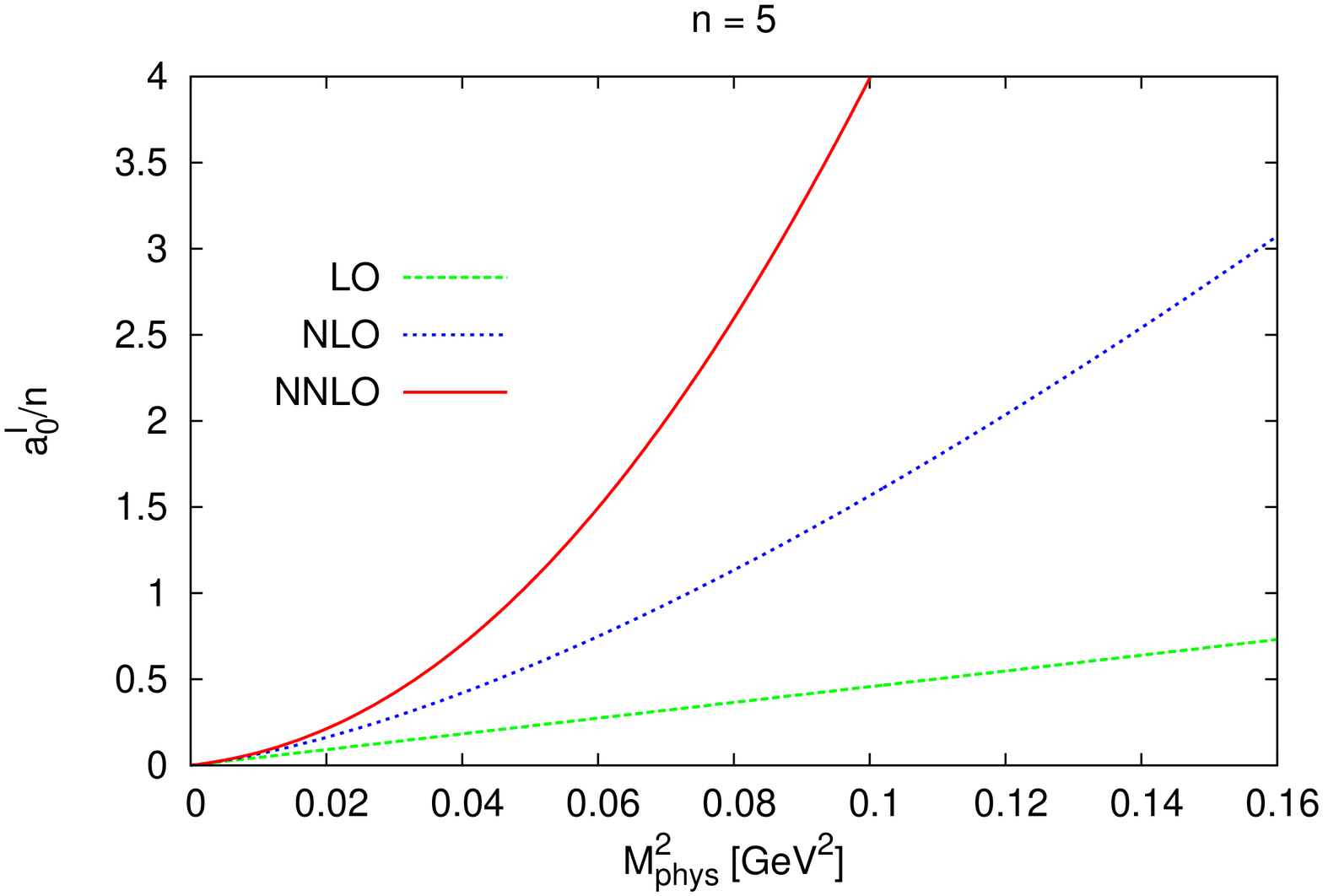}
\end{minipage}
\end{center}
\caption{\label{figaISUN} Scattering length $a^I_0/n$ for the
complex or QCD case,
$SU(n)\times SU(n)/SU(n)$.}
\end{figure}

The singlet case for the complex case is shown in Fig.~\ref{figaISUN}.
We have divided the scattering length by $n$ to make the lowest order
similar for all cases. Plotted are $n=2,\ldots,5$. One can see that
for a given $\mphys$ the convergence gets progressively worse for
larger $n$. For $n=2$ this corresponds to $a^0_0$.

\begin{figure}
\begin{center}
\begin{minipage}{7.3cm}
\includegraphics[angle=0,width=0.99\textwidth]{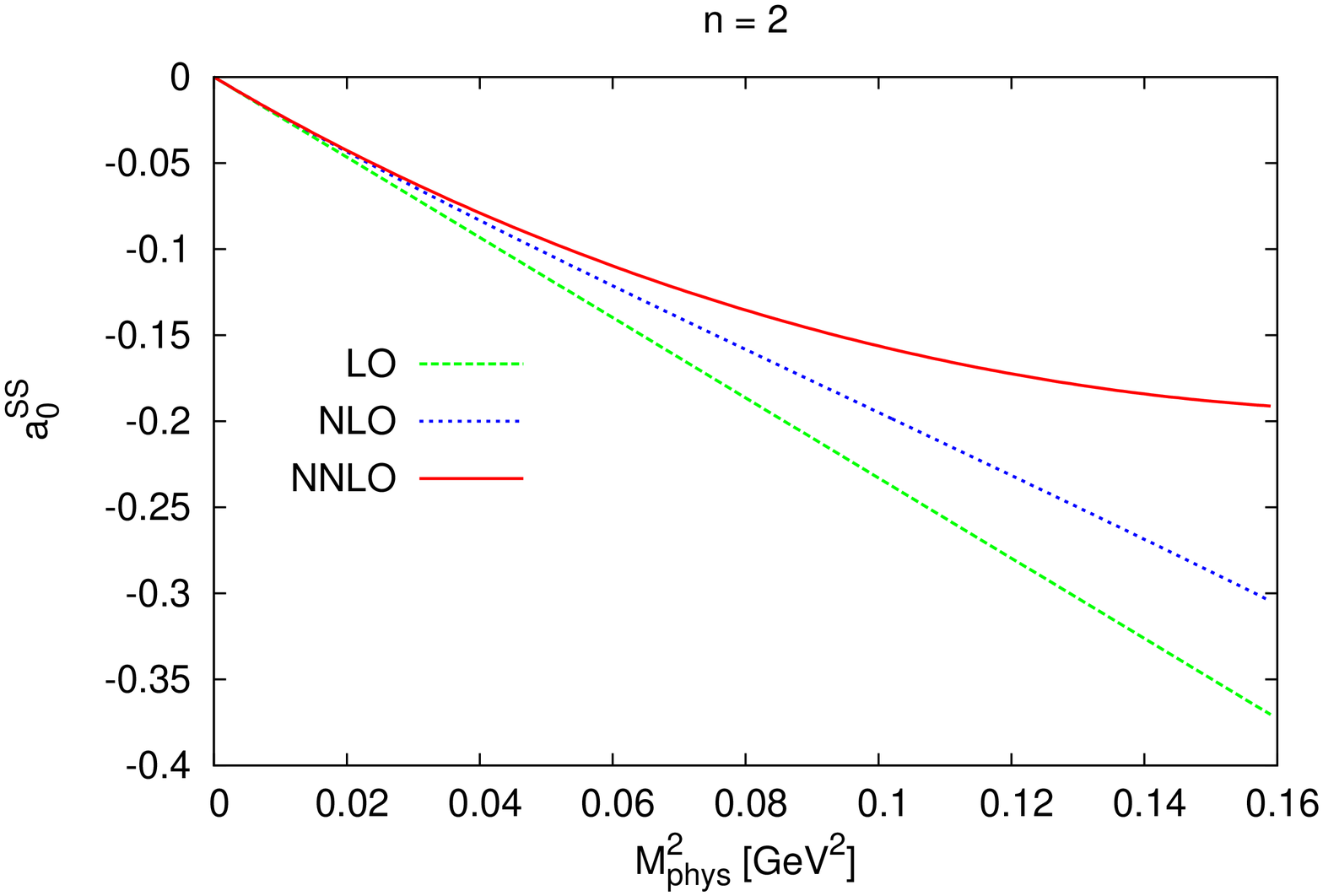}
\end{minipage}
\begin{minipage}{7.3cm}
\includegraphics[angle=0,width=0.99\textwidth]{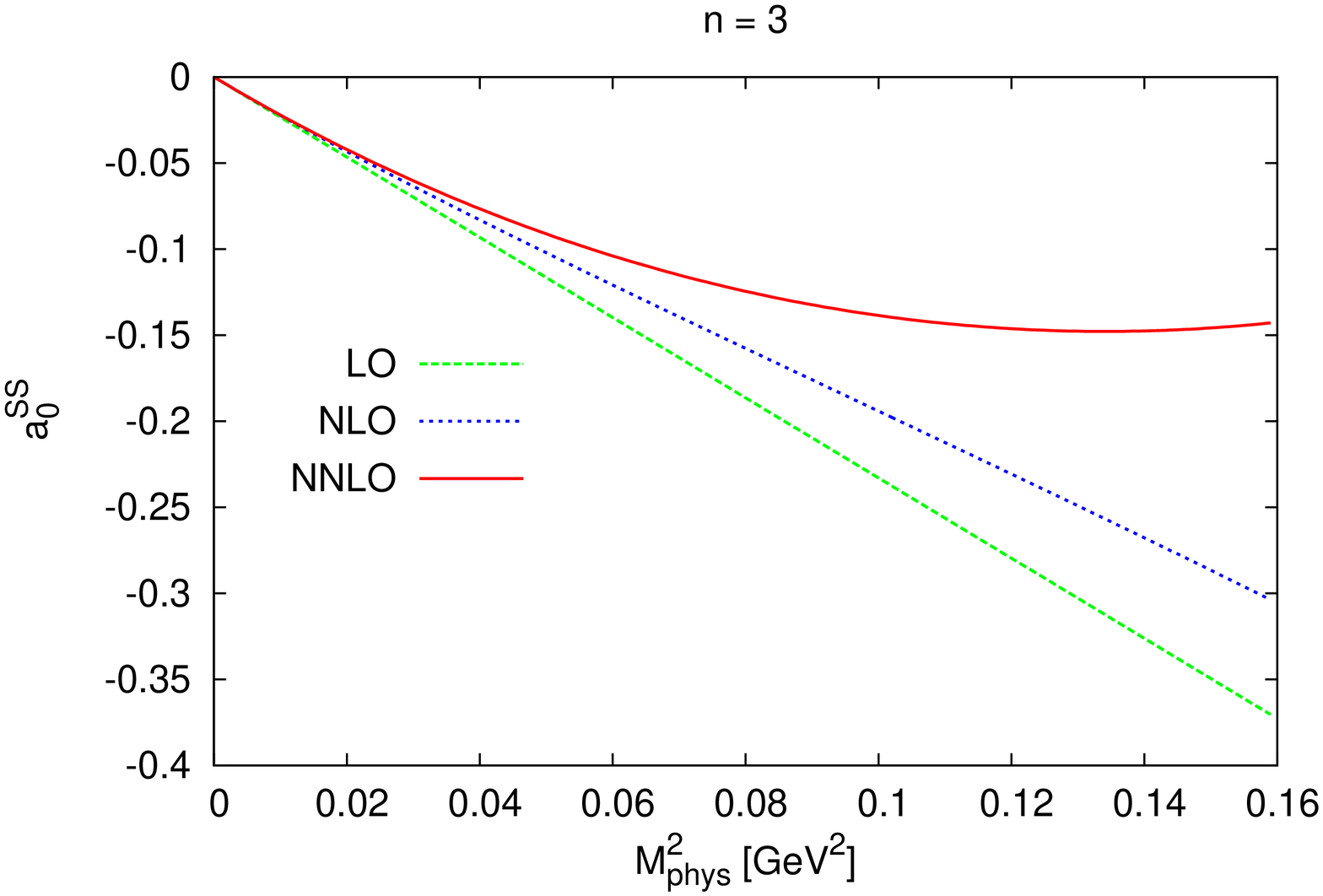}
\end{minipage}
\begin{minipage}{7.3cm}
\includegraphics[angle=0,width=0.99\textwidth]{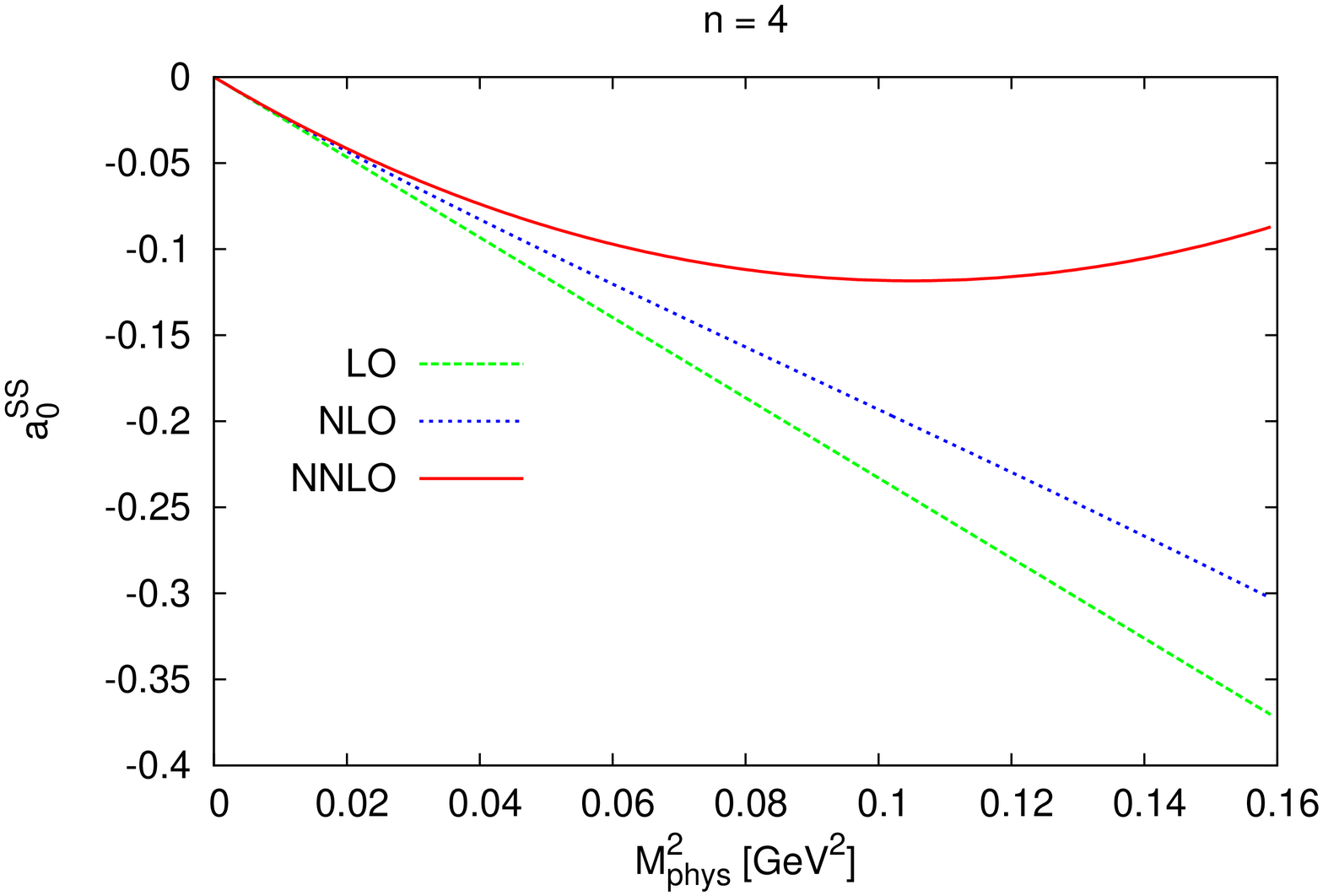}
\end{minipage}
\begin{minipage}{7.3cm}
\includegraphics[angle=0,width=0.99\textwidth]{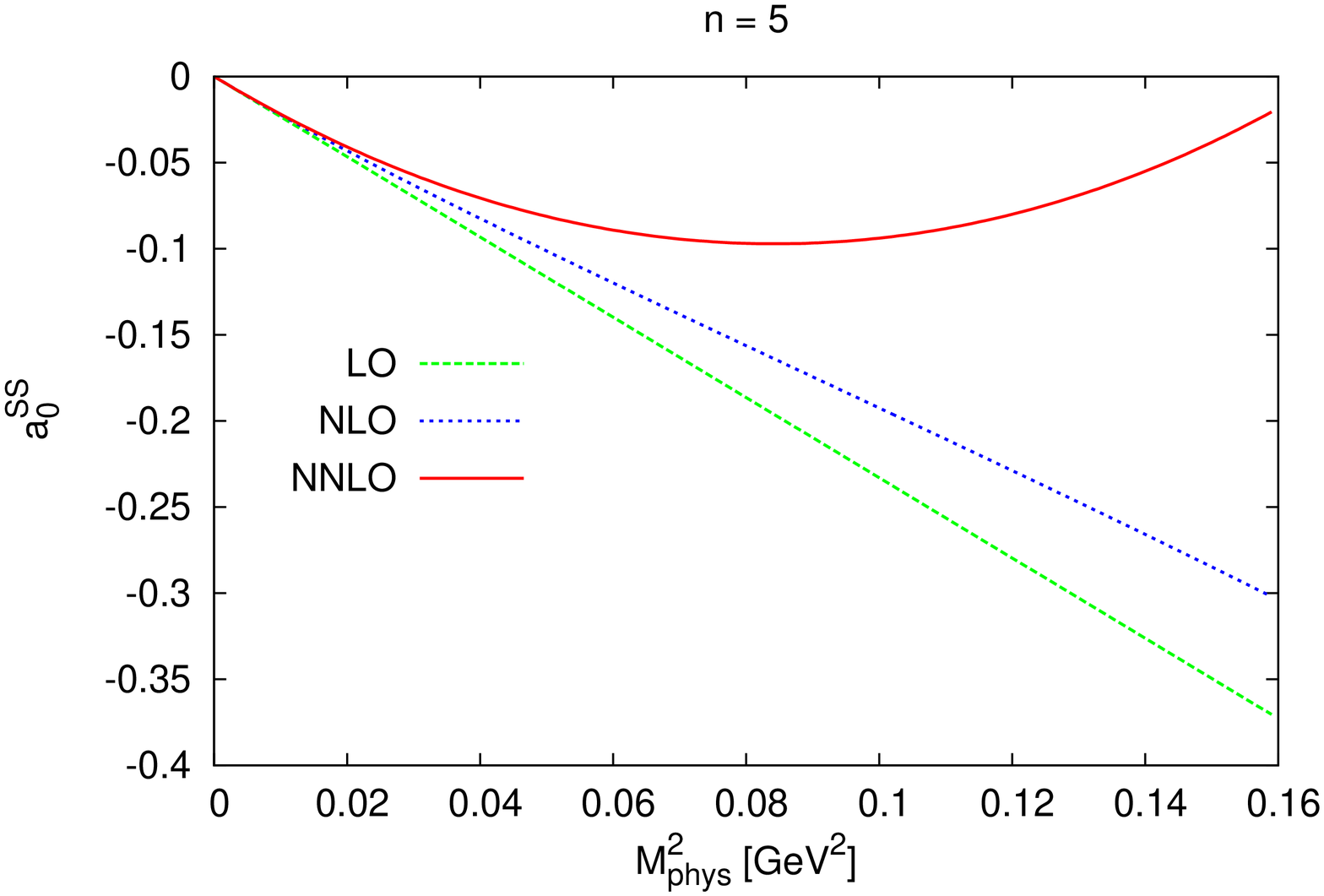}
\end{minipage}
\end{center}
\caption{\label{figaSSSUN} Scattering length $a^{SS}_0$ for the
complex or QCD case,
$SU(n)\times SU(n)/SU(n)$.}
\end{figure}

The fully symmetric case case for the complex case is shown in
Fig.~\ref{figaSSSUN}.
Plotted are $n=2,\ldots,5$. One can see that
for a given $\mphys$ the convergence gets progressively worse for
larger $n$. For $n=2$ this corresponds to $a^2_0$. The lowest order
is independent of $n$. The NLO order is only mildly dependent on $n$ while
the NNLO part grows fast with $n$.


\begin{figure}
\begin{center}
\begin{minipage}{7.3cm}
\includegraphics[angle=0,width=0.99\textwidth]{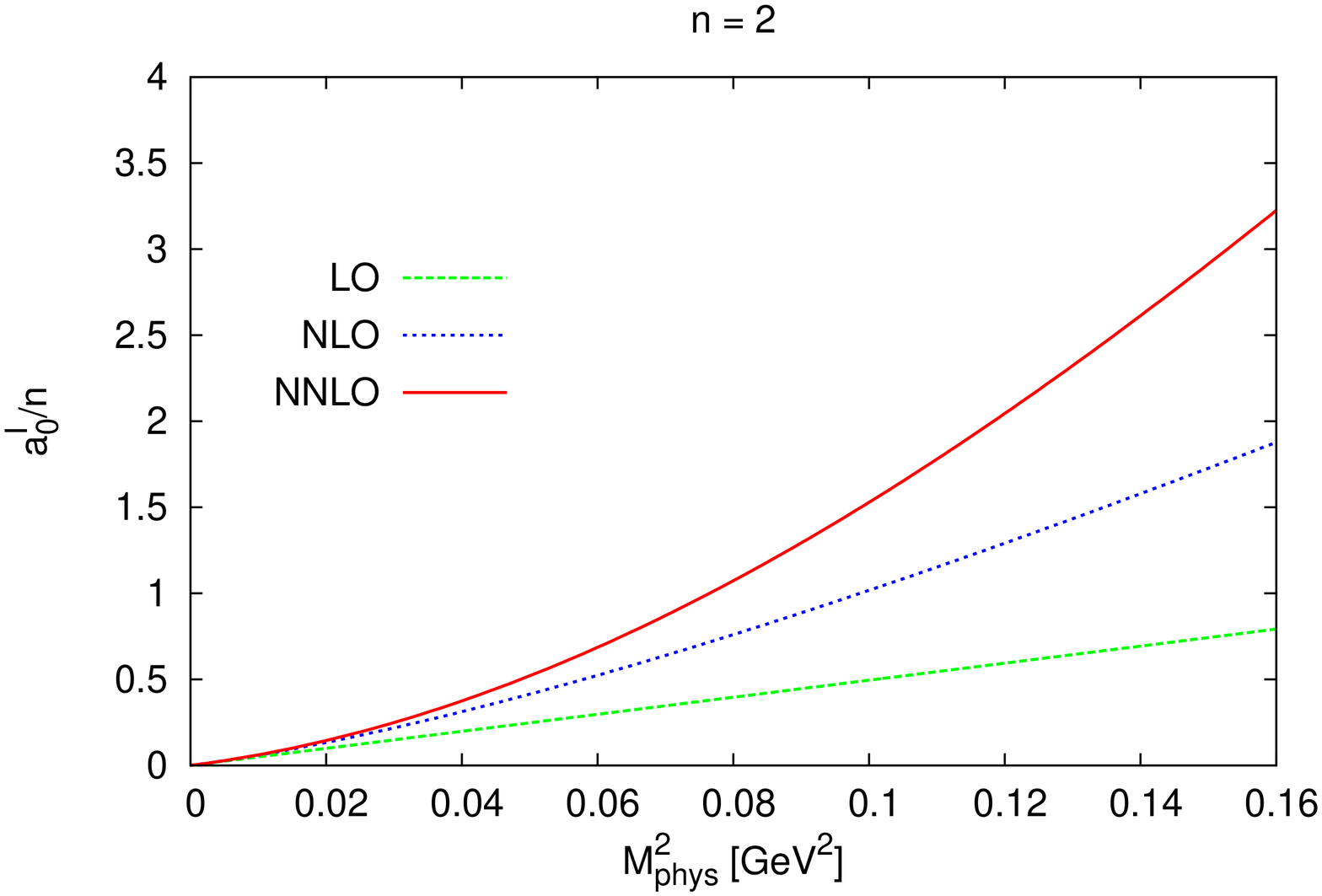}
\end{minipage}
\begin{minipage}{7.3cm}
\includegraphics[angle=0,width=0.99\textwidth]{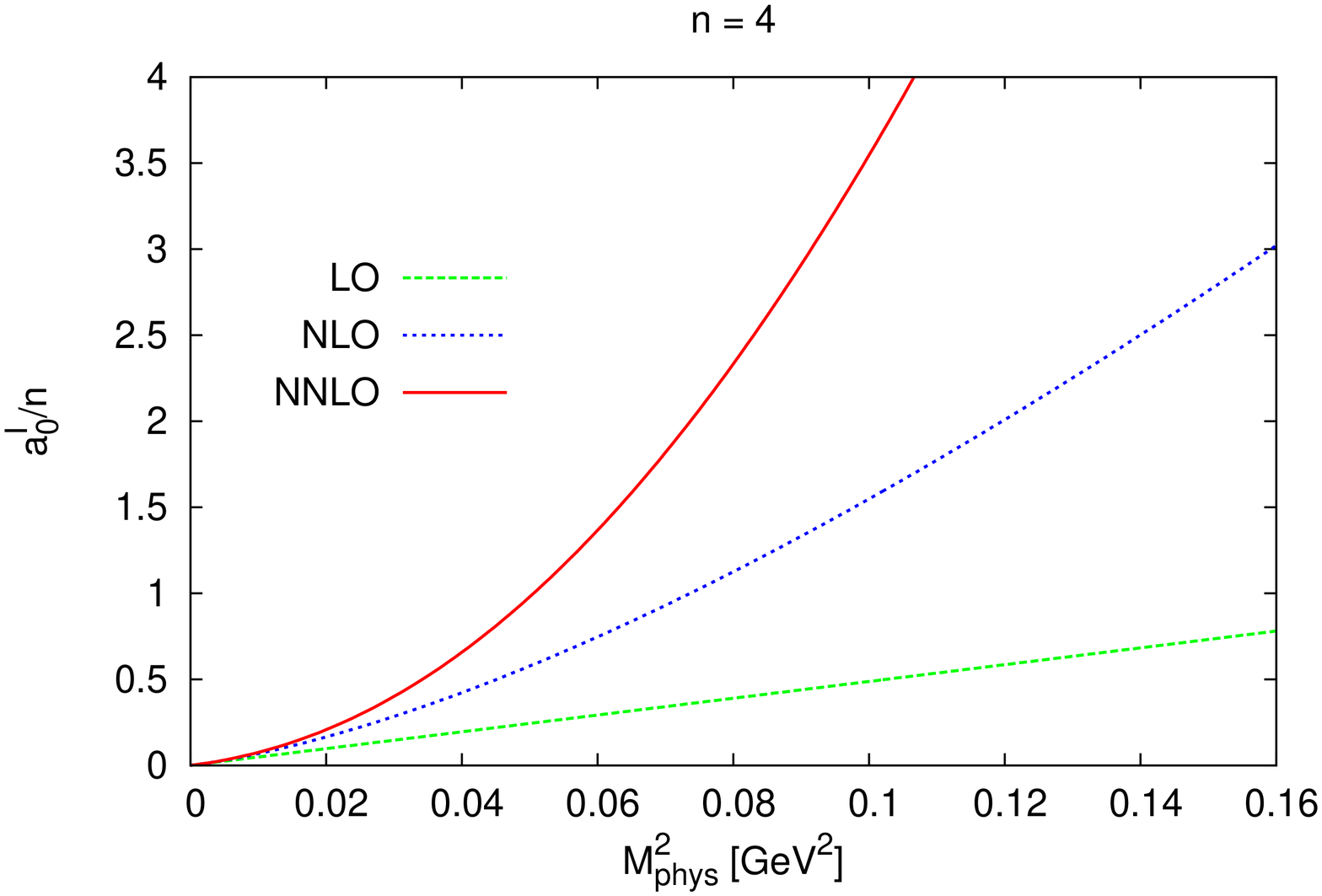}
\end{minipage}
\end{center}
\caption{\label{figaI0SON} Scattering length of $a^I_0/n$
for the real or adjoint case, $SU(2n)/SO(2n)$.}
\end{figure}

\begin{figure}
\begin{center}
\begin{minipage}{7.3cm}
\includegraphics[angle=0,width=0.99\textwidth]{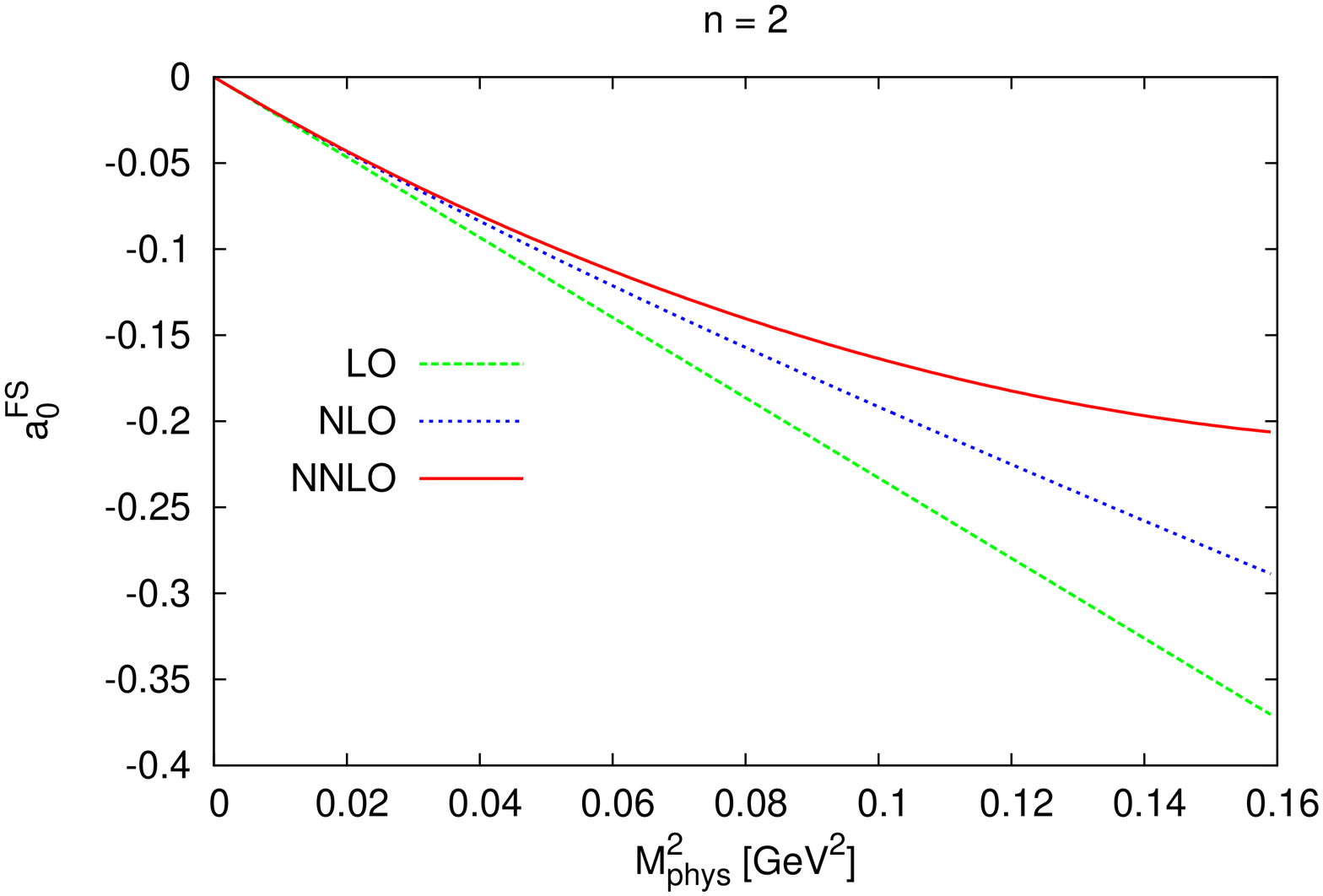}
\end{minipage}
\begin{minipage}{7.3cm}
\includegraphics[angle=0,width=0.99\textwidth]{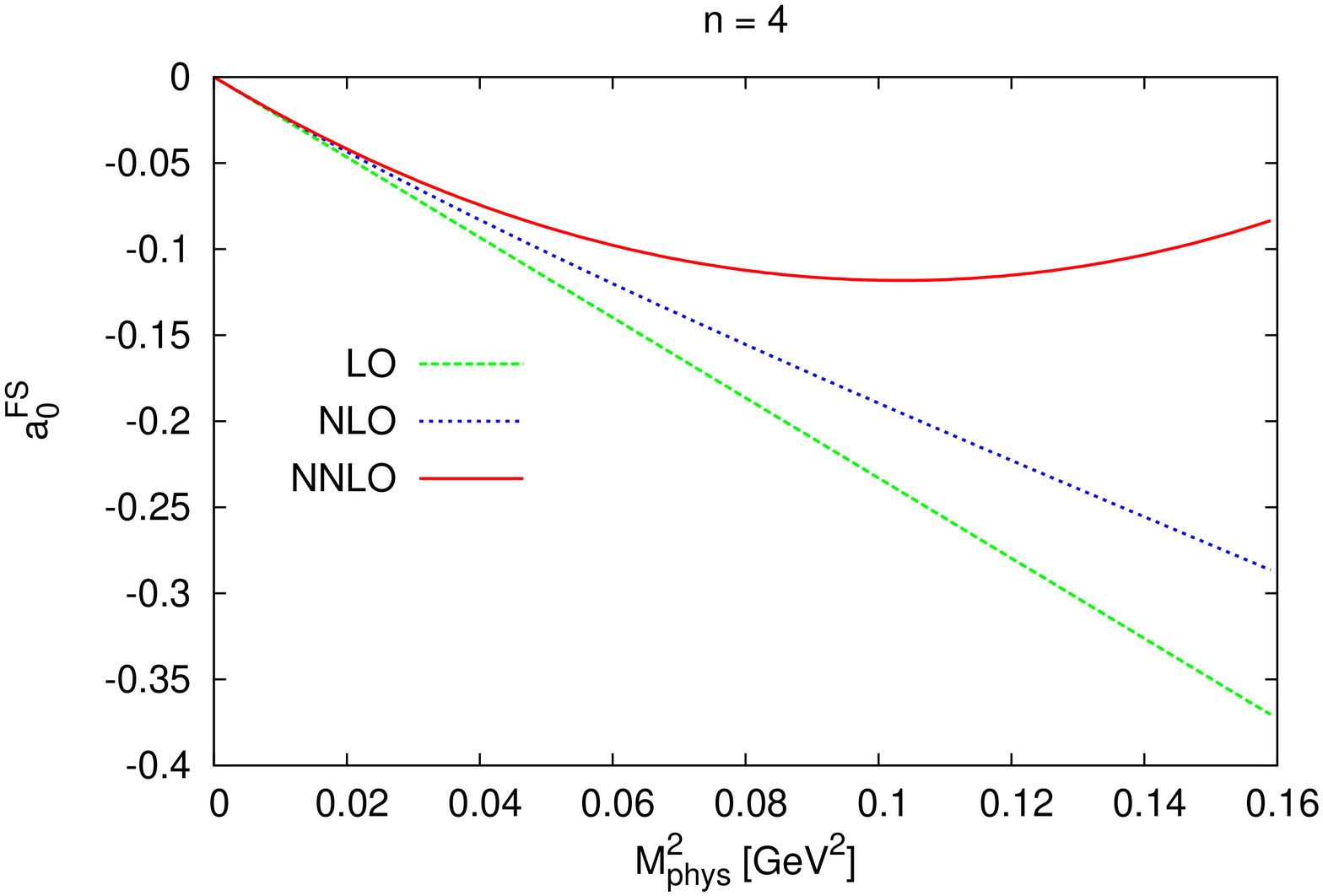}
\end{minipage}
\end{center}
\caption{\label{figaFS0SON} Scattering length of $a^{FS}_0$
for the real or adjoint case, $SU(2n)/SO(2n)$.}
\end{figure}

\begin{figure}
\begin{center}
\begin{minipage}{7.3cm}
\includegraphics[angle=0,width=0.99\textwidth]{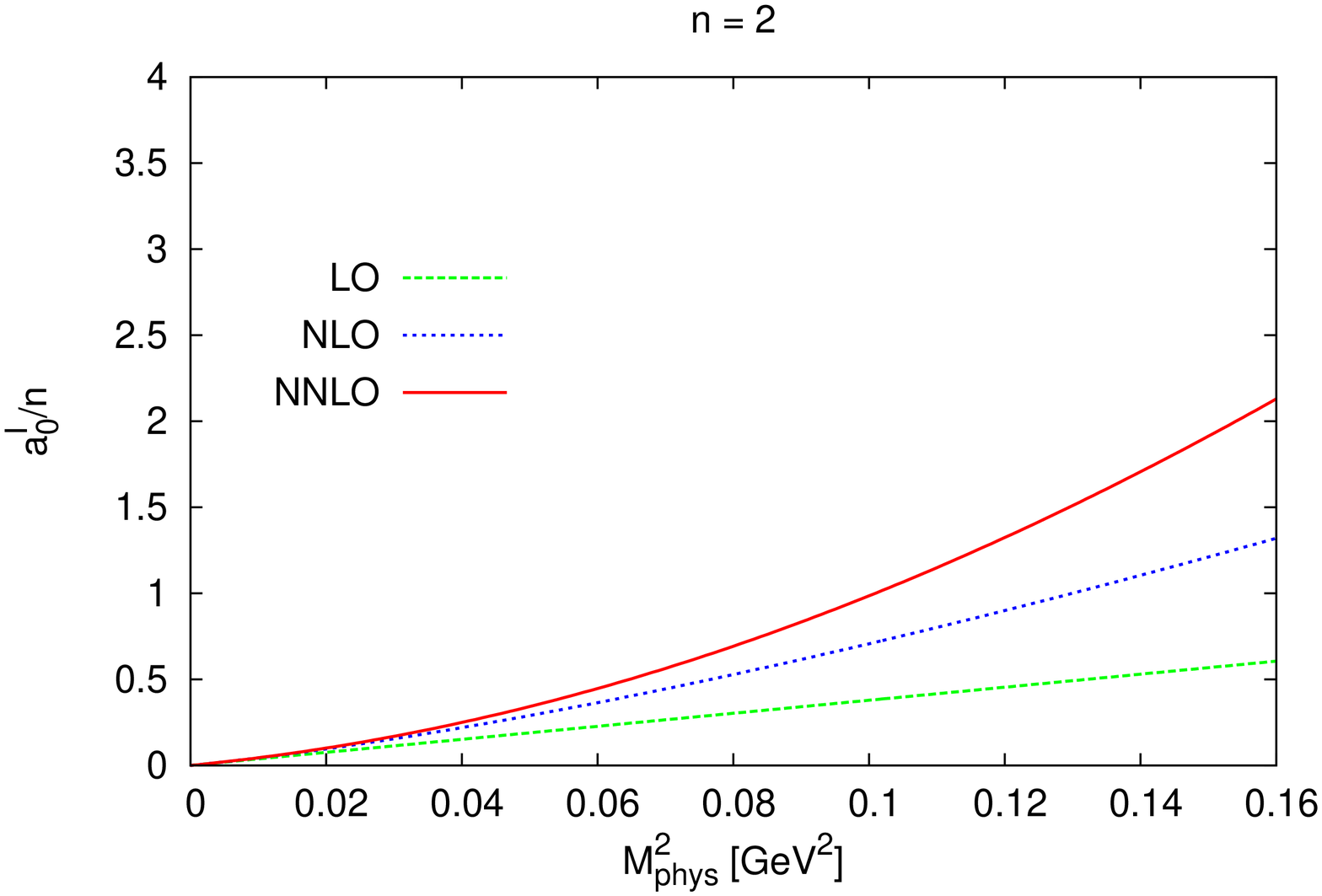}
\end{minipage}
\begin{minipage}{7.3cm}
\includegraphics[angle=0,width=0.99\textwidth]{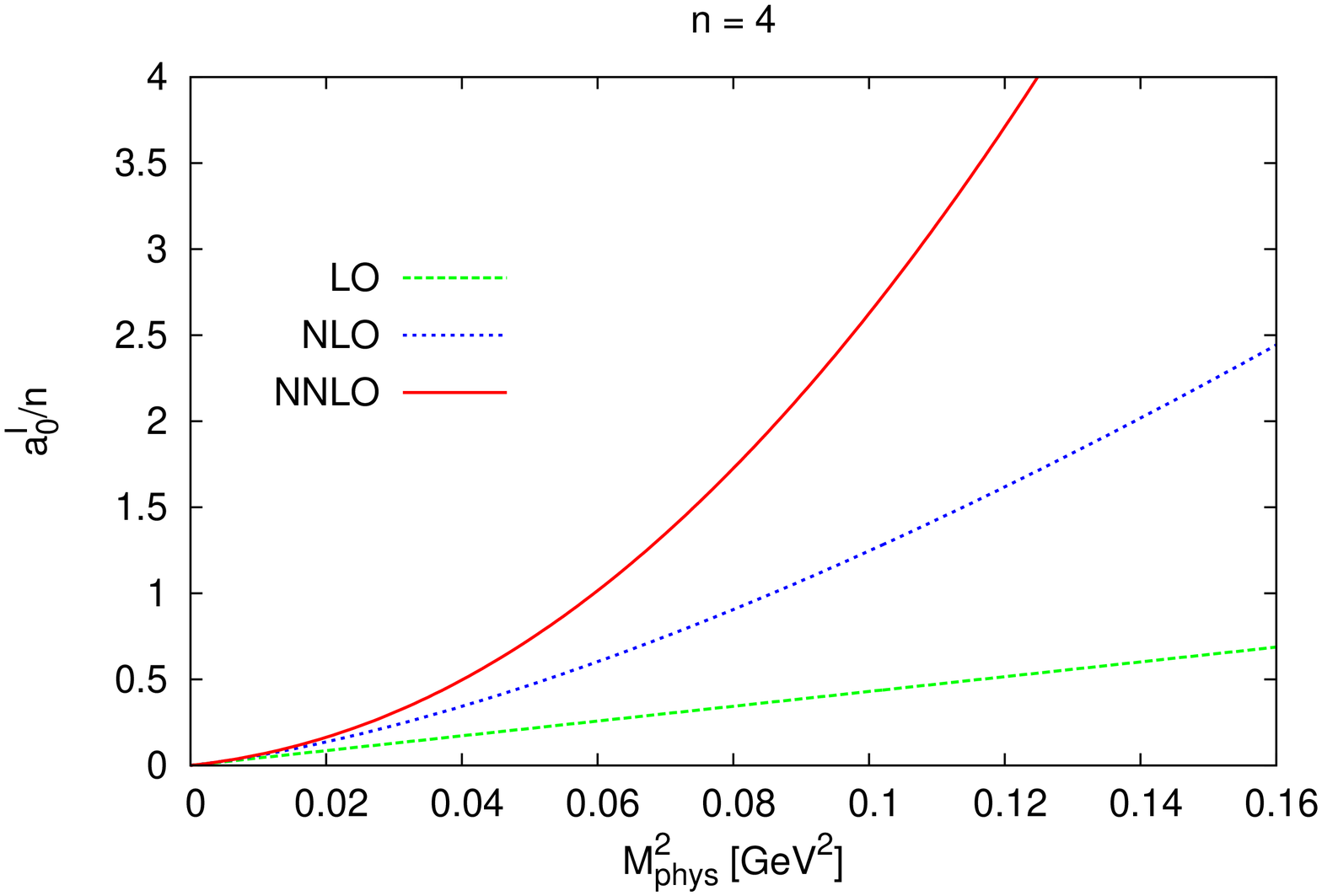}
\end{minipage}
\end{center}
\caption{\label{figaI0SPN} Scattering length of $a^I_0/n$ for
the pseudoreal or two-colour case, $SU(2n)/Sp(2n)$.}
\end{figure}

\begin{figure}
\begin{center}
\begin{minipage}{7.3cm}
\includegraphics[angle=0,width=0.99\textwidth]{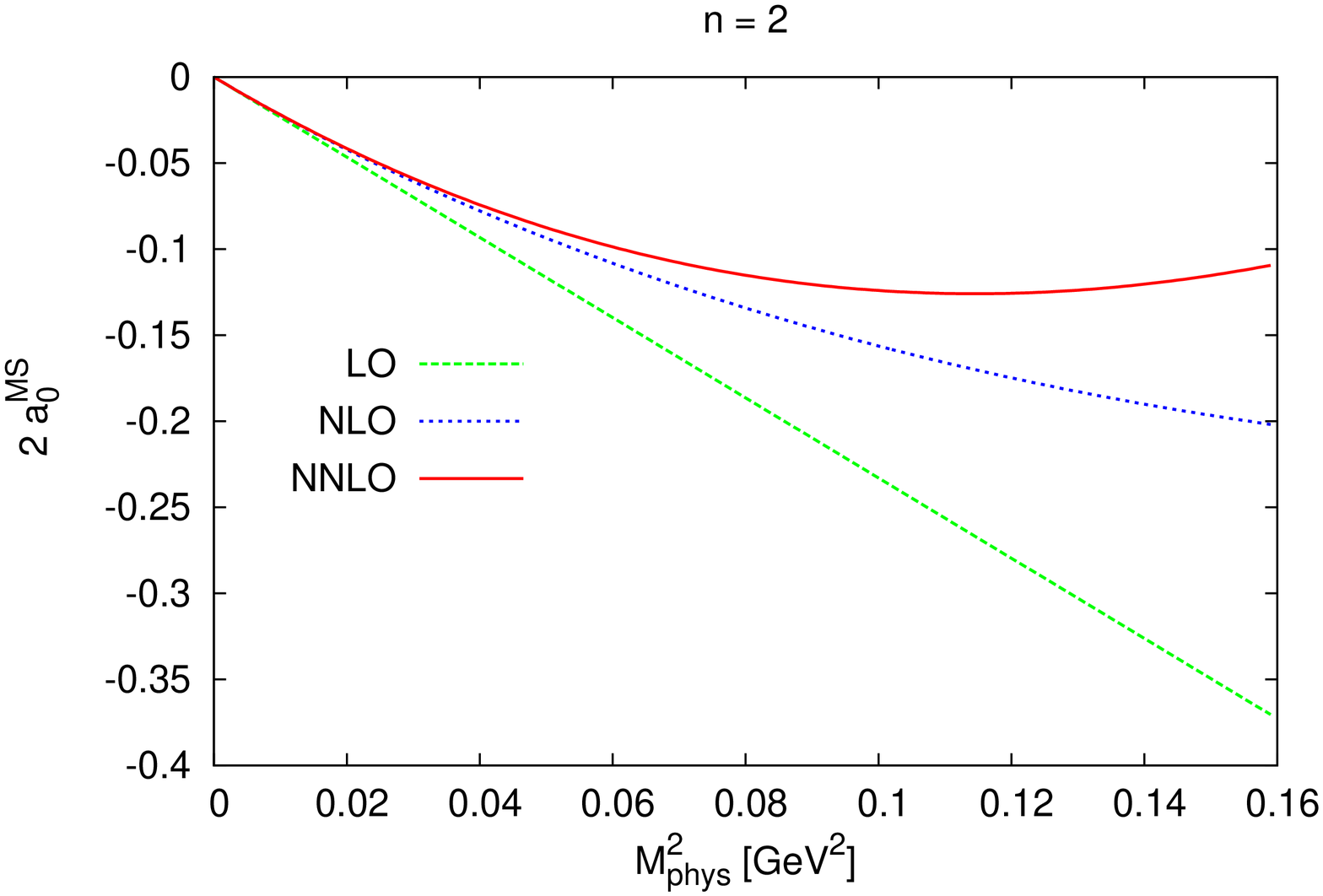}
\end{minipage}
\begin{minipage}{7.3cm}
\includegraphics[angle=0,width=0.99\textwidth]{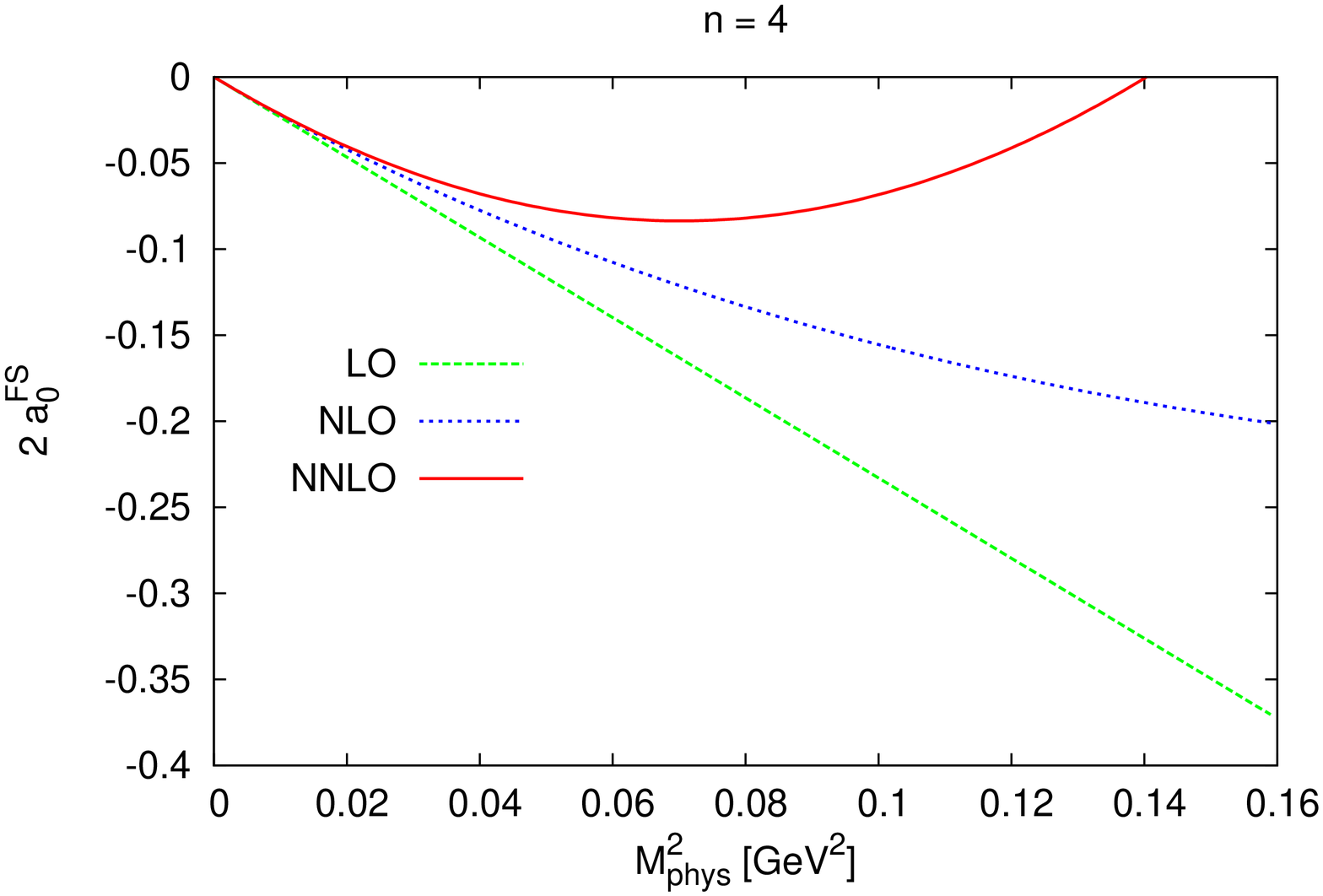}
\end{minipage}
\end{center}
\caption{\label{figaFASPN} Scattering length $2a^{MS}_0$ for
the pseudoreal or two-colour case, $SU(2n)/Sp(2n)$.
The factor of 2 included is because of the large $n$ relation
(\ref{largen2}).}
\end{figure}

\subsection{Large $n$ behaviour}

Looking at the lowest order expressions in App.~\ref{scattlength} we notice
immediately that the large $n$ behaviour for the scattering lengths fall in
three classes.

The scattering length is of order $n$ for
the singlet and symmetric and asymmetric representation and what is more they
are clearly related for the three cases:
\ba
\label{largen1}
a^I_0|_\mathrm{complex} &=&
a^I_0|_\mathrm{real} =
a^I_0|_\mathrm{pseudoreal} =_{LO} \frac{x_2}{\pi}\,\frac{n}{8}\,,
\\
\label{largena}
a^S_0|_\mathrm{complex} &=&
a^S_0|_\mathrm{real} =
a^A_0|_\mathrm{pseudoreal} =_{LO} \frac{x_2}{\pi}\,\frac{n}{16}\,,
\\
\label{largenb}
a^A_1|_\mathrm{complex} &=&
a^A_1|_\mathrm{real} =
a^S_1|_\mathrm{pseudoreal} =_{LO} \frac{x_2}{\pi}\,\frac{n}{48}\,,
\ea
The symbol $=_{LO}$ means equality at lowest order.
Do the relations (\ref{largen1}-\ref{largenb}) remain valid at higher orders?
If we choose $\fphys^2\propto n$ and $\mphys$
independent of $n$, we find that it is indeed the case for
(\ref{largena},\ref{largenb}). For (\ref{largen1}) it is true,
provided
we set the NLO LECs $L_i^r$ of the real and pseudoreal case to half those
of the complex case and the NNLO coefficients $K_i$ to 1/4 the complex case.
The contributions are nonzero at the three orders for all of them.
The subleading orders in $1/n$ are different.

A second class is those which are order $1$ in the coefficient of $x_2/\pi$
at lowest order. For these we find
\ba
\label{largen2}
a^{SS}_0|_\mathrm{complex} &=&
a^{FS}_0|_\mathrm{real} =
2 a^{MS}_0|_\mathrm{pseudoreal} =_{LO} \frac{x_2}{\pi}\,\frac{-1}{16}\,,
\\
\label{largenc}
a^{AA}_0|_\mathrm{complex} &=&
2 a^{MS}_0|_\mathrm{real} =
a^{FA}_0|_\mathrm{pseudoreal} =_{LO} \frac{x_2}{\pi}\,\frac{1}{16}\,.
\ea
We do indeed find
that the relations are also satisfied at NLO and NNLO. In fact none of the
scattering lengths in (\ref{largen2}) has a leading $n$ NLO correction
to the lowest order result. This can be clearly seen in Fig.~\ref{figaSSSUN}
where the NLO result is very similar in all plots.

The third class is the amplitudes that vanish at lowest order
\ba
\label{largen3}
a^{SA}_1|_\mathrm{complex} &=&
a^{AS}_1|_\mathrm{complex} =
2 a^{MA}_1|_\mathrm{real} =
2 a^{MA}_1|_\mathrm{pseudoreal} =_{LO} 0\,.
\ea
These are always suppressed by two powers of $n$ compared to the first set of
scattering lengths also at NLO and NNLO. The relations (\ref{largen3}) are
satisfied at NLO with same identifications of the LECs as above and almost at
NNLO. The only terms that do not satisfy the relation are proportional to
$L_4^r L_6^r$.

By comparing the plots shown one sees that the
large $n$ relations do predict the general behaviour but for $n=2$ and $n=4$
are not that accurate.

\section{Conclusions}
\label{conclusions}

In this work we have presented the calculation of general meson-meson
scattering for $n$ flavours in a
complex, real or pseudoreal representation of a strongly interaction gauge
group. These are also referred to as
QCD, Adjoint QCD and Two-color QCD and have as symmetry breaking patterns
$SU(n)\times SU(n)/SU(n)$, $SU(2n)/SO(2n)$ and $SU(2n)/Sp(2n)$

We first reviewed the effective field theories of these three different cases.
Those theories can be written in a very similar form as discussed
earlier \cite{paper1}. We have extended the methods used for $\pi\pi$ scattering
in ChPT \cite{BCEGS2} to all the present cases. At intermediate stages some more
integrals showed up, we have calculated them and they are tabulated in
an appendix.

The amplitude can in general be written in terms of two invariant amplitudes
which we called $B(s,t,u)$ and $C(s,t,u)$. These amplitudes can be written
in terms of simpler functions and we have given their fully analytical
expressions to NNLO.

Since the long term use of our work is the study of scattering on the lattice
for these alternative theories we have discussed the group theory involved and
all the possible intermediate channels. We have derived the amplitudes
in all these channel as a function of the invariant $B$ and $C$ functions.

The expressions for the different channels we have not shown explicitly but we
included expressions for scattering length of the lowest partial wave in
all channels. We presented a few representative numerical results for
the scattering lengths and discussed a series of relations between the different
theories in the limit of a large number of flavours $n$.

\section*{Acknowledgements}

This work is supported in part by the European Community-Research
Infrastructure Integrating Activity ``Study of Strongly Interacting Matter''
(HadronPhysics2, Grant Agreement n. 227431)
and the Swedish Research Council grants 621-2008-4074 and 621-2011-3326.
This work heavily used FORM \cite{Vermaseren:2000nd}.

\appendix

\section{Next-to-next-to leading order result}
\label{appNNLO}

\subsection{Complex or QCD}

\begin{eqnarray}
\lefteqn{B_S(s,t-u)=
   k_4(s)\Bigg\{\frac{n_f^2}{12}+\frac{2}{n_f^2}\Bigg\}(t-u)}&&
 \nonumber\\&&+ k_3(s)  \Bigg\{\frac{s^2
   n_f^2}{48}-\frac{s n_f^2}{18}-\frac{1}{48} s (t-u)n_f^2+\frac{(t-u)n_f^2}{36}
   -\frac{(t-u)}{6}+\frac{1}{3}-\frac{s}{3n_f^2}
\nonumber\\&&\hskip1cm
 +\frac{(t-u)}{3 n_f^2}-\frac{8}{3 n_f^2}\Bigg\}
 \nonumber\\&&+ k_2(s)  \Bigg\{\frac{n_f^2
   s^3}{64}+\frac{s^3}{16}-\frac{1}{576} n_f^2 (t-u)
   s^2-\frac{3 s^2}{8}+\frac{1}{72} n_f^2 (t-u)
   s-\frac{n_f^2
   (t-u)}{36}+\frac{2}{n_f^2}\Bigg\}
 \nonumber\\&&+  k_1(s)
   \Bigg\{\frac{n_f^2 s^3}{576}+\frac{s^3}{24}-\frac{17 n_f^2
   s^2}{288}+\frac{1}{576} n_f^2 (t-u)
   s^2-\frac{s^2}{4}+\frac{n_f^2 s}{12}-\frac{1}{96}
   n_f^2 (t-u) s
    \nonumber\\&&\hskip1cm
  -\frac{(t-u) s}{24}+\frac{s}{2 n_f^2}+\frac{7 s}{12}-\frac{n_f^2
   (t-u)}{48}+\frac{(t-u)}{4}+\frac{4}{n_f^2}-\frac{1}{2}\Bigg\}
 \nonumber\\&&+ \bar{J}(s)  \Bigg\{-\frac{5}{144} L n_f^2 s^3-\frac{5 L
   s^3}{24}+\frac{4 L^r_1 s^3}{3}+\frac{8 L^r_2
   s^3}{3}+\frac{2}{3} L^r_0 n_f s^3+\frac{4}{3} L^r_3
   n_f s^3
 \nonumber\\&&\hskip1cm
   +\frac{17}{576} n_f^2 \pi_{16} s^3
   +\frac{17 \pi_{16} s^3}{72}+\frac{7}{144} L n_f^2 s^2+\frac{4 Ls^2}{3}
    -\frac{8 L^r_1 s^2}{3}-\frac{28 L^r_2 s^2}{3}-4
   L^r_4 s^2
 \nonumber\\&&\hskip1cm
 -\frac{4}{3} L^r_0 n_f s^2-\frac{14}{3} L^r_3 n_f s^2
 +L^r_5 n_f s^2-\frac{35}{144}
   n_f^2 \pi_{16} s^2-\frac{13 \pi_{16} s^2}{9}+\frac{2}{3}
   L^r_1 (t-u) s^2
 \nonumber\\&&\hskip1cm
  -\frac{1}{3} L^r_2 (t-u)
   s^2-\frac{1}{3} L^r_0 n_f (t-u) s^2
  +\frac{1}{6} L^r_3 n_f (t-u) s^2-\frac{1}{432} n_f^2 \pi_{16}
   (t-u) s^2
 \nonumber\\&&\hskip1cm
-\frac{40 L^r_0 s^2}{3 n_f}-\frac{40 L^r_3s^2}{3 n_f}-\frac{5}{36} L n_f^2 s
   -\frac{5 Ls}{3}
  +\frac{16 L^r_1 s}{3}+\frac{32 L^r_2 s}{3}+16L^r_6 s+\frac{8 L^r_0 n_f s}{3}
   \nonumber\\&&\hskip1cm
 +\frac{16 L^r_3  n_f s}{3}-4 L^r_5 n_f s+8 L^r_8 n_f
   s+\frac{5}{12} n_f^2 \pi_{16} s
 +\frac{2 \pi_{16}
   s}{n_f^2}+\frac{65 \pi_{16} s}{36}
  \nonumber\\&&\hskip1cm
+\frac{1}{48} L n_f^2
   (t-u) s +\frac{L (t-u) s}{12}-\frac{8 L^r_1 (t-u)
   s}{3}+\frac{4 L^r_2 (t-u) s}{3}
 -\frac{4 L^r_4 (t-u) s}{3}
  \nonumber\\&&\hskip1cm
   +\frac{4}{3} L^r_0 n_f (t-u) s-\frac{2}{3}
   L^r_3 n_f (t-u) s-\frac{1}{3} L^r_5 n_f (t-u)s
  +\frac{17}{216} n_f^2 \pi_{16} (t-u) s
   \nonumber\\&&\hskip1cm
-\frac{5
   \pi_{16} (t-u) s}{72}+\frac{128 L^r_0 s}{3
   n_f}+\frac{128 L^r_3 s}{3 n_f}-\frac{L
   s}{n_f^2}+\frac{8 L}{3}+\frac{20 \pi_{16}}{n_f^2}
 -\frac{20\pi_{16}}{9}
 \nonumber\\&&\hskip1cm
-\frac{1}{12} L n_f^2 (t-u)-\frac{L (t-u)}{3}+\frac{16
   L^r_4 (t-u)}{3}+\frac{4 L^r_5 n_f (t-u)}{3}
  \nonumber\\&&\hskip1cm
 -\frac{11}{48}
   n_f^2 \pi_{16} (t-u)
-\frac{\pi_{16} (t-u)}{2 n_f^2}+\frac{10
   \pi_{16} (t-u)}{9}-\frac{160 L^r_0}{3 n_f}-\frac{160 L^r_3}{3
   n_f}+\frac{32 L^r_5}{n_f}
  \nonumber\\&&\hskip1cm
 -\frac{96 L^r_8}{n_f}-\frac{12
   L}{n_f^2}\Bigg\}
   \end{eqnarray}

\begin{eqnarray}
\lefteqn{B_T(t)=
   k_3(t) \left\{\frac{2 t}{3n_f^2}+\frac{t}{3}-\frac{4}{3n_f^2}
   -\frac{2}{3}\right\}
 +  k_2(t)
   \left\{-\frac{t^3}{8}+\frac{3 t^2}{4}-\frac{3
   t}{2}+1\right\}
}&&
 \nonumber\\&&+  k_1(t)
   \left\{-\frac{t^3}{12}+\frac{t^2}{2}-\frac{t}{n_f^2}-\frac{
   7 t}{6}+\frac{2}{n_f^2}+1\right\}
 \nonumber\\&&
 +  \bar{J}(t)
   \Bigg\{\frac{5 L t^3}{12}-\frac{8 L^r_1 t^3}{3}-\frac{16 L^r_2
   t^3}{3}-\frac{17 \pi_{16} t^3}{36}-\frac{8 L t^2}{3}+\frac{32
   L^r_1 t^2}{3}+\frac{88 L^r_2 t^2}{3}+8 L^r_4t^2
 \nonumber\\&&\hskip1cm
   +\frac{26 \pi_{16} t^2}{9}+\frac{19 L t}{3}-\frac{64
   L^r_1 t}{3}-\frac{176 L^r_2 t}{3}-16 L^r_4 t-32
   L^r_6 t-\frac{4 \pi_{16} t}{n_f^2}-\frac{64 \pi_{16}
   t}{9}
\nonumber\\&&\hskip1cm
  +\frac{2 L t}{n_f^2}
  -\frac{16 L}{3}+\frac{64
   L^r_1}{3}+\frac{128 L^r_2}{3}+64 L^r_6+\frac{8
   \pi_{16}}{n_f^2}+\frac{58 \pi_{16}}{9}-\frac{4 L}{n_f^2}\Bigg\}
\end{eqnarray}

\begin{eqnarray}
\lefteqn{C_S(s)=
  k_3(s) \Bigg\{\frac{n_f s^2}{12}-\frac{5 n_f
   s}{9}+\frac{2 s}{3 n_f}+\frac{16}{3n_f^3}\Bigg\}
}&&
 \nonumber\\&&
 +  k_2(s) \Bigg\{\frac{3 n_f s^3}{16}-\frac{6}{n_f^3}\Bigg\}
  +  k_1(s) \Bigg\{\frac{13 n_f
   s^3}{144}-\frac{23 n_f s^2}{72}+\frac{5 n_f
   s}{6}-\frac{2s}{n_f}-\frac{8}{n_f^3}\Bigg\}
 \nonumber\\&&
 +  \bar{J}(s)
   \Bigg\{\frac{8 L^r_0 s^3}{3}+\frac{16 L^r_3 s^3}{3}-\frac{5}{9}
   L n_f s^3+8 L^r_1 n_f s^3+\frac{8}{3} L^r_2
   n_f s^3+\frac{85}{144} n_f \pi_{16} s^3
    \nonumber\\&&\hskip1cm
  -\frac{16   L^r_0 s^2}{3}-\frac{56 L^r_3 s^2}{3}+4 L^r_5
   s^2+\frac{19}{36} L n_f s^2-32 L^r_1 n_f
   s^2-\frac{16}{3} L^r_2 n_f s^2
\nonumber\\&&\hskip1cm
  +16 L^r_4 n_f
   s^2
  -\frac{7}{6} n_f \pi_{16} s^2-\frac{16 L^r_1
   s^2}{n_f}-\frac{16 L^r_2 s^2}{3 n_f}+\frac{80
   L^r_0 s^2}{3 n_f^2}+\frac{80 L^r_3 s^2}{3
   n_f^2}+\frac{32 L^r_0 s}{3}
    \nonumber\\&&\hskip1cm
+\frac{64 L^r_3 s}{3}
-16 L^r_5 s+32 L^r_8 s-\frac{11 L n_f s}{9}+32
   L^r_1 n_f s+\frac{32 L^r_2 n_f s}{3}-32 L^r_4
   n_f s
    \nonumber\\&&\hskip1cm
  +32 L^r_6 n_f s
  +\frac{35 n_f \pi_{16}s}{9}-\frac{7 \pi_{16} s}{n_f}+\frac{4 L
   s}{n_f}+\frac{64 L^r_1 s}{n_f}+\frac{32 L^r_2
   s}{3 n_f}-\frac{32 L^r_4 s}{n_f}
    \nonumber\\&&\hskip1cm
  -\frac{256 L^r_0
   s}{3 n_f^2}
  -\frac{256 L^r_3 s}{3 n_f^2}
  -\frac{44
   \pi_{16}}{n_f^3}
  -\frac{4 L}{n_f}-\frac{64 L^r_1}{n_f}-\frac{64
   L^r_2}{3 n_f}+\frac{64 L^r_4}{n_f}-\frac{64
   L^r_6}{n_f}
    \nonumber\\&&\hskip1cm
+\frac{320 L^r_0}{3 n_f^2}+\frac{320 L^r_3}{3
   n_f^2}
-\frac{64 L^r_5}{n_f^2}+\frac{192
   L^r_8}{n_f^2}+\frac{28 L}{n_f^3}\Bigg\}
\end{eqnarray}

\begin{eqnarray}
\lefteqn{C_T(t)=
  k_3(t) \Bigg\{\frac{n_f t^2}{12}-\frac{2 n_f
   t}{9}-\frac{2 t}{3 n_f}+\frac{n_f}{9}+\frac{4}{3
   n_f}\Bigg\}
}&&
 \nonumber\\&&
+  k_1(t) \Bigg\{-\frac{5 n_f
   t^3}{144}+\frac{n_f t^2}{8}-\frac{n_f
   t}{36}+\frac{t}{n_f}-\frac{n_f}{6}-\frac{2}{n_f}\Bigg\}
 \nonumber\\&&+  \bar{J}(t) \Bigg\{-4 L^r_0 t^3-\frac{4 L^r_3
   t^3}{3}+\frac{5}{72} L n_f t^3-\frac{17}{144} n_f
   \pi_{16} t^3+24 L^r_0 t^2+\frac{16 L^r_3 t^2}{3}+4
   L^r_5 t^2
    \nonumber\\&&\hskip1cm
  -\frac{11}{18} L n_f t^2+\frac{1}{9} n_f
   \pi_{16} t^2-48 L^r_0 t-\frac{32 L^r_3 t}{3}-8
   L^r_5 t-16 L^r_8 t+\frac{31 L n_f
   t}{18}
\nonumber\\&&\hskip1cm
 +\frac{29 n_f \pi_{16} t}{36}
+\frac{4 \pi_{16}
   t}{n_f}-\frac{2 L t}{n_f}+32 L^r_0+\frac{32
   L^r_3}{3}+32 L^r_8-\frac{14 L n_f}{9}-\frac{10 n_f
   \pi_{16}}{9}
\nonumber\\&&\hskip1cm
-\frac{8 \pi_{16}}{n_f}+\frac{4 L}{n_f}\Bigg\}
\end{eqnarray}

\subsection{Real or adjoint}

\begin{eqnarray}
\lefteqn{ B_S(s,t-u) =
   k_4(s) \Bigg\{\frac{(t-u) n^2}{12}-\frac{(t-u)
   n}{12}-\frac{(t-u)}{6}-\frac{(t-u)}{2
   n}+\frac{(t-u)}{2 n^2}\Bigg\}
}&&
\nonumber\\&&
+k_3(s)
   \Bigg\{\frac{s^2 n^2}{48}-\frac{s n^2}{18}-\frac{1}{48} s
   (t-u) n^2+\frac{(t-u) n^2}{36}-\frac{s
   n}{24}-\frac{s (t-u) n}{24}
  +\frac{(t-u)
   n}{72}
\nonumber\\&&\hskip1cm
  -\frac{n}{9}
  -\frac{s^2}{48}+\frac{s}{18}-\frac{s
   (t-u)}{48}-\frac{7 (t-u)}{72}-\frac{1}{36}+\frac{s}{12
   n}-\frac{(t-u)}{12 n}
  +\frac{1}{2 n}-\frac{s}{12
   n^2}
\nonumber\\&&\hskip1cm
  +\frac{(t-u)}{12 n^2}-\frac{2}{3
   n^2}\Bigg\}
\nonumber\\&&
   +k_2(s) \Bigg\{\frac{n^2
   s^3}{64}-\frac{ns^3}{64}+\frac{3 s^3}{64}+\frac{3 n
   s^2}{32}-\frac{1}{576} n^2 (t-u) s^2-\frac{1}{288} n
   (t-u) s^2
\nonumber\\&&\hskip1cm
  -\frac{(t-u) s^2}{576}-\frac{9 s^2}{32}+\frac{1}{72}
   n^2 (t-u) s+\frac{n(t-u) s}{36}+\frac{(t-u)
   s}{72}+\frac{3 s}{16}-\frac{n^2 (t-u)}{36}
\nonumber\\&&\hskip1cm
  -\frac{n
   (t-u)}{18}-\frac{(t-u)}{36}-\frac{1}{2 n}+\frac{1}{2
   n^2}\Bigg\}
\nonumber\\&&
   +k_1(s) \Bigg\{\frac{n^2
   s^3}{576}+\frac{ns^3}{96}+\frac{11 s^3}{576}-\frac{17 n^2
   s^2}{288}-\frac{11 ns^2}{288}+\frac{1}{576} n^2 (t-u)
   s^2
-\frac{1}{144} n(t-u) s^2
\nonumber\\&&\hskip1cm
 -\frac{5 (t-u)
   s^2}{576}-\frac{3 s^2}{32}+\frac{n^2 s}{12}+\frac{n
   s}{144}-\frac{1}{96} n^2 (t-u) s+\frac{n(t-u)
   s}{32}
  +\frac{(t-u) s}{48}
  \nonumber\\&&\hskip1cm
  -\frac{s}{8 n}+\frac{s}{8
   n^2}+\frac{31 s}{144}+\frac{n}{6}-\frac{n^2
   (t-u)}{48}-\frac{n(t-u)}{24}+\frac{5
   (t-u)}{48}
  \nonumber\\&&\hskip1cm
  -\frac{3}{4
   n}+\frac{1}{n^2}+\frac{1}{24}\Bigg\}
   \nonumber\\&&
  +\bar J(t)
   \Bigg\{-\frac{5}{144} L n^2 s^3-\frac{19 L s^3}{144}+L^r_0
   s^3+\frac{4 L^r_1 s^3}{3}+\frac{8 L^r_2 s^3}{3}+\frac{L^r_3
   s^3}{3}+\frac{1}{96} L ns^3
  \nonumber\\&&\hskip1cm
 +\frac{2}{3} L^r_0 n
   s^3+\frac{4}{3} L^r_3 ns^3+\frac{17}{576} n^2
  \pi_{16} s^3-\frac{1}{288} n\pi_{16} s^3+\frac{29\pi_{16}
   s^3}{192}+\frac{7}{144} L n^2 s^2
  \nonumber\\&&\hskip1cm
  +\frac{31 L s^2}{36}-\frac{8
   L^r_0 s^2}{3}-\frac{8 L^r_1 s^2}{3}-\frac{28 L^r_2 s^2}{3}+2
   L^r_3 s^2-4 L^r_4 s^2-L^r_5 s^2-\frac{13}{144} L n
   s^2
  \nonumber\\&&\hskip1cm
 -\frac{4}{3} L^r_0 ns^2-\frac{14}{3} L^r_3 n
   s^2+L^r_5 ns^2-\frac{35}{144} n^2\pi_{16}
   s^2+\frac{19}{288} n\pi_{16} s^2
 -\frac{55\pi_{16}
   s^2}{72}
  \nonumber\\&&\hskip1cm
 +\frac{1}{48} L (t-u) s^2-\frac{1}{3} L^r_0 (t-u)
   s^2+\frac{2}{3} L^r_1 (t-u) s^2-\frac{1}{3} L^r_2 (t-u)
   s^2
  \nonumber\\&&\hskip1cm
  +\frac{1}{6} L^r_3 (t-u) s^2
 +\frac{1}{48} L n
   (t-u) s^2-\frac{1}{3} L^r_0 n(t-u) s^2+\frac{1}{6}
   L^r_3 n(t-u) s^2
  \nonumber\\&&\hskip1cm
 -\frac{1}{432} n^2\pi_{16}
   (t-u) s^2
 -\frac{47 n\pi_{16} (t-u) s^2}{1728}-\frac{43
  \pi_{16} (t-u) s^2}{1728}-\frac{20 L^r_0 s^2}{3
   n}
  \nonumber\\&&\hskip1cm
 -\frac{20 L^r_3 s^2}{3 n}-\frac{5}{36} L n^2
   s-\frac{97 L s}{72}+\frac{4 L^r_0 s}{3}
 +\frac{16 L^r_1
   s}{3}+\frac{32 L^r_2 s}{3}-8 L^r_3 s+2 L^r_5 s
  \nonumber\\&&\hskip1cm
+16 L^r_6
   s+4 L^r_8 s-\frac{17 L ns}{72}+\frac{8 L^r_0 n
   s}{3}+\frac{16 L^r_3 ns}{3}
 -4 L^r_5 ns+8
   L^r_8 ns
 \nonumber\\&&\hskip1cm
 +\frac{5}{12} n^2\pi_{16}
   s
+\frac{n\pi_{16} s}{24}-\frac{\pi_{16} s}{2
   n}
 +\frac{\pi_{16} s}{2 n^2}+\frac{13\pi_{16}
   s}{16}+\frac{1}{48} L n^2 (t-u) s
 -\frac{L (t-u)
   s}{24}
  \nonumber\\&&\hskip1cm
+\frac{4 L^r_0 (t-u) s}{3}
-\frac{8 L^r_1 (t-u)
   s}{3}+\frac{4 L^r_2 (t-u) s}{3}-\frac{2 L^r_3 (t-u)
   s}{3}-\frac{4 L^r_4 (t-u) s}{3}
  \nonumber\\&&\hskip1cm
-\frac{L^r_5 (t-u)
   s}{3}-\frac{1}{16} L n(t-u) s+\frac{4}{3} L^r_0 n
   (t-u) s-\frac{2}{3} L^r_3 n(t-u) s-\frac{1}{3}
   L^r_5 n(t-u) s
  \nonumber\\&&\hskip1cm +\frac{17}{216} n^2\pi_{16}
   (t-u) s+\frac{253}{864} n\pi_{16} (t-u) s+\frac{155
  \pi_{16} (t-u) s}{864}+\frac{L s}{4 n}+\frac{64 L^r_0
   s}{3 n}
  \nonumber\\&&\hskip1cm
 +\frac{64 L^r_3 s}{3 n}
-\frac{L s}{4
   n^2}+\frac{19 L}{18}+\frac{16 L^r_0}{3}+\frac{32
   L^r_3}{3}-8 L^r_5+16 L^r_8-\frac{5 L n}{18}+\frac{5
   n\pi_{16}}{6}
  \nonumber\\&&\hskip1cm
-\frac{4\pi_{16}}{n}+\frac{5
  \pi_{16}}{n^2}
 +\frac{2\pi_{16}}{9}-\frac{1}{12} L n^2
   (t-u)-\frac{L (t-u)}{6}+\frac{16 L^r_4 (t-u)}{3}
  \nonumber\\&&\hskip1cm
+\frac{4
   L^r_5 (t-u)}{3}-\frac{L n(t-u)}{12}
+\frac{4
   L^r_5 n(t-u)}{3}-\frac{11}{48} n^2\pi_{16}
   (t-u)
  \nonumber\\&&\hskip1cm
-\frac{59 n\pi_{16} (t-u)}{144}
+\frac{\pi_{16}
   (t-u)}{8 n}
 -\frac{\pi_{16} (t-u)}{8
   n^2}+\frac{3\pi_{16} (t-u)}{8}+\frac{5 L}{2
   n}-\frac{80 L^r_0}{3 n}
  \nonumber\\&&\hskip1cm
-\frac{80 L^r_3}{3
   n}
+\frac{16 L^r_5}{n}
-\frac{48
   L^r_8}{n}-\frac{3 L}{n^2}\Bigg\}
\end{eqnarray}

\begin{eqnarray}
\lefteqn{  B_T(t) =
  k_3(t) \Bigg\{\frac{nt^2}{24}+\frac{t^2}{24}-\frac{n
   t}{9}-\frac{t}{6 n}+\frac{t}{6
   n^2}+\frac{t}{18}+\frac{n}{18}+\frac{1}{3
   n}-\frac{1}{3 n^2}-\frac{5}{18}\Bigg\}
}&&
   \nonumber\\&&
 +k_2(t)
   \Bigg\{-\frac{3 t^3}{32}+\frac{9 t^2}{16}-\frac{9
   t}{8}+\frac{3}{4}\Bigg\}
   \nonumber\\&&
 +k_1(t) \Bigg\{-\frac{5 n
   t^3}{288}-\frac{11 t^3}{288}+\frac{nt^2}{16}+\frac{3
   t^2}{16}-\frac{nt}{72}+\frac{t}{4 n}-\frac{t}{4
   n^2}-\frac{31 t}{72}
 -\frac{n}{12}-\frac{1}{2
   n}
  \nonumber\\&&\hskip1cm
+\frac{1}{2 n^2}+\frac{5}{12}\Bigg\}
   \nonumber\\&&
 +\bar J(t)
   \Bigg\{\frac{19 L t^3}{72}-2 L^r_0 t^3-\frac{8 L^r_1
   t^3}{3}-\frac{16 L^r_2 t^3}{3}-\frac{2 L^r_3 t^3}{3}+\frac{5}{144}
   L nt^3-\frac{17}{288} n\pi_{16} t^3
  \nonumber\\&&\hskip1cm
-\frac{29
  \pi_{16} t^3}{96}-\frac{31 L t^2}{18}+12 L^r_0 t^2+\frac{32
   L^r_1 t^2}{3}+\frac{88 L^r_2 t^2}{3}+\frac{8 L^r_3 t^2}{3}+8
   L^r_4 t^2+2 L^r_5 t^2
  \nonumber\\&&\hskip1cm
 -\frac{11}{36} L nt^2+\frac{1}{18}
   n\pi_{16} t^2+\frac{55\pi_{16} t^2}{36}+\frac{151 L
   t}{36}-24 L^r_0 t-\frac{64 L^r_1 t}{3}-\frac{176 L^r_2
   t}{3}
  \nonumber\\&&\hskip1cm
-\frac{16 L^r_3 t}{3}
 -16 L^r_4 t-4 L^r_5 t-32 L^r_6
   t-8 L^r_8 t+\frac{31 L nt}{36}+\frac{29 n\pi_{16}
   t}{72}+\frac{\pi_{16} t}{n}-\frac{\pi_{16}
   t}{n^2}
  \nonumber\\&&\hskip1cm
-\frac{27\pi_{16} t}{8}
 -\frac{L t}{2 n}+\frac{L
   t}{2 n^2}-\frac{65 L}{18}+16 L^r_0+\frac{64
   L^r_1}{3}+\frac{128 L^r_2}{3}+\frac{16 L^r_3}{3}+64
   L^r_6
  \nonumber\\&&\hskip1cm
+16 L^r_8-\frac{7 L n}{9}
 -\frac{5 n
  \pi_{16}}{9}-\frac{2\pi_{16}}{n}+\frac{2
  \pi_{16}}{n^2}+\frac{55
  \pi_{16}}{18}+\frac{L}{n}-\frac{L}{n^2}\Bigg\}
\end{eqnarray}

\begin{eqnarray}
\lefteqn{    C_S(s) =
    k_3(s) \Bigg\{\frac{ns^2}{24}+\frac{s^2}{24}-\frac{5 n
   s}{18}+\frac{s}{6 n}-\frac{7 s}{36}-\frac{1}{3
   n^2}+\frac{2}{3 n^3}-\frac{1}{6}\Bigg\}
}&&
   \nonumber\\&&
 +k_2(s)
   \Bigg\{\frac{3 ns^3}{32}-\frac{s^3}{32}+\frac{3 s^2}{16}+\frac{1}{2
   n^2}-\frac{3}{4 n^3}\Bigg\}
   \nonumber\\&&
 +k_1(s)
   \Bigg\{\frac{13 ns^3}{288}+\frac{s^3}{288}-\frac{23 n
   s^2}{144}-\frac{s^2}{72}+\frac{5 ns}{12}-\frac{s}{2
   n}+\frac{s}{4}+\frac{1}{2
   n^2}-\frac{1}{n^3}+\frac{1}{4}\Bigg\}
   \nonumber\\&&
 +\bar J(s) \Bigg\{\frac{L
   s^3}{18}+\frac{4 L^r_0 s^3}{3}+\frac{8 L^r_3 s^3}{3}-\frac{5}{18}
   L ns^3+8 L^r_1 ns^3+\frac{8}{3} L^r_2 n
   s^3+\frac{85}{288} n\pi_{16} s^3
  \nonumber\\&&\hskip1cm
 -\frac{19\pi_{16}
   s^3}{288}-\frac{35 L s^2}{72}-\frac{8 L^r_0 s^2}{3}+8 L^r_1
   s^2+\frac{8 L^r_2 s^2}{3}-\frac{28 L^r_3 s^2}{3}+2 L^r_5
   s^2+\frac{19}{72} L ns^2
  \nonumber\\&&\hskip1cm
 -32 L^r_1 ns^2-\frac{16}{3}
   L^r_2 ns^2+16 L^r_4 ns^2-\frac{7}{12} n
  \pi_{16} s^2+\frac{3\pi_{16} s^2}{16}-\frac{8 L^r_1
   s^2}{n}-\frac{8 L^r_2 s^2}{3 n}
  \nonumber\\&&\hskip1cm
 +\frac{20 L^r_0
   s^2}{3 n^2}+\frac{20 L^r_3 s^2}{3 n^2}-\frac{L
   s}{9}+\frac{16 L^r_0 s}{3}-32 L^r_1 s-\frac{16 L^r_2
   s}{3}+\frac{32 L^r_3 s}{3}+16 L^r_4 s
  \nonumber\\&&\hskip1cm
  -8 L^r_5 s
 +16 L^r_8
   s-\frac{11 L ns}{18}+32 L^r_1 ns+\frac{32 L^r_2
   ns}{3}-32 L^r_4 ns+32 L^r_6 ns
  \nonumber\\&&\hskip1cm
+\frac{35
   n\pi_{16} s}{18}
 -\frac{7\pi_{16} s}{4 n}+\frac{89
  \pi_{16} s}{72}+\frac{L s}{n}+\frac{32 L^r_1
   s}{n}+\frac{16 L^r_2 s}{3 n}-\frac{16 L^r_4
   s}{n}
-\frac{64 L^r_0 s}{3 n^2}
  \nonumber\\&&\hskip1cm
-\frac{64 L^r_3 s}{3
   n^2}-\frac{L}{3}+32 L^r_1
 +\frac{32 L^r_2}{3}-32
   L^r_4+32 L^r_6+\frac{3\pi_{16}}{n^2}-\frac{11
  \pi_{16}}{2 n^3}+\frac{10
  \pi_{16}}{9}
  \nonumber\\&&\hskip1cm
-\frac{L}{n}-\frac{32 L^r_1}{n}-\frac{32
   L^r_2}{3 n}+\frac{32 L^r_4}{n}
 -\frac{32
   L^r_6}{n}-\frac{2 L}{n^2}+\frac{80 L^r_0}{3
   n^2}+\frac{80 L^r_3}{3 n^2}-\frac{16
   L^r_5}{n^2}
  \nonumber\\&&\hskip1cm
+\frac{48 L^r_8}{n^2}+\frac{7 L}{2
   n^3}\Bigg\}
\end{eqnarray}

\begin{eqnarray}
\lefteqn{  C_T(t) =
   k_3(t) \Bigg\{\frac{nt^2}{24}+\frac{t^2}{24}-\frac{n
   t}{9}-\frac{t}{6 n}-\frac{t}{9}+\frac{n}{18}+\frac{1}{3
   n}+\frac{1}{18}\Bigg\}
}&&
   \nonumber\\&&
 +k_2(t) \Bigg\{-\frac{t^3}{32}+\frac{3
   t^2}{16}-\frac{3 t}{8}+\frac{1}{4}\Bigg\}
   \nonumber\\&&
 +k_1(t) \Bigg\{-\frac{5
   nt^3}{288}+\frac{t^3}{288}+\frac{n
   t^2}{16}-\frac{t^2}{16}-\frac{nt}{72}+\frac{t}{4
   n}+\frac{11 t}{72}-\frac{n}{12}-\frac{1}{2
   n}-\frac{1}{12}\Bigg\}
   \nonumber\\&&
 +\bar J(t) \Bigg\{\frac{L t^3}{18}-2
   L^r_0 t^3-\frac{2 L^r_3 t^3}{3}+\frac{5}{144} L n
   t^3-\frac{17}{288} n\pi_{16} t^3-\frac{19\pi_{16}
   t^3}{288}-\frac{7 L t^2}{18}+12 L^r_0 t^2
  \nonumber\\&&\hskip1cm
 +\frac{8 L^r_3 t^2}{3}+2
   L^r_5 t^2-\frac{11}{36} L nt^2+\frac{1}{18} n
  \pi_{16} t^2+\frac{\pi_{16} t^2}{12}+\frac{37 L t}{36}-24 L^r_0
   t-\frac{16 L^r_3 t}{3}
  \nonumber\\&&\hskip1cm
-4 L^r_5 t
 -8 L^r_8 t+\frac{31 L
   nt}{36}+\frac{29 n\pi_{16} t}{72}+\frac{\pi_{16}
   t}{n}+\frac{13\pi_{16} t}{72}-\frac{L t}{2 n}-\frac{17
   L}{18}+16 L^r_0
  \nonumber\\&&\hskip1cm
+\frac{16 L^r_3}{3}+16 L^r_8
 -\frac{7 L
   n}{9}-\frac{5 n\pi_{16}}{9}-\frac{2
  \pi_{16}}{n}-\frac{\pi_{16}}{6}+\frac{L}{n}\Bigg\}
\end{eqnarray}

\subsection{Pseudo-real or two-colour}

\begin{eqnarray}
\lefteqn{    B_S(s,t-u) =
    k_4(s) \Bigg\{\frac{n^2}{12}+\frac{n}{12}-\frac{1}{6}+\frac{1}{2
   n}+\frac{1}{2 n^2}\Bigg\}(t-u)
}&&
   \nonumber\\&& +k_3(s)
   \Bigg\{\frac{s^2 n^2}{48}-\frac{s n^2}{18}-\frac{1}{48} s
   (t-u) n^2+\frac{(t-u) n^2}{36}+\frac{s
   n}{24}+\frac{s (t-u) n}{24}
  \nonumber\\&&\hskip1cm
 -\frac{(t-u)
   n}{72}+\frac{n}{9}-\frac{s^2}{48}+\frac{s}{18}-\frac{s
   (t-u)}{48}-\frac{7 (t-u)}{72}-\frac{1}{36}-\frac{s}{12
   n}+\frac{(t-u)}{12 n}
  \nonumber\\&&\hskip1cm
 -\frac{1}{2 n}-\frac{s}{12
   n^2}+\frac{(t-u)}{12 n^2}-\frac{2}{3
   n^2}\Bigg\}
   \nonumber\\&&
 +k_2(s) \Bigg\{\frac{n^2
   s^3}{64}+\frac{ns^3}{64}+\frac{3 s^3}{64}-\frac{3 n
   s^2}{32}-\frac{1}{576} n^2 (t-u) s^2+\frac{1}{288} n
   (t-u) s^2
  \nonumber\\&&\hskip1cm -\frac{(t-u) s^2}{576}-\frac{9 s^2}{32}+\frac{1}{72}
   n^2 (t-u) s-\frac{n(t-u) s}{36}+\frac{(t-u)
   s}{72}+\frac{3 s}{16}-\frac{n^2 (t-u)}{36}
  \nonumber\\&&\hskip1cm +\frac{n
   (t-u)}{18}-\frac{(t-u)}{36}+\frac{1}{2 n}+\frac{1}{2
   n^2}\Bigg\}
   \nonumber\\&&
 +k_1(s) \Bigg\{\frac{n^2
   s^3}{576}-\frac{ns^3}{96}+\frac{11 s^3}{576}-\frac{17 n^2
   s^2}{288}+\frac{11 ns^2}{288}+\frac{1}{576} n^2 (t-u)
   s^2
  \nonumber\\&&\hskip1cm +\frac{1}{144} n(t-u) s^2-\frac{5 (t-u)
   s^2}{576}-\frac{3 s^2}{32}+\frac{n^2 s}{12}-\frac{n
   s}{144}-\frac{1}{96} n^2 (t-u) s
  \nonumber\\&&\hskip1cm -\frac{n(t-u)
   s}{32}+\frac{(t-u) s}{48}+\frac{s}{8 n}+\frac{s}{8
   n^2}+\frac{31 s}{144}-\frac{n}{6}-\frac{n^2
   (t-u)}{48}
  \nonumber\\&&\hskip1cm +\frac{n(t-u)}{24}+\frac{5
   (t-u)}{48}+\frac{3}{4
   n}+\frac{1}{n^2}+\frac{1}{24}\Bigg\}
   \nonumber\\&&
 +\bar J(s)
   \Bigg\{-\frac{5}{144} L n^2 s^3-\frac{19 L s^3}{144}-L^r_0
   s^3+\frac{4 L^r_1 s^3}{3}+\frac{8 L^r_2 s^3}{3}-\frac{L^r_3
   s^3}{3}-\frac{1}{96} L ns^3
  \nonumber\\&&\hskip1cm +\frac{2}{3} L^r_0 n
   s^3+\frac{4}{3} L^r_3 ns^3+\frac{17}{576} n^2
  \pi_{16} s^3+\frac{1}{288} n\pi_{16} s^3+\frac{29\pi_{16}
   s^3}{192}+\frac{7}{144} L n^2 s^2
  \nonumber\\&&\hskip1cm +\frac{31 L s^2}{36}+\frac{8
   L^r_0 s^2}{3}-\frac{8 L^r_1 s^2}{3}-\frac{28 L^r_2 s^2}{3}-2
   L^r_3 s^2-4 L^r_4 s^2+L^r_5 s^2+\frac{13}{144} L n
   s^2
  \nonumber\\&&\hskip1cm -\frac{4}{3} L^r_0 ns^2-\frac{14}{3} L^r_3 n
   s^2+L^r_5 ns^2-\frac{35}{144} n^2\pi_{16}
   s^2-\frac{19}{288} n\pi_{16} s^2-\frac{55\pi_{16}
   s^2}{72}
  \nonumber\\&&\hskip1cm +\frac{1}{48} L (t-u) s^2+\frac{1}{3} L^r_0 (t-u)
   s^2+\frac{2}{3} L^r_1 (t-u) s^2-\frac{1}{3} L^r_2 (t-u)
   s^2-\frac{1}{6} L^r_3 (t-u) s^2
  \nonumber\\&&\hskip1cm -\frac{1}{48} L n
   (t-u) s^2-\frac{1}{3} L^r_0 n(t-u) s^2+\frac{1}{6}
   L^r_3 n(t-u) s^2-\frac{1}{432} n^2\pi_{16}
   (t-u) s^2
  \nonumber\\&&\hskip1cm +\frac{47 n\pi_{16} (t-u) s^2}{1728}-\frac{43
  \pi_{16} (t-u) s^2}{1728}-\frac{20 L^r_0 s^2}{3
   n}-\frac{20 L^r_3 s^2}{3 n}-\frac{5}{36} L n^2
   s-\frac{97 L s}{72}
  \nonumber\\&&\hskip1cm -\frac{4 L^r_0 s}{3}+\frac{16 L^r_1
   s}{3}+\frac{32 L^r_2 s}{3}+8 L^r_3 s-2 L^r_5 s+16 L^r_6
   s-4 L^r_8 s+\frac{17 L ns}{72}+\frac{8 L^r_0 n
   s}{3}
  \nonumber\\&&\hskip1cm +\frac{16 L^r_3 ns}{3}-4 L^r_5 ns+8
   L^r_8 ns+\frac{5}{12} n^2\pi_{16}
   s-\frac{n\pi_{16} s}{24}+\frac{\pi_{16} s}{2
   n}+\frac{\pi_{16} s}{2 n^2}+\frac{13\pi_{16}
   s}{16}
  \nonumber\\&&\hskip1cm +\frac{1}{48} L n^2 (t-u) s-\frac{L (t-u)
   s}{24}-\frac{4 L^r_0 (t-u) s}{3}-\frac{8 L^r_1 (t-u)
   s}{3}+\frac{4 L^r_2 (t-u) s}{3}
  \nonumber\\&&\hskip1cm +\frac{2 L^r_3 (t-u)
   s}{3}-\frac{4 L^r_4 (t-u) s}{3}+\frac{L^r_5 (t-u)
   s}{3}+\frac{1}{16} L n(t-u) s+\frac{4}{3} L^r_0 n
   (t-u) s
  \nonumber\\&&\hskip1cm -\frac{2}{3} L^r_3 n(t-u) s-\frac{1}{3}
   L^r_5 n(t-u) s+\frac{17}{216} n^2\pi_{16}
   (t-u) s-\frac{253}{864} n\pi_{16} (t-u) s
  \nonumber\\&&\hskip1cm +\frac{155
  \pi_{16} (t-u) s}{864}-\frac{L s}{4 n}+\frac{64 L^r_0
   s}{3 n}+\frac{64 L^r_3 s}{3 n}-\frac{L s}{4
   n^2}+\frac{19 L}{18}-\frac{16 L^r_0}{3}-\frac{32
   L^r_3}{3}+8 L^r_5
  \nonumber\\&&\hskip1cm -16 L^r_8+\frac{5 L n}{18}-\frac{5
   n\pi_{16}}{6}+\frac{4\pi_{16}}{n}+\frac{5
  \pi_{16}}{n^2}+\frac{2\pi_{16}}{9}-\frac{1}{12} L n^2
   (t-u)-\frac{L (t-u)}{6}
  \nonumber\\&&\hskip1cm +\frac{16 L^r_4 (t-u)}{3}-\frac{4
   L^r_5 (t-u)}{3}+\frac{L n(t-u)}{12}+\frac{4
   L^r_5 n(t-u)}{3}-\frac{11}{48} n^2\pi_{16}
   (t-u)
  \nonumber\\&&\hskip1cm +\frac{59 n\pi_{16} (t-u)}{144}-\frac{\pi_{16}
   (t-u)}{8 n}-\frac{\pi_{16} (t-u)}{8
   n^2}+\frac{3\pi_{16} (t-u)}{8}-\frac{5 L}{2
   n}-\frac{80 L^r_0}{3 n}-\frac{80 L^r_3}{3
   n}
  \nonumber\\&&\hskip1cm +\frac{16 L^r_5}{n}-\frac{48
   L^r_8}{n}-\frac{3 L}{n^2}\Bigg\}
\end{eqnarray}

\begin{eqnarray}
\lefteqn{  B_T(t) =
    k_3(t) \Bigg\{-\frac{nt^2}{24}+\frac{t^2}{24}+\frac{n
   t}{9}+\frac{t}{6 n}+\frac{t}{6
   n^2}+\frac{t}{18}-\frac{n}{18}-\frac{1}{3
   n}-\frac{1}{3 n^2}-\frac{5}{18}\Bigg\}
}&&
   \nonumber\\&& +k_2(t)
   \Bigg\{-\frac{3 t^3}{32}+\frac{9 t^2}{16}-\frac{9
   t}{8}+\frac{3}{4}\Bigg\}
   \nonumber\\&& +k_1(t) \Bigg\{\frac{5 n
   t^3}{288}-\frac{11 t^3}{288}-\frac{nt^2}{16}+\frac{3
   t^2}{16}+\frac{nt}{72}-\frac{t}{4 n}-\frac{t}{4
   n^2}-\frac{31 t}{72}+\frac{n}{12}
  \nonumber\\&&\hskip1cm +\frac{1}{2
   n}+\frac{1}{2 n^2}+\frac{5}{12}\Bigg\}
   \nonumber\\&&
 +\bar J(t)
   \Bigg\{\frac{19 L t^3}{72}+2 L^r_0 t^3-\frac{8 L^r_1
   t^3}{3}-\frac{16 L^r_2 t^3}{3}+\frac{2 L^r_3 t^3}{3}-\frac{5}{144}
   L nt^3+\frac{17}{288} n\pi_{16} t^3
  \nonumber\\&&\hskip1cm -\frac{29
  \pi_{16} t^3}{96}-\frac{31 L t^2}{18}-12 L^r_0 t^2+\frac{32
   L^r_1 t^2}{3}+\frac{88 L^r_2 t^2}{3}-\frac{8 L^r_3 t^2}{3}+8
   L^r_4 t^2-2 L^r_5 t^2
  \nonumber\\&&\hskip1cm +\frac{11}{36} L nt^2-\frac{1}{18}
   n\pi_{16} t^2+\frac{55\pi_{16} t^2}{36}+\frac{151 L
   t}{36}+24 L^r_0 t-\frac{64 L^r_1 t}{3}-\frac{176 L^r_2
   t}{3}+\frac{16 L^r_3 t}{3}
  \nonumber\\&&\hskip1cm -16 L^r_4 t+4 L^r_5 t-32 L^r_6
   t+8 L^r_8 t-\frac{31 L nt}{36}-\frac{29 n\pi_{16}
   t}{72}-\frac{\pi_{16} t}{n}-\frac{\pi_{16}
   t}{n^2}-\frac{27\pi_{16} t}{8}
  \nonumber\\&&\hskip1cm +\frac{L t}{2 n}+\frac{L
   t}{2 n^2}-\frac{65 L}{18}-16 L^r_0+\frac{64
   L^r_1}{3}+\frac{128 L^r_2}{3}-\frac{16 L^r_3}{3}+64
   L^r_6-16 L^r_8+\frac{7 L n}{9}
  \nonumber\\&&\hskip1cm +\frac{5 n
  \pi_{16}}{9}+\frac{2\pi_{16}}{n}+\frac{2
  \pi_{16}}{n^2}+\frac{55
  \pi_{16}}{18}-\frac{L}{n}-\frac{L}{n^2}\Bigg\}
\end{eqnarray}

\begin{eqnarray}
\lefteqn{ C_S(s) =
    k_3(s) \Bigg\{\frac{ns^2}{24}-\frac{s^2}{24}-\frac{5 n
   s}{18}+\frac{s}{6 n}+\frac{7 s}{36}+\frac{1}{3
   n^2}+\frac{2}{3 n^3}+\frac{1}{6}\Bigg\}
}&&
   \nonumber\\&& +k_2(s)
   \Bigg\{\frac{3 ns^3}{32}+\frac{s^3}{32}-\frac{3 s^2}{16}-\frac{1}{2
   n^2}-\frac{3}{4 n^3}\Bigg\}
   \nonumber\\&& +k_1(s)
   \Bigg\{\frac{13 ns^3}{288}-\frac{s^3}{288}-\frac{23 n
   s^2}{144}+\frac{s^2}{72}+\frac{5 ns}{12}-\frac{s}{2
   n}-\frac{s}{4}-\frac{1}{2
   n^2}-\frac{1}{n^3}-\frac{1}{4}\Bigg\}
   \nonumber\\&& +\bar J(s) \Bigg\{-\frac{L
   s^3}{18}+\frac{4 L^r_0 s^3}{3}+\frac{8 L^r_3 s^3}{3}-\frac{5}{18}
   L ns^3+8 L^r_1 ns^3+\frac{8}{3} L^r_2 n
   s^3+\frac{85}{288} n\pi_{16} s^3
  \nonumber\\&&\hskip1cm +\frac{19\pi_{16}
   s^3}{288}+\frac{35 L s^2}{72}-\frac{8 L^r_0 s^2}{3}-8 L^r_1
   s^2-\frac{8 L^r_2 s^2}{3}-\frac{28 L^r_3 s^2}{3}+2 L^r_5
   s^2+\frac{19}{72} L ns^2
  \nonumber\\&&\hskip1cm -32 L^r_1 ns^2-\frac{16}{3}
   L^r_2 ns^2+16 L^r_4 ns^2-\frac{7}{12} n
  \pi_{16} s^2-\frac{3\pi_{16} s^2}{16}-\frac{8 L^r_1
   s^2}{n}-\frac{8 L^r_2 s^2}{3 n}
  \nonumber\\&&\hskip1cm +\frac{20 L^r_0
   s^2}{3 n^2}+\frac{20 L^r_3 s^2}{3 n^2}+\frac{L
   s}{9}+\frac{16 L^r_0 s}{3}+32 L^r_1 s+\frac{16 L^r_2
   s}{3}+\frac{32 L^r_3 s}{3}-16 L^r_4 s-8 L^r_5 s
  \nonumber\\&&\hskip1cm +16 L^r_8
   s-\frac{11 L ns}{18}+32 L^r_1 ns+\frac{32 L^r_2
   ns}{3}-32 L^r_4 ns+32 L^r_6 ns+\frac{35
   n\pi_{16} s}{18}
  \nonumber\\&&\hskip1cm -\frac{7\pi_{16} s}{4 n}-\frac{89
  \pi_{16} s}{72}+\frac{L s}{n}+\frac{32 L^r_1
   s}{n}+\frac{16 L^r_2 s}{3 n}-\frac{16 L^r_4
   s}{n}-\frac{64 L^r_0 s}{3 n^2}-\frac{64 L^r_3 s}{3
   n^2}+\frac{L}{3}-32 L^r_1
  \nonumber\\&&\hskip1cm -\frac{32 L^r_2}{3}+32
   L^r_4-32 L^r_6-\frac{3\pi_{16}}{n^2}-\frac{11
  \pi_{16}}{2 n^3}-\frac{10
  \pi_{16}}{9}-\frac{L}{n}-\frac{32 L^r_1}{n}-\frac{32
   L^r_2}{3 n}
  \nonumber\\&&\hskip1cm +\frac{32 L^r_4}{n}-\frac{32
   L^r_6}{n}+\frac{2 L}{n^2}+\frac{80 L^r_0}{3
   n^2}+\frac{80 L^r_3}{3 n^2}-\frac{16
   L^r_5}{n^2}+\frac{48 L^r_8}{n^2}+\frac{7 L}{2
   n^3}\Bigg\}
\end{eqnarray}

\begin{eqnarray}
\lefteqn{ C_T(t) =
  k_3(t) \Bigg\{\frac{nt^2}{24}-\frac{t^2}{24}-\frac{n
   t}{9}-\frac{t}{6 n}+\frac{t}{9}+\frac{n}{18}+\frac{1}{3
   n}-\frac{1}{18}\Bigg\}
}&&
   \nonumber\\&&
 +k_2(t) \Bigg\{\frac{t^3}{32}-\frac{3
   t^2}{16}+\frac{3 t}{8}-\frac{1}{4}\Bigg\}
   \nonumber\\&&
 +k_1(t) \Bigg\{-\frac{5
   nt^3}{288}-\frac{t^3}{288}+\frac{n
   t^2}{16}+\frac{t^2}{16}-\frac{nt}{72}+\frac{t}{4
   n}-\frac{11 t}{72}-\frac{n}{12}-\frac{1}{2
   n}+\frac{1}{12}\Bigg\}
   \nonumber\\&&
 +\bar J(t) \Bigg\{-\frac{L t^3}{18}-2
   L^r_0 t^3-\frac{2 L^r_3 t^3}{3}+\frac{5}{144} L n
   t^3-\frac{17}{288} n\pi_{16} t^3+\frac{19\pi_{16}
   t^3}{288}+\frac{7 L t^2}{18}
  \nonumber\\&&\hskip1cm +12 L^r_0 t^2+\frac{8 L^r_3 t^2}{3}+2
   L^r_5 t^2-\frac{11}{36} L nt^2+\frac{1}{18} n
  \pi_{16} t^2-\frac{\pi_{16} t^2}{12}-\frac{37 L t}{36}-24 L^r_0
   t
  \nonumber\\&&\hskip1cm -\frac{16 L^r_3 t}{3}-4 L^r_5 t-8 L^r_8 t+\frac{31 L
   nt}{36}+\frac{29 n\pi_{16} t}{72}+\frac{\pi_{16}
   t}{n}-\frac{13\pi_{16} t}{72}-\frac{L t}{2 n}
  \nonumber\\&&\hskip1cm +\frac{17
   L}{18}+16 L^r_0+\frac{16 L^r_3}{3}+16 L^r_8-\frac{7 L
   n}{9}-\frac{5 n\pi_{16}}{9}-\frac{2
  \pi_{16}}{n}+\frac{\pi_{16}}{6}+\frac{L}{n}\Bigg\}
\end{eqnarray}

\section{Polynomial parts}
\label{appdivergences}

Divergent parts can be put here

\subsection{Complex or QCD}

The coefficients of polynomials part for $B_P$ at NNLO.
\begin{eqnarray*}
\gamma_1&=& 32 K^r_{13}+32 K^r_{14} n-96 K^r_{17}-96 K^r_{18} n+96
K^r_{25}+32 K^r_{26} n
+64 K^r_{3}-64 K^r_{37}\\
&&+96 K^r_{39} +32 K^r_{40} n+\frac{29 L^2 n^2}{36}+\frac{19
L^2}{n^2}+\frac{L^2}{3} -\frac{80 L L^r_{0} n}{3}+\frac{64 L
L^r_{0}}{3 n}
-64 L L^r_{1}\\
&&-\frac{224 L L^r_{2}}{3}-8 L L^r_{3} n+\frac{64 L L^r_{3}}{3
n}+\frac{64 L L^r_{4}}{3}
+\frac{40 L L^r_{5} n}{3}-\frac{96 L L^r_{5}}{n}-160 L L^r_{6}\\
&&-32 L L^r_{7}-64 L L^r_{8} n
+\frac{224 L L^r_{8}}{n}+256 L^r_{4} L^r_{8} n+256 L^r_{5} L^r_{8}-512 L^r_{6} L^r_{8} n-512 (L^r_{8})^2\\
&&+\pi_{16}^2 \left(\frac{n^2 \pi ^2}{27}+\frac{1645 n^2}{1728}-\frac{35}{2 n^2}+\frac{4 \pi^2}{9}-\frac{181}{54}\right)\\
&&+\pi_{16} \Bigg(\frac{229 L n^2}{216}+\frac{4 L}{n^2}-\frac{26
L}{9}-\frac{80 L^r_0 n}{9}+\frac{256 L^r_0}{9 n}-\frac{32
L^r_1}{3}-\frac{368 L^r_2}{9}\\
&&-\frac{8 L^r_3 n}{3}+\frac{256 L^r_3}{9 n}+\frac{256
L^r_4}{9}+\frac{64 L^r_5 n}{9}-\frac{64 L^r_5}{n}-128
L^r_6-32 L^r_8 n+\frac{192 L^r_8}{n}\Bigg)\\
\gamma_2 &=& -32 K^r_{13}-32 K^r_{14} n+64 K^r_{17}+64 K^r_{18} n-16
K^r_{19}-8 K^r_{20} n
-16 K^r_{23}-32 K^r_{28}\\
&&-96 K^r_{3}-16 K^r_{33}+32 K^r_{37}-\frac{17}{36} L^2 n^2
-\frac{3 L^2}{2 n^2}-\frac{13 L^2}{3}+24 L L^r_{0} n+\frac{16 L L^r_{0}}{n}\\
&&+\frac{176 L L^r_{1}}{3}+\frac{248 L L^r_{2}}{3}+\frac{4 L L^r_{3}
n}{3}
+\frac{16 L L^r_{3}}{n}+\frac{32 L L^r_{4}}{3}+\frac{20 L L^r_{5} n}{3}+48 L L^r_{6}\\
&&+8 L L^r_{8} n-32 L^r_{4} L^r_{5} n-32 (L^r_{5})^2+64 L^r_{5}
L^r_{6} n+64 L^r_{5} L^r_{8}
\\
&&+\pi_{16} \Bigg(-\frac{445 L n^2}{432}-\frac{37 L}{36} +8 L^r_0
n-\frac{16 L^r_0}{n}+\frac{80 L^r_1}{9}+\frac{512 L^r_2}{9}
+\frac{16 L^r_3 n}{9}\\
&&-\frac{16 L^r_3}{n}+\frac{32 L^r_4}{9}+\frac{8 L^r_5 n}{9}
+48 L^r_6+8 L^r_8 n \Bigg)\\
&& +\pi_{16}^2 \left(-\frac{5}{72} n^2 \pi ^2-\frac{3865
n^2}{10368}+\frac{3}{n^2}
-\frac{13 \pi ^2}{36}-\frac{25}{432}\right)\\
\gamma_3 &=&2 K^r_{11}+8 K^r_{13}+8 K^r_{14} n-16 K^r_{17}-12
K^r_{18} n+16 K^r_{28}
+48 K^r_{3}-4 K^r_{31}+8 K^r_{5}\\
&&+2 K^r_{7}+2 K^r_{8} n+\frac{29 L^2 n^2}{288}+\frac{27 L^2}{16}
-8 L L^r_{0} n+\frac{20 L L^r_{0}}{3 n}-\frac{56 L L^r_{1}}{3}-40 L L^r_{2}\\
&&-\frac{8 L L^r_{3} n}{3}+\frac{20 L L^r_{3}}{3 n}-8 L L^r_{4}-L
L^r_{5} n
\\
&& +\pi_{16} \Bigg(\frac{49 L n^2}{216}+\frac{29 L}{16}-\frac{7
L^r_0 n}{3} +\frac{56 L^r_0}{9 n}-\frac{62 L^r_1}{9}-31
L^r_2-\frac{25 L^r_3 n}{18}\\
&&+\frac{56 L^r_3}{9 n}
-\frac{20 L^r_4}{3}-\frac{2 L^r_5 n}{3}\Bigg)\\
&&+\pi_{16}^2 \left(\frac{n^2 \pi ^2}{72}+\frac{445
n^2}{5184}+\frac{3 \pi ^2}{16}
-\frac{23}{192}\right)\\
\gamma_4 &=& 2 K^r_{11}+4 K^r_{18} n+4 K^r_{31}+8 K^r_{5}+2
K^r_{7}+2 K^r_{8} n
+\frac{11 L^2 n^2}{288}-\frac{7 L^2}{48}-\frac{4 L L^r_{0} n}{3}\\
&&+\frac{20 L L^r_{0}}{3 n}-\frac{4 L L^r_{2}}{3}-2 L L^r_{3}
n+\frac{20 L L^r_{3}}{3 n}+\frac{8 L L^r_{4}}{3}-\frac{L L^r_{5}
n}{3}
\\
&&+\pi_{16} \Bigg(\frac{29 L n^2}{216}+\frac{17 L}{144}-\frac{13
L^r_0 n}{9}+\frac{56 L^r_0}{9 n}-\frac{10 L^r_1}{3}-\frac{31
L^r_2}{9}-\frac{11 L^r_3 n}{6}\\
&&+\frac{56 L^r_3}{9 n}+\frac{20 L^r_4}{9}-\frac{4 L^r_5 n}{9}\Bigg)\\
&&+\pi_{16}^2 \left(-\frac{1}{108} n^2 \pi ^2+\frac{421 n^2}{3456}-\frac{\pi ^2}{144}-\frac{115}{1728}\right)\\
\gamma_5 &=&K^r_{1}-8 K^r_{3}-4 K^r_{5}-\frac{5 L^2
n^2}{1152}-\frac{15 L^2}{64}+\frac{5 L L^r_{0} n}{12}+\frac{5 L
L^r_{1}}{2}+\frac{25 L L^r_{2}}{4}+\frac{5 L L^r_{3} n}{24}
\\&& +\pi_{16} \left(-\frac{19 L n^2}{2304}-\frac{13 L}{32}+\frac{2 L^r_0 n}{9}+\frac{7 L^r_1}{3}+\frac{35 L^r_2}{6}+\frac{5 L^r_3 n}{18}\right)\\
&&+\pi_{16}^2 \left(\frac{n^2 \pi ^2}{3456}-\frac{1015 n^2}{165888}-\frac{\pi ^2}{32}+\frac{23}{384}\right)\\
\gamma_6 &=&3 K^r_{1}-4 K^r_{5}-\frac{5}{384} L^2 n^2-\frac{5
L^2}{64}+\frac{7 L L^r_{0} n}{12}+\frac{5 L L^r_{1}}{6}+\frac{25 L
L^r_{2}}{12}+\frac{23 L L^r_{3} n}{24}
\\&&
 +\pi_{16} \left(-\frac{203 L n^2}{6912}-\frac{13 L}{96}+\frac{4 L^r_0 n}{9}+\frac{7 L^r_1}{9}+\frac{35 L^r_2}{18}+\frac{17 L^r_3 n}{18}\right)\\
&&+\pi_{16}^2 \left(-\frac{n^2 \pi ^2}{3456}-\frac{1933
n^2}{165888}-\frac{\pi ^2}{96}+\frac{23}{1152}\right)
\end{eqnarray*}
The coefficients of polynomials part for $C_P$ at NNLO.
\begin{eqnarray*}
\delta_1 &=& 64 K^r_{10} n-128 K^r_{18}+128 K^r_{2}-64 K^r_{20}-128
K^r_{21}-128 K^r_{22} n
+128 K^r_{26}+192 K^r_{27} n\\
&&-128 K^r_{35}+128 K^r_{40}+64 K^r_{9} -\frac{14 L^2}{n^3}+\frac{2
L^2}{n}-\frac{192 L L^r_{0}}{n^2}
-64 L L^r_{1} n+\frac{128 L L^r_{1}}{n}\\
&&+\frac{32 L L^r_{2}}{n}-\frac{192 L L^r_{3}}{n^2}+64 L L^r_{4}
n-\frac{96 L L^r_{4}}{n}
+\frac{96 L L^r_{5}}{n^2}+16 L L^r_{5}-64 L L^r_{6} n+\frac{64 L L^r_{6}}{n}\\
&&+\frac{64 L L^r_{7}}{n}-\frac{192 L L^r_{8}}{n^2}-32 L L^r_{8}
-256 (L^r_{4})^2 n-256 L^r_{4} L^r_{5}+1024 L^r_{4} L^r_{6} n+512 L^r_{4} L^r_{8}\\
&&+512 L^r_{5} L^r_{6}-1024 (L^r_{6})^2 n-1024 L^r_{6} L^r_{8}
\\
&&+\pi_{16} \left(-\frac{12 L}{n^3}+\frac{4 L}{n}-\frac{64
L^r_0}{n^2}+\frac{64 L^r_1}{n}-\frac{64 L^r_3}{n^2}
-\frac{64 L^r_4}{n}+\frac{64 L^r_5}{n^2}+\frac{64 L^r_6}{n}-\frac{192 L^r_8}{n^2}\right)\\
&&+\pi_{16}^2\left(\frac{10}{n^3}+\frac{3}{n}\right)
\\
\delta_2 &=& -64 K^r_{10} n+128 K^r_{18}-192 K^r_{2}+32 K^r_{20}+64
K^r_{21}+64 K^r_{22} n
-32 K^r_{32}+64 K^r_{35}\\
&&-32 K^r_{38}-64 K^r_{9}+\frac{37 L^2 n}{36}-\frac{3 L^2}{n}
+\frac{416 L L^r_{0}}{3 n^2}-16 L L^r_{0}-\frac{96 L L^r_{1}}{n}-16 L L^r_{2} n\\
&&-\frac{64 L L^r_{2}}{3 n}+\frac{416 L L^r_{3}}{3 n^2} -\frac{128 L
L^r_{3}}{3}+16 L L^r_{4} n+\frac{32 L L^r_{4}}{n}
+24 L L^r_{5}-32 L L^r_{6} n\\
&&-48 L L^r_{8}+128 (L^r_{4})^2 n+128 L^r_{4} L^r_{5}-256 L^r_{4}
L^r_{6} n-256 L^r_{4} L^r_{8}
\\
&&+\pi_{16} \Bigg(-\frac{31 L n}{12}-\frac{L}{n}+\frac{608 L^r_0}{9
n^2}-32 L^r_1 n-\frac{64 L^r_1}{n}-\frac{16 L^r_2}{9 n}
+\frac{608 L^r_3}{9 n^2}-\frac{176 L^r_3}{9}\\
&&+32 L^r_4 n+\frac{32 L^r_4}{n}+24 L^r_5-32 L^r_6 n-48 L^r_8\Bigg)\\
&&+\pi_{16}^2 \left(-\frac{2 n \pi ^2}{27}-\frac{373
n}{1296}-\frac{\pi ^2}{3 n}+\frac{25}{4 n}\right)
\\
\delta_3 &=&16 K^r_{10} n+4 K^r_{15}+4 K^r_{16} n-32 K^r_{18}+96
K^r_{2}-8 K^r_{29}
+16 K^r_{32}+16 K^r_{9}-\frac{13 L^2 n}{36}\\
&&-\frac{80 L L^r_{0}}{3 n^2}+\frac{20 L L^r_{0}}{3} +32 L L^r_{1}
n+\frac{16 L L^r_{1}}{n}+\frac{20 L L^r_{2} n}{3}
+\frac{16 L L^r_{2}}{3 n}-\frac{80 L L^r_{3}}{3 n^2}+\frac{88 L L^r_{3}}{3}\\
&&-16 L L^r_{4} n-6 L L^r_{5}
\\
&&+\pi_{16} \Bigg(\frac{11 L n}{12}-\frac{224 L^r_0}{9 n^2}+\frac{8
L^r_0}{9}+32 L^r_1 n+\frac{16 L^r_1}{n}+\frac{8 L^r_2 n}{9}
+\frac{40 L^r_2}{9 n}\\
&&-\frac{224 L^r_3}{9 n^2}+\frac{166 L^r_3}{9}-16 L^r_4 n-6 L^r_5\Bigg)\\
&&+\pi_{16}^2 \left(\frac{625 n}{1296}-\frac{25 n \pi ^2}{432}\right)\\
\delta_4 &=&4 K^r_{15}+4 K^r_{16} n+8 K^r_{29}+\frac{5 L^2 n}{24}-4
L L^r_{0}-4 L L^r_{2} n-2 L L^r_{5}
\\
&&+\pi_{16} \left(\frac{L n}{8}+2 L^r_3-2 L^r_5\right) +\pi_{16}^2 \left(\frac{n \pi ^2}{144}+\frac{7 n}{32}\right)\\
\delta_5 &=&-16 K^r_{2}+2 K^r_{4}+2 K^r_{6}+\frac{55 L^2
n}{192}-\frac{11 L L^r_{0}}{3}-8 L L^r_{1} n-\frac{8 L L^r_{2}
n}{3}-\frac{17 L L^r_{3}}{3}
\\
&& +\pi_{16} \left(\frac{101 L n}{192}-\frac{29 L^r_0}{9}-8 L^r_1
n-\frac{20 L^r_2 n}{9}-\frac{97 L^r_3}{18}\right)
+\pi_{16}^2 \left(\frac{19 n \pi ^2}{576}-\frac{115 n}{6912}\right)\\
\delta_6 &=&6 K^r_{4}-2 K^r_{6}+\frac{5 L^2 n}{192}-3 L L^r_{0}-L
L^r_{3}
\\
&&+\pi_{16} \left(\frac{Ln}{64}-3 L^r_0-\frac{5 L^r_3}{6}\right)+\pi_{16}^2 \left(\frac{5 n \pi ^2}{576}-\frac{437 n}{6912}\right)\\
\end{eqnarray*}

\subsection{Real or adjoint}

The coefficients of polynomials part for $B_P$ at NNLO.
\begin{eqnarray*}
\gamma_1&=&32 K^r_{13}+64 K^r_{14} n-96 K^r_{17}-192 K^r_{18} n
+96 K^r_{25}+64 K^r_{26} n+64 K^r_{3}-64 K^r_{37}\\
&&+96 K^r_{39}+64 K^r_{40} n+\frac{29 L^2 n^2}{36}+\frac{19 L^2}{4
n^2}
+\frac{83 L^2 n}{36}-\frac{17 L^2}{4 n}+\frac{19 L^2}{12}-\frac{80 L L^r_{0} n}{3}\\
&&+\frac{32 L L^r_{0}}{3 n}-\frac{80 L L^r_{0}}{3}-64 L
L^r_{1}-\frac{224 L L^r_{2}}{3}
-8 L L^r_{3} n+\frac{32 L L^r_{3}}{3 n}-\frac{56 L L^r_{3}}{3}+\frac{64 L L^r_{4}}{3}\\
&&+\frac{40 L L^r_{5} n}{3}-\frac{48 L L^r_{5}}{n}+\frac{64 L
L^r_{5}}{3}
-160 L L^r_{6}-32 L L^r_{7}-64 L L^r_{8} n+\frac{112 L L^r_{8}}{n}-80 L L^r_{8}\\
&&+512 L^r_{4} L^r_{8} n+256 L^r_{5} L^r_{8}-1024 L^r_{6} L^r_{8}
n-512 (L^r_{8})^2
\\
&&+\pi_{16} \Bigg(\frac{229 L n^2}{216}+\frac{L}{n^2}+\frac{623 L
n}{216}-\frac{L}{n}+\frac{155 L}{108}-\frac{80 L^r_0 n}{9}+\frac{128
L^r_0}{9 n}
-\frac{224 L^r_0}{9}\\
&&-\frac{32 L^r_1}{3}-\frac{368 L^r_2}{9}-\frac{8 L^r_3
n}{3}+\frac{128 L^r_3}{9 n}
-\frac{56 L^r_3}{9}+\frac{256 L^r_4}{9}+\frac{64 L^r_5 n}{9}-\frac{32 L^r_5}{n}\\
&&+\frac{136 L^r_5}{9}-128 L^r_6-32 L^r_8 n+\frac{96 L^r_8}{n}-64 L^r_8\Bigg)\\
&&+\pi_{16}^2 \left(\frac{n^2 \pi ^2}{27}+\frac{1645
n^2}{1728}-\frac{35}{8 n^2}+\frac{10763 n}{5184}+\frac{27}{8
n}+\frac{13 \pi ^2}{54}-\frac{2149}{1296}\right)
\\
\gamma_2&=&-32 K^r_{13}-64 K^r_{14} n+64 K^r_{17}+128 K^r_{18} n-16
K^r_{19}
-16 K^r_{20} n-16 K^r_{23}-32 K^r_{28}\\
&&-96 K^r_{3}-16 K^r_{33}+32 K^r_{37}-\frac{17}{36} L^2 n^2
-\frac{3 L^2}{8 n^2}-\frac{85 L^2 n}{72}+\frac{3 L^2}{8 n}-\frac{43 L^2}{16}+24 L L^r_{0} n\\
&&+\frac{8 L L^r_{0}}{n}+\frac{80 L L^r_{0}}{3}+\frac{176 L
L^r_{1}}{3} +\frac{248 L L^r_{2}}{3}+\frac{4 L L^r_{3} n}{3}+\frac{8
L L^r_{3}}{n}
+\frac{32 L L^r_{3}}{3}+\frac{32 L L^r_{4}}{3}\\
&&+\frac{20 L L^r_{5} n}{3}+\frac{14 L L^r_{5}}{3}+48 L L^r_{6}
+8 L L^r_{8} n+12 L L^r_{8}-64 L^r_{4} L^r_{5} n-32 (L^r_{5})^2\\
&&+128 L^r_{5} L^r_{6} n+64 L^r_{5} L^r_{8}\\
&&+\pi_{16} \Bigg(-\frac{445 L n^2}{432}-\frac{317 L
n}{108}-\frac{743 L}{216}+8 L^r_0 n-\frac{8 L^r_0}{n}+\frac{272
L^r_0}{9}
+\frac{80 L^r_1}{9}+\frac{512 L^r_2}{9}\\
&&+\frac{16 L^r_3 n}{9}-\frac{8 L^r_3}{n}+\frac{38
L^r_3}{9}+\frac{32 L^r_4}{9}
+\frac{8 L^r_5 n}{9}+\frac{26 L^r_5}{9}+48 L^r_6+8 L^r_8 n+12 L^r_8\Bigg)\\
&&+\pi_{16}^2 \left(-\frac{5}{72} n^2 \pi ^2-\frac{3865
n^2}{10368}+\frac{3}{4 n^2}-\frac{35 n \pi ^2}{216}-\frac{1837
n}{3456}-\frac{3}{4 n}-\frac{91 \pi ^2}{432}-\frac{853}{1296}\right)
\\
\gamma_3&=&
 2 K^r_{11}+8 K^r_{13}+16 K^r_{14} n-16 K^r_{17}-24 K^r_{18} n
+16 K^r_{28}+48 K^r_{3}-4 K^r_{31}+8 K^r_{5}\\
&&+2 K^r_{7}+4 K^r_{8} n+\frac{29 L^2 n^2}{288}+\frac{23 L^2 n}{72}
+\frac{101 L^2}{96}-8 L L^r_{0} n+\frac{10 L L^r_{0}}{3 n}-16 L L^r_{0}\\
&&-\frac{56 L L^r_{1}}{3}-40 L L^r_{2}-\frac{8 L L^r_{3} n}{3}+\frac{10 L L^r_{3}}{3 n}-6 L L^r_{3}-8 L L^r_{4}-L L^r_{5} n-2 L L^r_{5}\\
&&+\pi_{16} \Bigg(\frac{49 L n^2}{216}+\frac{503 L
n}{864}+\frac{2867 L}{1728}-\frac{7 L^r_0 n}{3}+\frac{28 L^r_0}{9 n}
-\frac{43 L^r_0}{3}-\frac{62 L^r_1}{9}-31 L^r_2\\
&&-\frac{25 L^r_3 n}{18}+\frac{28 L^r_3}{9 n}-3 L^r_3-\frac{20
L^r_4}{3}-\frac{2 L^r_5 n}{3}
-\frac{5 L^r_5}{3}\Bigg)\\
&&+\pi_{16}^2 \left(\frac{n^2 \pi ^2}{72}+\frac{445
n^2}{5184}+\frac{67 n \pi ^2}{864} -\frac{59 n}{288}+\frac{185 \pi
^2}{1728}+\frac{2705}{20736}\right)
\\
\gamma_4&=& 2 K^r_{11}+8 K^r_{18} n+4 K^r_{31}+8 K^r_{5}+2 K^r_{7}+4
K^r_{8} n
+\frac{11 L^2 n^2}{288}-\frac{L^2 n}{72}-\frac{7 L^2}{96}-\frac{4 L L^r_{0} n}{3}\\
&&+\frac{10 L L^r_{0}}{3 n}-\frac{4 L L^r_{0}}{3}-\frac{4 L L^r_{2}}{3}-2 L L^r_{3} n+\frac{10 L L^r_{3}}{3 n}-\frac{4 L L^r_{3}}{3}+\frac{8 L L^r_{4}}{3}-\frac{L L^r_{5} n}{3}+\frac{2 L L^r_{5}}{3}\\
&&+\pi_{16} \Bigg(\frac{29 L n^2}{216}-\frac{23 L n}{864}-\frac{137
L}{1728}-\frac{13 L^r_0 n}{9}+\frac{28 L^r_0}{9 n}
-\frac{13 L^r_0}{9}-\frac{10 L^r_1}{3}-\frac{31 L^r_2}{9}\\
&&-\frac{11 L^r_3 n}{6}+\frac{28 L^r_3}{9 n}-\frac{19 L^r_3}{9}
+\frac{20 L^r_4}{9}-\frac{4 L^r_5 n}{9}+\frac{5 L^r_5}{9}\Bigg)\\
&&+\pi_{16}^2 \left(-\frac{1}{108} n^2 \pi ^2+\frac{421
n^2}{3456}-\frac{n \pi ^2}{288}-\frac{289 n}{10368}-\frac{11 \pi
^2}{1728}-\frac{1091}{20736}\right)
\\
\gamma_5&=&K^r_{1}-8 K^r_{3}-4 K^r_{5}-\frac{5 L^2
n^2}{1152}-\frac{55 L^2 n}{2304} -\frac{5 L^2}{32}+\frac{5 L L^r_{0}
n}{12}+\frac{5 L L^r_{0}}{2}
+\frac{5 L L^r_{1}}{2}\\
&&+\frac{25 L L^r_{2}}{4}+\frac{5 L L^r_{3} n}{24}+\frac{5 L L^r_{3}}{8}\\
&&+ \pi_{16} \left(-\frac{19 L n^2}{2304}-\frac{13 L
n}{768}-\frac{307 L}{1152}+\frac{2 L^r_0 n}{9}+\frac{7
L^r_0}{3}+\frac{7 L^r_1}{3}
+\frac{35 L^r_2}{6}+\frac{5 L^r_3 n}{18}+\frac{7 L^r_3}{12}\right)\\
&&+\pi_{16}^2 \left(\frac{n^2 \pi ^2}{3456}-\frac{1015 n^2}{165888}-\frac{29 n \pi ^2}{3456}+\frac{10313 n}{165888}-\frac{19 \pi ^2}{1152}+\frac{71}{13824}\right)\\
\gamma_6&=& 3 K^r_{1}-4 K^r_{5}-\frac{5}{384} L^2 n^2+\frac{L^2
n}{768}-\frac{5 L^2}{96}
+\frac{7 L L^r_{0} n}{12}+\frac{5 L L^r_{0}}{6}+\frac{5 L L^r_{1}}{6}\\
&&+\frac{25 L L^r_{2}}{12}+\frac{23 L L^r_{3} n}{24}+\frac{5 L L^r_{3}}{24}\\
&&+ \pi_{16} \left(-\frac{203 L n^2}{6912}+\frac{65 L
n}{6912}-\frac{307 L}{3456}+\frac{4 L^r_0 n}{9}
+\frac{7 L^r_0}{9}+\frac{7 L^r_1}{9}+\frac{35 L^r_2}{18}+\frac{17 L^r_3 n}{18}+\frac{7 L^r_3}{36}\right)\\
&&+\pi_{16}^2 \left(-\frac{n^2 \pi ^2}{3456}-\frac{1933
n^2}{165888}-\frac{11 n \pi ^2}{3456}+\frac{5299 n}{165888}-\frac{19
\pi ^2}{3456}+\frac{71}{41472}\right)
\end{eqnarray*}
The coefficients of polynomials part for $C_P$ at NNLO.
\begin{eqnarray*}
\delta_1 &=&128 K^r_{10} n-128 K^r_{18}+128 K^r_{2}-64 K^r_{20}-128
K^r_{21}
-256 K^r_{22} n+128 K^r_{26}+384 K^r_{27} n\\
&&-128 K^r_{35}+128 K^r_{40}+64 K^r_{9}-\frac{7 L^2}{4 n^3}
+\frac{L^2}{n^2}+\frac{L^2}{2 n}+\frac{L^2}{2}-\frac{48 L
L^r_{0}}{n^2}
-64 L L^r_{1} n+\frac{64 L L^r_{1}}{n}\\
&&-64 L L^r_{1}+\frac{16 L L^r_{2}}{n}-16 L L^r_{2}-\frac{48 L
L^r_{3}}{n^2}
+64 L L^r_{4} n-\frac{48 L L^r_{4}}{n}+48 L L^r_{4}+\frac{24 L L^r_{5}}{n^2}\\
&&+8 L L^r_{5}-64 L L^r_{6} n+\frac{32 L L^r_{6}}{n}-32 L L^r_{6}
+\frac{32 L L^r_{7}}{n}-\frac{48 L L^r_{8}}{n^2}-16 L L^r_{8}-512 (L^r_{4})^2 n\\
&&-256 L^r_{4} L^r_{5}+2048 L^r_{4} L^r_{6} n+512 L^r_{4} L^r_{8}+512 L^r_{5} L^r_{6}-2048 (L^r_{6})^2 n-1024 L^r_{6} L^r_{8}\\
&&+\pi_{16} \Bigg(-\frac{3 L}{2
n^3}+\frac{L}{n^2}+\frac{L}{n}-L-\frac{16 L^r_0}{n^2}+\frac{32
L^r_1}{n}-32 L^r_1
-\frac{16 L^r_3}{n^2}-\frac{32 L^r_4}{n}+32 L^r_4\\
&&+\frac{16 L^r_5}{n^2}+\frac{32 L^r_6}{n}-32 L^r_6-\frac{48 L^r_8}{n^2}\Bigg)\\
&&+\pi_{16}^2\left(\frac{5}{4 n^3}-\frac{1}{2 n^2}+\frac{3}{4
n}-\frac{1}{6}\right)
\\
\delta_2 &=& -128 K^r_{10} n+128 K^r_{18}-192 K^r_{2}+32 K^r_{20}+64
K^r_{21}
+128 K^r_{22} n-32 K^r_{32}+64 K^r_{35}\\
&&-32 K^r_{38}-64 K^r_{9}+\frac{37 L^2 n}{72}-\frac{3 L^2}{4 n}
-\frac{7 L^2}{36}+\frac{104 L L^r_{0}}{3 n^2}-8 L L^r_{0}-\frac{48 L L^r_{1}}{n}+48 L L^r_{1}\\
&&-16 L L^r_{2} n-\frac{32 L L^r_{2}}{3 n}+\frac{32 L L^r_{2}}{3}
+\frac{104 L L^r_{3}}{3 n^2}-\frac{64 L L^r_{3}}{3}+16 L L^r_{4}
n+\frac{16 L L^r_{4}}{n}
-16 L L^r_{4}\\
&&+12 L L^r_{5}-32 L L^r_{6} n-24 L L^r_{8}+256 (L^r_{4})^2 n+128 L^r_{4} L^r_{5}-512 L^r_{4} L^r_{6} n-256 L^r_{4} L^r_{8}\\
 &&+\pi_{16} \Bigg(-\frac{31 L n}{24}-\frac{L}{4 n}-\frac{5 L}{18}+\frac{152 L^r_0}{9 n^2}-32 L^r_1 n
-\frac{32 L^r_1}{n}+32 L^r_1-\frac{8 L^r_2}{9 n}+\frac{8 L^r_2}{9}\\
&&+\frac{152 L^r_3}{9 n^2}-\frac{88 L^r_3}{9}+32 L^r_4 n
+\frac{16 L^r_4}{n}-16 L^r_4+12 L^r_5-32 L^r_6 n-24 L^r_8\Bigg)\\
&&+\pi_{16}^2 \left(-\frac{n \pi ^2}{27}-\frac{\pi ^2}{12
n}-\frac{373 n}{2592}+\frac{25}{16 n}-\frac{5 \pi
^2}{216}-\frac{565}{648}\right)
\\
\delta_3 &=&32 K^r_{10} n+4 K^r_{15}+8 K^r_{16} n-32 K^r_{18}+96
K^r_{2}
-8 K^r_{29}+16 K^r_{32}+16 K^r_{9}-\frac{13 L^2 n}{72}\\
&&+\frac{49 L^2}{144}-\frac{20 L L^r_{0}}{3 n^2}+\frac{10 L
L^r_{0}}{3}+32 L L^r_{1} n
+\frac{8 L L^r_{1}}{n}-8 L L^r_{1}+\frac{20 L L^r_{2} n}{3}+\frac{8 L L^r_{2}}{3 n}\\
&&-\frac{8 L L^r_{2}}{3}-\frac{20 L L^r_{3}}{3 n^2}+\frac{44 L L^r_{3}}{3}-16 L L^r_{4} n-3 L L^r_{5}\\
&&+\pi_{16} \Bigg(\frac{11 L n}{24}+\frac{115 L}{144} -\frac{56
L^r_0}{9 n^2}+\frac{4 L^r_0}{9}+32 L^r_1 n
+\frac{8 L^r_1}{n}-8 L^r_1+\frac{8 L^r_2 n}{9}+\frac{20 L^r_2}{9 n}\\
&&-\frac{20 L^r_2}{9}-\frac{56 L^r_3}{9 n^2}+\frac{83 L^r_3}{9}-16 L^r_4 n-3 L^r_5\Bigg)\\
&&+\pi_{16}^2 \left(-\frac{25 n \pi ^2}{864}+\frac{625 n}{2592}+\frac{17 \pi ^2}{864}+\frac{1451}{5184}\right)\\
\delta_4 &=& 4 K^r_{15}+8 K^r_{16} n+8 K^r_{29}+\frac{5 L^2 n}{48}+\frac{L^2}{24}-2 L L^r_{0}-4 L L^r_{2} n-L L^r_{5}\\
&&+ \pi_{16} \left(\frac{L n}{16}+\frac{7 L}{48}+L^r_3-L^r_5\right) \\
&&+\pi_{16}^2 \left(\frac{n \pi ^2}{288}+\frac{7 n}{64}+\frac{\pi
^2}{288}+\frac{11}{192}\right)
\\
\delta_5 &=&-16 K^r_{2}+2 K^r_{4}+2 K^r_{6}+\frac{55 L^2 n}{384}-\frac{L^2}{48}-\frac{11 L L^r_{0}}{6}-8 L L^r_{1} n-\frac{8 L L^r_{2} n}{3}-\frac{17 L L^r_{3}}{6}\\
&&+\pi_{16} \left(\frac{101 L n}{384}-\frac{13 L}{384}-\frac{29 L^r_0}{18}-8 L^r_1 n-\frac{20 L^r_2 n}{9}-\frac{97 L^r_3}{36}\right)\\
&&+\pi_{16}^2 \left(\frac{19 n \pi ^2}{1152}-\frac{115
n}{13824}+\frac{\pi ^2}{1152}-\frac{349}{13824}\right)
\\
\delta_6 &=&6 K^r_{4}-2 K^r_{6}+\frac{5 L^2 n}{384}+\frac{L^2}{48}-\frac{3 L L^r_{0}}{2}-\frac{L L^r_{3}}{2}\\
&& +\pi_{16} \left(\frac{L n}{128}+\frac{13 L}{384}-\frac{3
L^r_0}{2}-\frac{5
  L^r_3}{12}\right)\\
&&+\pi_{16}^2 \left(\frac{5 n \pi ^2}{1152}-\frac{437
  n}{13824}-\frac{\pi ^2}{1152}+\frac{349}{13824}\right)
\end{eqnarray*}

\subsection{Pseudo-real or two-colour}

The coefficients of polynomials part for $B_P$ at NNLO.
\begin{eqnarray*}
\gamma_1&=&32 K^r_{13}+64 K^r_{14} n-96 K^r_{17}-192 K^r_{18} n
+96 K^r_{25}+64 K^r_{26} n+64 K^r_{3}-64 K^r_{37}\\
&&+96 K^r_{39}+64 K^r_{40} n+\frac{29 L^2 n^2}{36}-\frac{83 L^2
n}{36}
+\frac{19 L^2}{4 n^2}+\frac{17 L^2}{4 n}+\frac{19 L^2}{12}-\frac{80 L L^r_{0} n}{3}\\
&&+\frac{32 L L^r_{0}}{3 n}+\frac{80 L L^r_{0}}{3}-64 L L^r_{1}
-\frac{224 L L^r_{2}}{3}-8 L L^r_{3} n+\frac{32 L L^r_{3}}{3 n}
+\frac{56 L L^r_{3}}{3}+\frac{64 L L^r_{4}}{3}\\
&&+\frac{40 L L^r_{5} n}{3}-\frac{48 L L^r_{5}}{n}-\frac{64 L
L^r_{5}}{3}
-160 L L^r_{6}-32 L L^r_{7}-64 L L^r_{8} n+\frac{112 L L^r_{8}}{n}\\
&&+80 L L^r_{8}+512 L^r_{4} L^r_{8} n+256 L^r_{5} L^r_{8}-1024 L^r_{6} L^r_{8} n-512 (L^r_{8})^2\\
&&+\pi_{16} \Bigg(\frac{229 L n^2}{216}-\frac{623 L
n}{216}+\frac{L}{n^2}+\frac{L}{n}+\frac{155 L}{108}-\frac{80 L^r_0
n}{9}
+\frac{128 L^r_0}{9 n}+\frac{224 L^r_0}{9}\\
&&-\frac{32 L^r_1}{3}-\frac{368 L^r_2}{9}-\frac{8 L^r_3
n}{3}+\frac{128 L^r_3}{9 n}+\frac{56 L^r_3}{9}
+\frac{256 L^r_4}{9}+\frac{64 L^r_5 n}{9}\\
&&-\frac{32 L^r_5}{n}-\frac{136 L^r_5}{9}-128 L^r_6-32 L^r_8 n+\frac{96 L^r_8}{n}+64 L^r_8\Bigg)\\
&&+\pi_{16}^2 \left(\frac{n^2 \pi ^2}{27}+\frac{1645
n^2}{1728}-\frac{10763 n}{5184}-\frac{35}{8 n^2}-\frac{27}{8
n}+\frac{13 \pi ^2}{54}-\frac{2149}{1296}\right)
\\
\gamma_2&=& -32 K^r_{13}-64 K^r_{14} n+64 K^r_{17}+128 K^r_{18} n-16
K^r_{19}
-16 K^r_{20} n-16 K^r_{23}-32 K^r_{28}\\
&&-96 K^r_{3}-16 K^r_{33}+32 K^r_{37}-\frac{17}{36} L^2 n^2+\frac{85
L^2 n}{72}
-\frac{3 L^2}{8 n^2}-\frac{3 L^2}{8 n}-\frac{43 L^2}{16}+24 L L^r_{0} n\\
&&+\frac{8 L L^r_{0}}{n}-\frac{80 L L^r_{0}}{3}+\frac{176 L
L^r_{1}}{3} +\frac{248 L L^r_{2}}{3}+\frac{4 L L^r_{3} n}{3}+\frac{8
L L^r_{3}}{n}
-\frac{32 L L^r_{3}}{3}+\frac{32 L L^r_{4}}{3}\\
&&+\frac{20 L L^r_{5} n}{3}-\frac{14 L L^r_{5}}{3}+48 L L^r_{6}+8 L
L^r_{8} n
-12 L L^r_{8}-64 L^r_{4} L^r_{5} n\\
&&-32 (L^r_{5})^2+128 L^r_{5} L^r_{6} n+64 L^r_{5} L^r_{8}\\
&&+  \pi_{16} \Bigg(-\frac{445 L n^2}{432}+\frac{317 L
n}{108}-\frac{743 L}{216}+8 L^r_0 n-\frac{8 L^r_0}{n}
-\frac{272 L^r_0}{9}+\frac{80 L^r_1}{9}\\
&&+\frac{512 L^r_2}{9}+\frac{16 L^r_3 n}{9}-\frac{8 L^r_3}{n}
-\frac{38 L^r_3}{9}+\frac{32 L^r_4}{9}+\frac{8 L^r_5 n}{9}-\frac{26 L^r_5}{9}+48 L^r_6+8 L^r_8 n-12 L^r_8\Bigg)\\
&&+\pi_{16}^2 \left(-\frac{5}{72} n^2 \pi ^2-\frac{3865
n^2}{10368}+\frac{35 n \pi ^2}{216}+\frac{1837 n}{3456}+\frac{3}{4
n^2} +\frac{3}{4 n}-\frac{91 \pi ^2}{432}-\frac{853}{1296}\right)
\\
\gamma_3&=& 2 K^r_{11}+8 K^r_{13}+16 K^r_{14} n-16 K^r_{17}-24
K^r_{18} n
+16 K^r_{28}+48 K^r_{3}-4 K^r_{31}+8 K^r_{5}\\
&&+2 K^r_{7}+4 K^r_{8} n+\frac{29 L^2 n^2}{288}-\frac{23 L^2 n}{72}
+\frac{101 L^2}{96}-8 L L^r_{0} n+\frac{10 L L^r_{0}}{3 n}+16 L L^r_{0}\\
&&-\frac{56 L L^r_{1}}{3}-40 L L^r_{2}-\frac{8 L L^r_{3} n}{3}+\frac{10 L L^r_{3}}{3 n}+6 L L^r_{3}-8 L L^r_{4}-L L^r_{5} n+2 L L^r_{5}\\
&&+\pi_{16} \Bigg(\frac{49 L n^2}{216}-\frac{503 L
n}{864}+\frac{2867 L}{1728}-\frac{7 L^r_0 n}{3}+\frac{28 L^r_0}{9 n}
+\frac{43 L^r_0}{3}-\frac{62 L^r_1}{9}-31 L^r_2\\
&&-\frac{25 L^r_3 n}{18}+\frac{28 L^r_3}{9 n}+3 L^r_3-\frac{20 L^r_4}{3}-\frac{2 L^r_5 n}{3}+\frac{5 L^r_5}{3}\Bigg)\\
&& +\pi_{16}^2 \left(\frac{n^2 \pi ^2}{72}+\frac{445
n^2}{5184}-\frac{67 n \pi ^2}{864} +\frac{59 n}{288}+\frac{185 \pi
^2}{1728}+\frac{2705}{20736}\right)
\\
\gamma_4&=& 2 K^r_{11}+8 K^r_{18} n+4 K^r_{31}+8 K^r_{5}+2 K^r_{7}
+4 K^r_{8} n+\frac{11 L^2 n^2}{288}+\frac{L^2 n}{72}-\frac{7 L^2}{96}-\frac{4 L L^r_{0} n}{3}\\
&&+\frac{10 L L^r_{0}}{3 n}+\frac{4 L L^r_{0}}{3}-\frac{4 L L^r_{2}}{3}-2 L L^r_{3} n+\frac{10 L L^r_{3}}{3 n}+\frac{4 L L^r_{3}}{3}+\frac{8 L L^r_{4}}{3}-\frac{L L^r_{5} n}{3}-\frac{2 L L^r_{5}}{3}\\
&&+ \pi_{16} \Bigg(\frac{29 L n^2}{216}+\frac{23 L n}{864}-\frac{137
L}{1728}-\frac{13 L^r_0 n}{9}+\frac{28 L^r_0}{9 n}
+\frac{13 L^r_0}{9}-\frac{10 L^r_1}{3}-\frac{31 L^r_2}{9}\\
&&-\frac{11 L^r_3 n}{6}+\frac{28 L^r_3}{9 n}
+\frac{19 L^r_3}{9}+\frac{20 L^r_4}{9}-\frac{4 L^r_5 n}{9}-\frac{5 L^r_5}{9}\Bigg)\\
&&+\pi_{16}^2 \left(-\frac{1}{108} n^2 \pi ^2+\frac{421
n^2}{3456}+\frac{n \pi ^2}{288}+\frac{289 n}{10368}-\frac{11 \pi
^2}{1728}-\frac{1091}{20736}\right)
\\
\gamma_5&=& K^r_{1}-8 K^r_{3}-4 K^r_{5}-\frac{5 L^2
n^2}{1152}+\frac{55 L^2 n}{2304}
-\frac{5 L^2}{32}+\frac{5 L L^r_{0} n}{12}-\frac{5 L L^r_{0}}{2}\\
&&+\frac{5 L L^r_{1}}{2}+\frac{25 L L^r_{2}}{4}+\frac{5 L L^r_{3} n}{24}-\frac{5 L L^r_{3}}{8}\\
&&+\pi_{16} \left(-\frac{19 L n^2}{2304}+\frac{13 L
n}{768}-\frac{307 L}{1152}+\frac{2 L^r_0 n}{9}-\frac{7
L^r_0}{3}+\frac{7 L^r_1}{3}
+\frac{35 L^r_2}{6}+\frac{5 L^r_3 n}{18}-\frac{7 L^r_3}{12}\right)\\
&&+\pi_{16}^2 \left(\frac{n^2 \pi ^2}{3456}-\frac{1015
n^2}{165888}+\frac{29 n \pi ^2}{3456}-\frac{10313
n}{165888}-\frac{19 \pi ^2}{1152}+\frac{71}{13824}\right)
\\
\gamma_6&=& 3 K^r_{1}-4 K^r_{5}-\frac{5}{384} L^2 n^2-\frac{L^2
n}{768}-\frac{5 L^2}{96}
+\frac{7 L L^r_{0} n}{12}-\frac{5 L L^r_{0}}{6}+\frac{5 L L^r_{1}}{6}\\
&&+\frac{25 L L^r_{2}}{12}+\frac{23 L L^r_{3} n}{24}-\frac{5 L L^r_{3}}{24}\\
&&+ \pi_{16} \left(-\frac{203 L n^2}{6912}-\frac{65 L
n}{6912}-\frac{307 L}{3456}+\frac{4 L^r_0 n}{9}-\frac{7 L^r_0}{9}
+\frac{7 L^r_1}{9}+\frac{35 L^r_2}{18}+\frac{17 L^r_3 n}{18}-\frac{7 L^r_3}{36}\right)\\
&&+\pi_{16}^2 \left(-\frac{n^2 \pi ^2}{3456}-\frac{1933
n^2}{165888}+\frac{11 n \pi ^2}{3456}-\frac{5299 n}{165888}-\frac{19
\pi ^2}{3456}+\frac{71}{41472}\right)
\end{eqnarray*}
The coefficients of polynomials part for $C_P$ at NNLO.
\begin{eqnarray*}
\delta_1&=& 128 K^r_{10} n-128 K^r_{18}+128 K^r_{2}-64 K^r_{20}
-128 K^r_{21}-256 K^r_{22} n+128 K^r_{26}\\
&&+384 K^r_{27} n-128 K^r_{35} +128 K^r_{40}+64 K^r_{9}-\frac{7
L^2}{4 n^3}-\frac{L^2}{n^2}
+\frac{L^2}{2 n}-\frac{L^2}{2}-\frac{48 L L^r_{0}}{n^2}\\
&&-64 L L^r_{1} n+\frac{64 L L^r_{1}}{n}+64 L L^r_{1}+\frac{16 L
L^r_{2}}{n}+16 L L^r_{2}
-\frac{48 L L^r_{3}}{n^2}+64 L L^r_{4} n\\
&&-\frac{48 L L^r_{4}}{n}-48 L L^r_{4}+\frac{24 L L^r_{5}}{n^2}+8 L
L^r_{5}-64 L L^r_{6} n+\frac{32 L L^r_{6}}{n}
+32 L L^r_{6}+\frac{32 L L^r_{7}}{n}\\
&&-\frac{48 L L^r_{8}}{n^2}-16 L L^r_{8}-512 (L^r_{4})^2 n-256 L^r_{4} L^r_{5}+2048 L^r_{4} L^r_{6} n+512 L^r_{4} L^r_{8}\\
&&+512 L^r_{5} L^r_{6}-2048 (L^r_{6})^2 n-1024 L^r_{6} L^r_{8}
\\&&+ \pi_{16} \Bigg(-\frac{3 L}{2 n^3}-\frac{L}{n^2}+\frac{L}{n}+L-\frac{16 L^r_0}{n^2}+\frac{32 L^r_1}{n}+32 L^r_1
-\frac{16 L^r_3}{n^2}-\frac{32 L^r_4}{n}\\
&&-32 L^r_4+\frac{16 L^r_5}{n^2}+\frac{32 L^r_6}{n}+32 L^r_6-\frac{48 L^r_8}{n^2}\Bigg)\\
&&+\left(\frac{5}{4 n^3}+\frac{1}{2 n^2}+\frac{3}{4
n}+\frac{1}{6}\right) \pi_{16}^2
\\
\delta_2&=&
 -128 K^r_{10} n+128 K^r_{18}-192 K^r_{2}+32 K^r_{20}
+64 K^r_{21}+128 K^r_{22} n-32 K^r_{32}+64 K^r_{35}\\
&&-32 K^r_{38}-64 K^r_{9}+\frac{37 L^2 n}{72}-\frac{3 L^2}{4
n}+\frac{7 L^2}{36}
+\frac{104 L L^r_{0}}{3 n^2}-8 L L^r_{0}-\frac{48 L L^r_{1}}{n}-48 L L^r_{1}\\
&&-16 L L^r_{2} n-\frac{32 L L^r_{2}}{3 n}-\frac{32 L L^r_{2}}{3}
+\frac{104 L L^r_{3}}{3 n^2}-\frac{64 L L^r_{3}}{3}+16 L L^r_{4} n
+\frac{16 L L^r_{4}}{n}+16 L L^r_{4}\\
&&+12 L L^r_{5}-32 L L^r_{6} n-24 L L^r_{8}+256 (L^r_{4})^2 n+128 L^r_{4} L^r_{5}-512 L^r_{4} L^r_{6} n-256 L^r_{4} L^r_{8}\\
&&+ \pi_{16} \Bigg(-\frac{31 L n}{24}-\frac{L}{4 n}+\frac{5
L}{18}+\frac{152 L^r_0}{9 n^2}-32 L^r_1 n-\frac{32 L^r_1}{n}-32
L^r_1
-\frac{8 L^r_2}{9 n}\\
&&-\frac{8 L^r_2}{9}+\frac{152 L^r_3}{9 n^2}-\frac{88 L^r_3}{9}+32 L^r_4 n+\frac{16 L^r_4}{n}+16 L^r_4+12 L^r_5-32 L^r_6 n-24 L^r_8\Bigg)\\
&&+\pi_{16}^2 \left(-\frac{n \pi ^2}{27}-\frac{373
n}{2592}-\frac{\pi ^2}{12 n}+\frac{25}{16 n}+\frac{5 \pi
^2}{216}+\frac{565}{648}\right)
\\
\delta_3&=& 32 K^r_{10} n+4 K^r_{15}+8 K^r_{16} n-32 K^r_{18}+96
K^r_{2}-8 K^r_{29}
+16 K^r_{32}+16 K^r_{9}-\frac{13 L^2 n}{72}\\
&&-\frac{49 L^2}{144}-\frac{20 L L^r_{0}}{3 n^2}+\frac{10 L
L^r_{0}}{3}
+32 L L^r_{1} n+\frac{8 L L^r_{1}}{n}+8 L L^r_{1}+\frac{20 L L^r_{2} n}{3}\\
&&+\frac{8 L L^r_{2}}{3 n}+\frac{8 L L^r_{2}}{3}-\frac{20 L L^r_{3}}{3 n^2}+\frac{44 L L^r_{3}}{3}-16 L L^r_{4} n-3 L L^r_{5}\\
&&+\pi_{16} \Bigg(\frac{11 L n}{24}-\frac{115 L}{144}-\frac{56
L^r_0}{9 n^2}+\frac{4 L^r_0}{9}+32 L^r_1 n
+\frac{8 L^r_1}{n}+8 L^r_1+\frac{8 L^r_2 n}{9}+\frac{20 L^r_2}{9 n}\\
&&+\frac{20 L^r_2}{9}-\frac{56 L^r_3}{9 n^2}+\frac{83 L^r_3}{9}-16 L^r_4 n-3 L^r_5\Bigg)\\
&&+\pi_{16}^2 \left(-\frac{25 n \pi ^2}{864}+\frac{625
n}{2592}-\frac{17 \pi ^2}{864}-\frac{1451}{5184}\right)
\\
\delta_4&=&
4 K^r_{15}+8 K^r_{16} n+8 K^r_{29}+\frac{5 L^2 n}{48}-\frac{L^2}{24}-2 L L^r_{0}-4 L L^r_{2} n-L L^r_{5}\\
&&+ \pi_{16} \left(\frac{L n}{16}-\frac{7 L}{48}+L^r_3-L^r_5\right)
+\pi_{16}^2 \left(\frac{n \pi ^2}{288}+\frac{7 n}{64}-\frac{\pi
^2}{288}-\frac{11}{192}\right)
\\
\delta_5&=&
-16 K^r_{2}+2 K^r_{4}+2 K^r_{6}+\frac{55 L^2 n}{384}+\frac{L^2}{48}-\frac{11 L L^r_{0}}{6}-8 L L^r_{1} n-\frac{8 L L^r_{2} n}{3}-\frac{17 L L^r_{3}}{6}\\
&&+\pi_{16} \left(\frac{101 L n}{384}+\frac{13 L}{384}-\frac{29 L^r_0}{18}-8 L^r_1 n-\frac{20 L^r_2 n}{9}-\frac{97 L^r_3}{36}\right)\\
&&+\pi_{16}^2 \left(\frac{19 n \pi ^2}{1152}-\frac{115
n}{13824}-\frac{\pi ^2}{1152}+\frac{349}{13824}\right)
\\
\delta_6&=&
 6 K^r_{4}-2 K^r_{6}+\frac{5 L^2 n}{384}-\frac{L^2}{48}-\frac{3 L L^r_{0}}{2}-\frac{L L^r_{3}}{2}\\
&&+\pi_{16} \left(\frac{L n}{128}-\frac{13 L}{384}-\frac{3 L^r_0}{2}-\frac{5 L^r_3}{12}\right) \\
&&+\pi_{16}^2 \left(\frac{5 n \pi ^2}{1152}-\frac{437
n}{13824}+\frac{\pi ^2}{1152}-\frac{349}{13824}\right)
\\
\end{eqnarray*}

\section{Scattering lengths}
\label{scattlength}

\subsection{Complex or QCD case}

\ba
\lefteqn{\pi a^I_0 =
        x_2 \, \Bigg(
          - \frac{1}{16 n}
          + \frac{n}{8}
          \Bigg)}&&
\nonumber\\&&
       + x_2^2 \, \Bigg(
          - \frac{2}{n}\,\alpha_4
          - \frac{1}{n}\,\alpha_3
          - \frac{1}{4 n}\,\alpha_2
          - \frac{3}{16 n}\,\alpha_1
          + \beta_4
          - \frac{1}{2}\,\beta_3
          - \frac{1}{8}\,\beta_2
          + \frac{1}{32}\,\beta_1
\nonumber\\&&\hskip1cm
          + 2\,n\,\alpha_4
          + \frac{n}{8}\,\alpha_1
          + \frac{n^2}{2}\,\beta_3
          + \frac{n^2}{8}\,\beta_2
          + \frac{n^2}{32}\,\beta_1
          \Bigg)
\nonumber\\&&
       + \pi_{16}\,x_2^2 \, \Bigg(
          - \frac{1}{2}
          + \frac{1}{8 n^2}
          + \frac{n^2}{2}
          \Bigg)
\nonumber\\&&
       + x_2^3 \, \Bigg(
          - \frac{4}{n}\,\gamma_5
          - \frac{2}{n}\,\gamma_4
          - \frac{1}{n}\,\gamma_3
          - \frac{1}{4 n}\,\gamma_2
          - \frac{3}{16 n}\,\gamma_1
          - 2\,\delta_5
          + \delta_4
          - \frac{1}{2}\,\delta_3
          - \frac{1}{8}\,\delta_2
\nonumber\\&&\hskip1cm
          + \frac{1}{32}\,\delta_1
          + 2\,n\,\gamma_4
          + \frac{n}{8}\,\gamma_1
          + 2\,n^2\,\delta_5
          + \frac{n^2}{2}\,\delta_3
          + \frac{n^2}{8}\,\delta_2
          + \frac{n^2}{32}\,\delta_1
          \Bigg)
\nonumber\\&&
       + \pi_{16}\,x_2^3 \, \Bigg(
           \frac{7}{4 n^3}\,L
          + \frac{12}{n^2}\,L^r_{8}
          - \frac{4}{n^2}\,L^r_{5}
          + \frac{12}{n^2}\,L^r_{3}
          + \frac{12}{n^2}\,L^r_{0}
          - \frac{5}{n}\,L
          - \frac{4}{n}\,L^r_{6}
\nonumber\\&&\hskip1cm
          + \frac{12}{n}\,L^r_{4}
          - \frac{4}{n}\,L^r_{2}
          - \frac{4}{n}\,L^r_{1}
          - 32\,L^r_{8}
          + 8\,L^r_{5}
          - 32\,L^r_{3}
          - 32\,L^r_{0}
          + \frac{17 n}{4}\,L
\nonumber\\&&\hskip1cm
          + 4\,n\,L^r_{6}
          - 28\,n\,L^r_{4}
          + 4\,n\,L^r_{2}
          + 4\,n\,L^r_{1}
          + 16\,n^2\,L^r_{8}
          + 16\,n^2\,L^r_{3}
\nonumber\\&&\hskip1cm
          + 16\,n^2\,L^r_{0}
          - \frac{5 n^3}{2}\,L
          + 8\,n^3\,L^r_{6}
          + 8\,n^3\,L^r_{4}
          + 8\,n^3\,L^r_{2}
          + 8\,n^3\,L^r_{1}
          \Bigg)
\nonumber\\&&
       + \pi_{16}^2\,x_2^3 \, \Bigg[
          - \frac{3}{n^3}
          + \frac{7}{n}
          - \frac{49 n}{24}
          + \frac{n^3}{12}
        +\pi^2 \Bigg(
            \frac{1}{2 n^3}
          - \frac{7}{6 n}
          + \frac{53 n}{144}
          - \frac{5 n^3}{72}
          \Bigg) \Bigg]\,,
\\
\lefteqn{ \pi a^S_0 =
       + \,x_2 \, \Bigg(
          - \frac{1}{8 n}
          + \frac{n}{16}
          \Bigg)}&&
\nonumber\\&&
       + x_2^2 \, \Bigg(
          - \frac{4}{n}\,\alpha_4
          - \frac{2}{n}\,\alpha_3
          - \frac{1}{2 n}\,\alpha_2
          - \frac{3}{8 n}\,\alpha_1
          + \beta_4
          + \frac{1}{16}\,\beta_1
          + n\,\alpha_4
          + \frac{n}{16}\,\alpha_1
          \Bigg)
\nonumber\\&&
       + \pi_{16}\,x_2^2 \, \Bigg(
          - \frac{1}{2}
          + \frac{1}{2 n^2}
          + \frac{n^2}{8}
          \Bigg)
\nonumber\\&&
       + x_2^3 \, \left(
          - \frac{8}{n}\,\gamma_5
          - \frac{4}{n}\,\gamma_4
          - \frac{2}{n}\,\gamma_3
          - \frac{1}{2 n}\,\gamma_2
          - \frac{3}{8 n}\,\gamma_1
          + \delta_4
          + \frac{1}{16}\,\delta_1
          + n\,\gamma_4
          + \frac{n}{16}\,\gamma_1
          \right)
\nonumber\\&&
       + \pi_{16}\,x_2^3 \, \Bigg(
          \frac{7}{n^3}\,L
          + \frac{48}{n^2}\,L^r_{8}
          - \frac{16}{n^2}\,L^r_{5}
          + \frac{48}{n^2}\,L^r_{3}
          + \frac{48}{n^2}\,L^r_{0}
          - \frac{11}{2 n}\,L
          - \frac{16}{n}\,L^r_{6}
\nonumber\\&&\hskip1cm
          + \frac{16}{n}\,L^r_{4}
          - \frac{16}{n}\,L^r_{2}
          - \frac{16}{n}\,L^r_{1}
          - 32\,L^r_{8}
          + 8\,L^r_{5}
          - 32\,L^r_{3}
          - 32\,L^r_{0}
          + \frac{3 n}{2}\,L
\nonumber\\&&\hskip1cm
          + 8\,n\,L^r_{6}
          - 8\,n\,L^r_{4}
          + 8\,n\,L^r_{2}
          + 8\,n\,L^r_{1}
          + 4\,n^2\,L^r_{8}
          + 4\,n^2\,L^r_{3}
          + 4\,n^2\,L^r_{0}
\nonumber\\&&\hskip1cm
          - \frac{n^3}{4}\,L
          \Bigg)
\nonumber\\&&
       + \pi_{16}^2\,x_2^3 \, \Bigg[
          - \frac{10}{n^3}
          + \frac{9}{2 n}
          + \frac{5 n}{6}
          - \frac{7 n^3}{24}
       + \pi^2 \Bigg(
           \frac{5}{3 n^3}
          - \frac{3}{4 n}
          - \frac{n}{9}
          + \frac{5 n^3}{144}
          \Bigg) \Bigg]\,,
\\
\lefteqn{ \pi a^A_1 =
       + x_2 \, \Bigg(
            \frac{n}{48}
          \Bigg)}&&
\nonumber\\&&
       + x_2^2 \, \Bigg(
            \frac{1}{3}\,\beta_4
          + \frac{1}{24}\,\beta_2
          - \frac{n}{3}\,\alpha_4
          - \frac{n}{24}\,\alpha_2
          \Bigg)
\nonumber\\&&
       + \pi_{16}\,x_2^2 \, \Bigg(
          - \frac{1}{72}
          + \frac{1}{72 n^2}
          - \frac{n^2}{432}
          \Bigg)
\nonumber\\&&
       + x_2^3 \, \Bigg(
            \frac{2}{3}\,\delta_6
          + \frac{1}{3}\,\delta_4
          + \frac{1}{24}\,\delta_2
          - \frac{2 n}{3}\,\gamma_6
          - \frac{n}{3}\,\gamma_4
          - \frac{n}{24}\,\gamma_2
          \Bigg)
\nonumber\\&&
       + \pi_{16}\,x_2^3 \, \Bigg(
            \frac{7}{36 n^3}\,L
          + \frac{4}{3 n^2}\,L^r_{8}
          - \frac{4}{9 n^2}\,L^r_{5}
          + \frac{20}{27 n^2}\,L^r_{3}
          + \frac{20}{27 n^2}\,L^r_{0}
          - \frac{1}{9 n}\,L
\nonumber\\&&\hskip1cm
          - \frac{4}{9 n}\,L^r_{6}
          + \frac{4}{9 n}\,L^r_{4}
          - \frac{4}{27 n}\,L^r_{2}
          - \frac{4}{9 n}\,L^r_{1}
          - \frac{8}{9}\,L^r_{8}
          + \frac{2}{9}\,L^r_{5}
          - \frac{4}{9}\,L^r_{3}
          - \frac{16}{27}\,L^r_{0}
\nonumber\\&&\hskip1cm
          + \frac{49 n}{648}\,L
          - \frac{4 n}{9}\,L^r_{6}
          - \frac{4 n}{27}\,L^r_{4}
          - \frac{8 n}{27}\,L^r_{2}
          - \frac{4 n}{27}\,L^r_{1}
          - \frac{n^2}{27}\,L^r_{5}
          + \frac{n^3}{432}\,L
          \Bigg)
\nonumber\\&&
       + \pi_{16}^2\,x_2^3 \, \Bigg[
            \frac{1}{12 n^3}
          + \frac{1}{8 n}
          - \frac{7 n}{54}
          - \frac{25 n^3}{5184}
       + \pi^2 \Bigg(
            \frac{1}{54 n^3}
          - \frac{5}{216 n}
          + \frac{25 n}{1296}
          - \frac{n^3}{1296}
          \Bigg)\Bigg]\,,
\\
\lefteqn{ \pi a^{SA}_1 =
       + x_2^2 \, \Bigg(
           \frac{1}{3}\,\beta_4
          + \frac{1}{24}\,\beta_2
          \Bigg)
       + \pi_{16}\,x_2^2 \, \Bigg(
          - \frac{1}{144}
          + \frac{1}{72 n^2}
          \Bigg)}&&
\nonumber\\&&
       + x_2^3 \, \Bigg(
            \frac{2}{3}\,\delta_6
          + \frac{1}{3}\,\delta_4
          + \frac{1}{24}\,\delta_2
          \Bigg)
\nonumber\\&&
       + \pi_{16}\,x_2^3 \, \Bigg(
            \frac{7}{36 n^3}\,L
          + \frac{4}{3 n^2}\,L^r_{8}
          - \frac{4}{9 n^2}\,L^r_{5}
          + \frac{20}{27 n^2}\,L^r_{3}
         + \frac{20}{27 n^2}\,L^r_{0}
          - \frac{1}{18 n}\,L
\nonumber\\&&\hskip0.65cm
          - \frac{4}{9 n}\,L^r_{6}
          + \frac{4}{9 n}\,L^r_{4}
          - \frac{4}{27 n}\,L^r_{2}
          - \frac{4}{9 n}\,L^r_{1}
          - \frac{2}{9}\,L^r_{8}
          - \frac{2}{27}\,L^r_{3}
          - \frac{2}{9}\,L^r_{0}
          + \frac{7 n}{648}\,L
          \Bigg)
\nonumber\\&&
       + \pi_{16}^2\,x_2^3 \, \Bigg(
            \frac{1}{12 n^3}
          + \frac{n}{324}
          \Bigg)
       + \pi^2\,\pi_{16}^2\,x_2^3 \, \Bigg(
            \frac{1}{54 n^3}
          - \frac{1}{216 n}
          - \frac{n}{2592}
          \Bigg)\,,
\\
\lefteqn{ \pi a^{AS}_1 = \pi a^{SA}_1}&&
\\
\lefteqn{\pi a^{SS}_0 =
        - \frac{1}{16} x_2 }&&
\nonumber\\&&
       + x_2^2 \, \Bigg(
            \alpha_3
          + \frac{1}{4}\,\alpha_2
          + \frac{1}{16}\,\alpha_1
          + \beta_4
          + \frac{1}{16}\,\beta_1
       + \frac{1}{8}\pi_{16} \Bigg)
\nonumber\\&&
       + x_2^3 \, \Bigg(
            \delta_4
          + \frac{1}{16}\,\delta_1
          + 4\,\gamma_5
          + \gamma_3
          + \frac{1}{4}\,\gamma_2
          + \frac{1}{16}\,\gamma_1
          \Bigg)
\nonumber\\&&
       + \pi_{16}\,x_2^3 \, \Bigg(
            \frac{1}{2 n^2}\,L
          - \frac{1}{2 n}\,L
          + \frac{1}{2}\,L
          - 4\,L^r_{8}
          - 8\,L^r_{6}
          + 4\,L^r_{5}
          + 8\,L^r_{4}
          - 4\,L^r_{3}
\nonumber\\&&\hskip1cm
          - 8\,L^r_{2}
          - 8\,L^r_{1}
          - 4\,L^r_{0}
          \Bigg)
\nonumber\\&&
       + \pi_{16}^2\,x_2^3 \, \Bigg[
          - \frac{1}{2}
          - \frac{1}{n^2}
          + \frac{1}{n}
          - \frac{n}{24}
       + \pi^2\, \Bigg(
            \frac{1}{12}
          + \frac{1}{6 n^2}
          - \frac{1}{6 n}
          + \frac{7 n}{144}
          \Bigg)\Bigg]\,,
\\
\lefteqn{\pi a^{AA}_0 =
       + \frac{1}{16}x_2 }&&
\nonumber\\&&
       + x_2^2 \, \Bigg(
          - \alpha_3
          - \frac{1}{4}\,\alpha_2
          - \frac{1}{16}\,\alpha_1
          + \beta_4
          + \frac{1}{16}\,\beta_1
       +  \frac{ \pi_{16}}{8}
          \Bigg)
\nonumber\\&&
       + x_2^3 \, \Bigg(
            \delta_4
          + \frac{1}{16}\,\delta_1
          - 4\,\gamma_5
          - \gamma_3
          - \frac{1}{4}\,\gamma_2
          - \frac{1}{16}\,\gamma_1
          \Bigg)
\nonumber\\&&
       + \pi_{16}\,x_2^3 \, \Bigg(
          - \frac{1}{2 n^2}\,L
          - \frac{1}{2 n}\,L
          - \frac{1}{2}\,L
          - 4\,L^r_{8}
          + 8\,L^r_{6}
          + 4\,L^r_{5}
          - 8\,L^r_{4}
          - 4\,L^r_{3}
\nonumber\\&&\hskip1cm
          + 8\,L^r_{2}
          + 8\,L^r_{1}
          - 4\,L^r_{0}
          \Bigg)
\nonumber\\&&
       + \pi_{16}^2\,x_2^3 \, \Bigg[
            \frac{1}{2}
          + \frac{1}{n^2}
          + \frac{1}{n}
          - \frac{n}{24}
        + \pi^2\, \Bigg(
          - \frac{1}{12}
          - \frac{1}{6 n^2}
          - \frac{1}{6 n}
          + \frac{7 n}{144}
          \Bigg)\Bigg]\,.
\ea

\subsection{Real or adjoint case}

\ba
\lefteqn{\pi a^I_0 =
       + x_2 \, \Bigg(
            \frac{1}{32}
          - \frac{1}{32 n}
          + \frac{n}{8}
          \Bigg)}&&
\nonumber\\&&
       + x_2^2 \, \Bigg(
          - \frac{1}{n}\,\alpha_4
          - \frac{1}{2 n}\,\alpha_3
          - \frac{1}{8 n}\,\alpha_2
          - \frac{3}{32 n}\,\alpha_1
          + \alpha_4
          + \frac{1}{2}\,\alpha_3
          + \frac{1}{8}\,\alpha_2
          + \frac{3}{32}\,\alpha_1
\nonumber\\&&\hskip1cm
          + \beta_4
          - \frac{1}{2}\,\beta_3
          - \frac{1}{8}\,\beta_2
          + \frac{1}{32}\,\beta_1
          + 2\,n\,\alpha_4
          + \frac{n}{8}\,\alpha_1
          + \frac{n}{2}\,\beta_3
          + \frac{n}{8}\,\beta_2
          + \frac{n}{32}\,\beta_1
\nonumber\\&&\hskip1cm
          + n^2\,\beta_3
          + \frac{n^2}{4}\,\beta_2
          + \frac{n^2}{16}\,\beta_1
          \Bigg)
\nonumber\\&&
       + \pi_{16}\,x_2^2 \, \Bigg(
          - \frac{7}{32}
          + \frac{1}{32 n^2}
          - \frac{1}{16 n}
          + \frac{n}{4}
          + \frac{n^2}{2}
          \Bigg)
\nonumber\\&&
       + x_2^3 \, \Bigg(
          - \frac{2}{n}\,\gamma_5
          - \frac{1}{n}\,\gamma_4
          - \frac{1}{2 n}\,\gamma_3
          - \frac{1}{8 n}\,\gamma_2
          - \frac{3}{32 n}\,\gamma_1
          - 2\,\delta_5
          + \delta_4
          - \frac{1}{2}\,\delta_3
          - \frac{1}{8}\,\delta_2
\nonumber\\&&\hskip1cm
          + \frac{1}{32}\,\delta_1
          + 2\,\gamma_5
          + \gamma_4
          + \frac{1}{2}\,\gamma_3
          + \frac{1}{8}\,\gamma_2
          + \frac{3}{32}\,\gamma_1
          + 2\,n\,\delta_5
          + \frac{n}{2}\,\delta_3
          + \frac{n}{8}\,\delta_2
\nonumber\\&&\hskip1cm
          + \frac{n}{32}\,\delta_1
          + 2\,n\,\gamma_4
          + \frac{n}{8}\,\gamma_1
          + 4\,n^2\,\delta_5
          + n^2\,\delta_3
          + \frac{n^2}{4}\,\delta_2
          + \frac{n^2}{16}\,\delta_1
          \Bigg)
\nonumber\\&&
       + \pi_{16}\,x_2^3 \, \Bigg(
            \frac{7}{32 n^3}\,L
          - \frac{15}{32 n^2}\,L
          + \frac{3}{n^2}\,L^r_{8}
          - \frac{1}{n^2}\,L^r_{5}
          + \frac{3}{n^2}\,L^r_{3}
          + \frac{3}{n^2}\,L^r_{0}
\nonumber\\&&\hskip1cm
          - \frac{29}{32 n}\,L
          - \frac{6}{n}\,L^r_{8}
          - \frac{2}{n}\,L^r_{6}
          + \frac{2}{n}\,L^r_{5}
          + \frac{6}{n}\,L^r_{4}
          - \frac{6}{n}\,L^r_{3}
          - \frac{2}{n}\,L^r_{2}
          - \frac{2}{n}\,L^r_{1}
\nonumber\\&&\hskip1cm
          - \frac{6}{n}\,L^r_{0}
          + \frac{55}{32}\,L
          - 13\,L^r_{8}
          + 3\,L^r_{5}
          - 8\,L^r_{4}
          - 13\,L^r_{3}
          - 13\,L^r_{0}
          + \frac{21 n}{16}\,L
\nonumber\\&&\hskip1cm
          + 16\,n\,L^r_{8}
          + 6\,n\,L^r_{6}
          - 4\,n\,L^r_{5}
          - 26\,n\,L^r_{4}
          + 16\,n\,L^r_{3}
          + 6\,n\,L^r_{2}
          + 6\,n\,L^r_{1}
\nonumber\\&&\hskip1cm
          + 16\,n\,L^r_{0}
          - \frac{19 n^2}{8}\,L
          + 16\,n^2\,L^r_{8}
          + 12\,n^2\,L^r_{6}
          + 12\,n^2\,L^r_{4}
          + 16\,n^2\,L^r_{3}
\nonumber\\&&\hskip1cm
          + 12\,n^2\,L^r_{2}
          + 12\,n^2\,L^r_{1}
          + 16\,n^2\,L^r_{0}
          - \frac{5 n^3}{2}\,L
          + 16\,n^3\,L^r_{6}
          + 16\,n^3\,L^r_{4}
\nonumber\\&&\hskip1cm
          + 16\,n^3\,L^r_{2}
          + 16\,n^3\,L^r_{1}
          \Bigg)
\nonumber\\&&
       + \pi_{16}^2\,x_2^3 \, \Bigg(
          - \frac{85}{48}
          - \frac{3}{8 n^3}
          + \frac{3}{4 n^2}
          + \frac{5}{4 n}
          - \frac{3 n}{8}
          + \frac{29 n^2}{48}
          + \frac{n^3}{12}
          \Bigg)
\nonumber\\&&
       + \pi^2\,\pi_{16}^2\,x_2^3 \, \Bigg(
            \frac{89}{288}
          + \frac{1}{16 n^3}
          - \frac{1}{8 n^2}
          - \frac{5}{24 n}
          + \frac{n}{16}
          - \frac{49 n^2}{288}
          - \frac{5 n^3}{72}
          \Bigg)\,,
\\
\lefteqn{\pi a^A_1 =
       + x_2 \, \Bigg(
            \frac{1}{48}
          + \frac{n}{48}
          \Bigg)}&&
\nonumber\\&&
       + x_2^2 \, \Bigg(
          - \frac{1}{3}\,\alpha_4
          - \frac{1}{24}\,\alpha_2
          + \frac{1}{3}\,\beta_4
          + \frac{1}{24}\,\beta_2
          - \frac{n}{3}\,\alpha_4
          - \frac{n}{24}\,\alpha_2
          \Bigg)
\nonumber\\&&
       + \pi_{16}\,x_2^2 \, \Bigg(
          - \frac{11}{864}
          + \frac{1}{288 n^2}
          - \frac{1}{288 n}
          - \frac{7 n}{864}
          - \frac{n^2}{432}
          \Bigg)
\nonumber\\&&
       + x_2^3 \, \Bigg(
            \frac{2}{3}\,\delta_6
          + \frac{1}{3}\,\delta_4
          + \frac{1}{24}\,\delta_2
          - \frac{2}{3}\,\gamma_6
          - \frac{1}{3}\,\gamma_4
          - \frac{1}{24}\,\gamma_2
          - \frac{2 n}{3}\,\gamma_6
          - \frac{n}{3}\,\gamma_4
          - \frac{n}{24}\,\gamma_2
          \Bigg)
\nonumber\\&&
       + \pi_{16}\,x_2^3 \, \Bigg(
            \frac{7}{288 n^3}\,L
          - \frac{1}{36 n^2}\,L
          + \frac{1}{3 n^2}\,L^r_{8}
          - \frac{1}{9 n^2}\,L^r_{5}
          + \frac{5}{27 n^2}\,L^r_{3}
          + \frac{5}{27 n^2}\,L^r_{0}
\nonumber\\&&\hskip1cm
          - \frac{5}{288 n}\,L
          - \frac{1}{3 n}\,L^r_{8}
          - \frac{2}{9 n}\,L^r_{6}
          + \frac{1}{9 n}\,L^r_{5}
          + \frac{2}{9 n}\,L^r_{4}
          - \frac{5}{27 n}\,L^r_{3}
          - \frac{2}{27 n}\,L^r_{2}
\nonumber\\&&\hskip1cm
          - \frac{2}{9 n}\,L^r_{1}
          - \frac{5}{27 n}\,L^r_{0}
          + \frac{67}{1296}\,L
          - \frac{4}{9}\,L^r_{8}
          - \frac{2}{9}\,L^r_{6}
          + \frac{1}{54}\,L^r_{5}
          - \frac{10}{27}\,L^r_{4}
\nonumber\\&&\hskip1cm
          - \frac{5}{27}\,L^r_{3}
          - \frac{2}{9}\,L^r_{2}
          + \frac{2}{27}\,L^r_{1}
          - \frac{10}{27}\,L^r_{0}
          + \frac{125 n}{2592}\,L
          - \frac{4 n}{9}\,L^r_{6}
          - \frac{7 n}{54}\,L^r_{5}
\nonumber\\&&\hskip1cm
          - \frac{4 n}{27}\,L^r_{4}
          + \frac{n}{27}\,L^r_{3}
          - \frac{8 n}{27}\,L^r_{2}
          - \frac{4 n}{27}\,L^r_{1}
          - \frac{2 n}{27}\,L^r_{0}
          + \frac{7 n^2}{864}\,L
          - \frac{n^2}{27}\,L^r_{5}
\nonumber\\&&\hskip1cm
          + \frac{n^3}{432}\,L
          \Bigg)
\nonumber\\&&
       + \pi_{16}^2\,x_2^3 \, \Bigg(
          - \frac{13}{96}
          + \frac{1}{96 n^3}
          + \frac{7}{288 n^2}
          - \frac{281 n}{1728}
          - \frac{151 n^2}{2592}
          - \frac{25 n^3}{5184}
          \Bigg)
\nonumber\\&&
       + \pi^2 \,\pi_{16}^2\,x_2^3 \, \Bigg(
           \frac{173}{10368}
          + \frac{1}{432 n^3}
          - \frac{5}{864 n^2}
          - \frac{1}{576 n}
          + \frac{169 n}{10368}
          + \frac{n^2}{432}
          - \frac{n^3}{1296}
          \Bigg)\,,
\nonumber\\&&
\\
\lefteqn{\pi a^S_0 =
       + x_2 \, \Bigg(
            \frac{1}{32}
          - \frac{1}{16 n}
          + \frac{n}{16}
          \Bigg)}&&
\nonumber\\&&
       + x_2^2 \, \Bigg(
          - \frac{2}{n}\,\alpha_4
          - \frac{1}{n}\,\alpha_3
          - \frac{1}{4 n}\,\alpha_2
          - \frac{3}{16 n}\,\alpha_1
          + \alpha_4
          + \frac{1}{2}\,\alpha_3
          + \frac{1}{8}\,\alpha_2
          + \frac{3}{32}\,\alpha_1
\nonumber\\&&\hskip1cm
          + \beta_4
          + \frac{1}{16}\,\beta_1
          + n\,\alpha_4
          + \frac{n}{16}\,\alpha_1
          \Bigg)
\nonumber\\&&
       + \pi_{16}\,x_2^2 \, \Bigg(
          - \frac{7}{32}
          + \frac{1}{8 n^2}
          - \frac{1}{8 n}
          + \frac{n}{8}
          + \frac{n^2}{8}
          \Bigg)
\nonumber\\&&
       + x_2^3 \, \Bigg(
          - \frac{4}{n}\,\gamma_5
          - \frac{2}{n}\,\gamma_4
          - \frac{1}{n}\,\gamma_3
          - \frac{1}{4 n}\,\gamma_2
          - \frac{3}{16 n}\,\gamma_1
          + \delta_4
          + \frac{1}{16}\,\delta_1
          + 2\,\gamma_5
\nonumber\\&&\hskip1cm
          + \gamma_4
          + \frac{1}{2}\,\gamma_3
          + \frac{1}{8}\,\gamma_2
          + \frac{3}{32}\,\gamma_1
          + n\,\gamma_4
          + \frac{n}{16}\,\gamma_1
          \Bigg)
\nonumber\\&&
       + \pi_{16}\,x_2^3 \, \Bigg(
          + \frac{7}{8 n^3}\,L
          - \frac{19}{16 n^2}\,L
          + \frac{12}{n^2}\,L^r_{8}
          - \frac{4}{n^2}\,L^r_{5}
          + \frac{12}{n^2}\,L^r_{3}
          + \frac{12}{n^2}\,L^r_{0}
          - \frac{13}{16 n}\,L
\nonumber\\&&\hskip1cm
          - \frac{12}{n}\,L^r_{8}
          - \frac{8}{n}\,L^r_{6}
          + \frac{4}{n}\,L^r_{5}
          + \frac{8}{n}\,L^r_{4}
          - \frac{12}{n}\,L^r_{3}
          - \frac{8}{n}\,L^r_{2}
          - \frac{8}{n}\,L^r_{1}
          - \frac{12}{n}\,L^r_{0}
\nonumber\\&&\hskip1cm
          + \frac{41}{32}\,L
          - 13\,L^r_{8}
          + 4\,L^r_{6}
          + 3\,L^r_{5}
          - 4\,L^r_{4}
          - 13\,L^r_{3}
          + 4\,L^r_{2}
          + 4\,L^r_{1}
          - 13\,L^r_{0}
\nonumber\\&&\hskip1cm
          + \frac{3 n}{8}\,L
          + 8\,n\,L^r_{8}
          + 8\,n\,L^r_{6}
          - 2\,n\,L^r_{5}
          - 8\,n\,L^r_{4}
          + 8\,n\,L^r_{3}
          + 8\,n\,L^r_{2}
\nonumber\\&&\hskip1cm
          + 8\,n\,L^r_{1}
          + 8\,n\,L^r_{0}
          - \frac{n^2}{2}\,L
          + 4\,n^2\,L^r_{8}
          + 4\,n^2\,L^r_{3}
          + 4\,n^2\,L^r_{0}
          - \frac{n^3}{4}\,L
          \Bigg)
\nonumber\\&&
       + \pi_{16}^2\,x_2^3 \, \Bigg(
          - \frac{37}{48}
          - \frac{5}{4 n^3}
          + \frac{13}{8 n^2}
          + \frac{3}{8 n}
          + \frac{31 n}{48}
          - \frac{3 n^2}{16}
          - \frac{7 n^3}{24}
          \Bigg)
\nonumber\\&&
       + \pi^2\,\pi_{16}^2\,x_2^3 \, \Bigg(
            \frac{41}{288}
          + \frac{5}{24 n^3}
          - \frac{13}{48 n^2}
          - \frac{1}{16 n}
          - \frac{29 n}{288}
          + \frac{n^2}{96}
          + \frac{5 n^3}{144}
          \Bigg)\,,
\\
\lefteqn{\pi a^{FS}_0 =
       + x_2 \, \Bigg(
          - \frac{1}{16}
          \Bigg)}&&
\nonumber\\&&
       + x_2^2 \, \Bigg(
            \alpha_3
          + \frac{1}{4}\,\alpha_2
          + \frac{1}{16}\,\alpha_1
          + \beta_4
          + \frac{1}{16}\,\beta_1
          \Bigg)
       + \frac{1}{8}\pi_{16}\,x_2^2
\nonumber\\&&
       + x_2^3 \, \Bigg(
          + \delta_4
          + \frac{1}{16}\,\delta_1
          + 4\,\gamma_5
          + \gamma_3
          + \frac{1}{4}\,\gamma_2
          + \frac{1}{16}\,\gamma_1
          \Bigg)
\nonumber\\&&
       + \pi_{16}\,x_2^3 \, \Bigg(
            \frac{1}{8 n^2}\,L
          - \frac{1}{4 n}\,L
          + \frac{3}{8}\,L
          - 4\,L^r_{8}
          - 8\,L^r_{6}
          + 4\,L^r_{5}
          + 8\,L^r_{4}
          - 4\,L^r_{3}
\nonumber\\&&\hskip1cm
          - 8\,L^r_{2}
          - 8\,L^r_{1}
          - 4\,L^r_{0}
          \Bigg)
\nonumber\\&&
       + \pi_{16}^2\,x_2^3 \, \Bigg[
          - \frac{17}{24}
          - \frac{1}{4 n^2}
          + \frac{1}{2 n}
          - \frac{n}{24}
       + \pi^2 \Bigg(
            \frac{13}{144}
          + \frac{1}{24 n^2}
          - \frac{1}{12 n}
          + \frac{7 n}{144}
          \Bigg)\Bigg]\,,
\\
\lefteqn{\pi a^{MA}_1 =
       + x_2^2 \, \Bigg[
            \frac{1}{3}\,\beta_4
          + \frac{1}{24}\,\beta_2
       + \pi_{16} \, \Bigg(
          - \frac{1}{288}
          + \frac{1}{288 n^2}
          \Bigg)\Bigg]}&&
\nonumber\\&&
       + \,x_2^3 \, \Bigg(
            \frac{2}{3}\,\delta_6
          + \frac{1}{3}\,\delta_4
          + \frac{1}{24}\,\delta_2
          \Bigg)
\nonumber\\&&
       + \pi_{16}\,x_2^3 \, \Bigg(
            \frac{7}{288 n^3}\,L
          - \frac{1}{72 n^2}\,L
          + \frac{1}{3 n^2}\,L^r_{8}
          - \frac{1}{9 n^2}\,L^r_{5}
          + \frac{5}{27 n^2}\,L^r_{3}
          + \frac{5}{27 n^2}\,L^r_{0}
\nonumber\\&&\hskip1cm
          - \frac{1}{72 n}\,L
          - \frac{2}{9 n}\,L^r_{6}
          + \frac{2}{9 n}\,L^r_{4}
          - \frac{2}{27 n}\,L^r_{2}
          - \frac{2}{9 n}\,L^r_{1}
          + \frac{11}{2592}\,L
          - \frac{1}{9}\,L^r_{8}
\nonumber\\&&\hskip1cm
          + \frac{2}{9}\,L^r_{6}
          - \frac{2}{9}\,L^r_{4}
          - \frac{1}{27}\,L^r_{3}
          + \frac{2}{27}\,L^r_{2}
          + \frac{2}{9}\,L^r_{1}
          - \frac{1}{9}\,L^r_{0}
          + \frac{7 n}{1296}\,L
          \Bigg)
\nonumber\\&&
       + \pi_{16}^2\,x_2^3 \, \Bigg(
            \frac{17}{2592}
          + \frac{1}{96 n^3}
          - \frac{1}{144 n^2}
          + \frac{n}{648}
          \Bigg)
\nonumber\\&&
       + \pi^2\,\pi_{16}^2\,x_2^3 \, \Bigg(
          - \frac{1}{1296}
          + \frac{1}{432 n^3}
          - \frac{1}{864 n^2}
          - \frac{1}{864 n}
          - \frac{n}{5184}
          \Bigg)\,,
\\
\lefteqn{\pi a^{MS}_0 =
       + \frac{1}{32}x_2 }&&
\nonumber\\&&
       + x_2^2 \, \Bigg(
          - \frac{1}{2}\,\alpha_3
          - \frac{1}{8}\,\alpha_2
          - \frac{1}{32}\,\alpha_1
          + \beta_4
          + \frac{1}{16}\,\beta_1
        + \frac{1}{32}\pi_{16}  \Bigg)
\nonumber\\&&
       + x_2^3 \, \Bigg(
            \delta_4
          + \frac{1}{16}\,\delta_1
          - 2\,\gamma_5
          - \frac{1}{2}\,\gamma_3
          - \frac{1}{8}\,\gamma_2
          - \frac{1}{32}\,\gamma_1
          \Bigg)
\nonumber\\&&
       + \pi_{16}\,x_2^3 \, \Bigg(
          - \frac{1}{16 n^2}\,L
          - \frac{1}{16 n}\,L
          - \frac{3}{32}\,L
          - L^r_{8}
          + 4\,L^r_{6}
          + L^r_{5}
          - 4\,L^r_{4}
          - L^r_{3}
\nonumber\\&&\hskip1cm
          + 4\,L^r_{2}
          + 4\,L^r_{1}
          - L^r_{0}
          \Bigg)
\nonumber\\&&
       + \pi_{16}^2\,x_2^3 \, \Bigg[
            \frac{1}{96}
          + \frac{1}{8 n^2}
          + \frac{1}{8 n}
          - \frac{n}{96}
        + \pi^2 \, \Bigg(
          - \frac{5}{576}
          - \frac{1}{48 n^2}
          - \frac{1}{48 n}
          + \frac{7 n}{576}
          \Bigg)\Bigg]\,.
\ea

\subsection{Pseudo-real or two-colour case}

\ba
\lefteqn{\pi a^I_0 =
       + x_2 \, \Bigg(
          - \frac{1}{32}
          - \frac{1}{32 n}
          + \frac{n}{8}
          \Bigg)}&&
\nonumber\\&&
       + x_2^2 \, \Bigg(
          - \frac{1}{n}\,\alpha_4
          - \frac{1}{2 n}\,\alpha_3
          - \frac{1}{8 n}\,\alpha_2
          - \frac{3}{32 n}\,\alpha_1
          - \alpha_4
          - \frac{1}{2}\,\alpha_3
          - \frac{1}{8}\,\alpha_2
          - \frac{3}{32}\,\alpha_1
\nonumber\\&&\hskip1cm
          + \beta_4
          - \frac{1}{2}\,\beta_3
          - \frac{1}{8}\,\beta_2
          + \frac{1}{32}\,\beta_1
          + 2\,n\,\alpha_4
          + \frac{n}{8}\,\alpha_1
          - \frac{n}{2}\,\beta_3
          - \frac{n}{8}\,\beta_2
          - \frac{n}{32}\,\beta_1
\nonumber\\&&\hskip1cm
          + n^2\,\beta_3
          + \frac{n^2}{4}\,\beta_2
          + \frac{n^2}{16}\,\beta_1
          \Bigg)
\nonumber\\&&
       + \pi_{16}\,x_2^2 \, \Bigg(
          - \frac{7}{32}
          + \frac{1}{32 n^2}
          + \frac{1}{16 n}
          - \frac{n}{4}
          + \frac{n^2}{2}
          \Bigg)
\nonumber\\&&
       + x_2^3 \, \Bigg(
          - \frac{2}{n}\,\gamma_5
          - \frac{1}{n}\,\gamma_4
          - \frac{1}{2 n}\,\gamma_3
          - \frac{1}{8 n}\,\gamma_2
          - \frac{3}{32 n}\,\gamma_1
          - 2\,\delta_5
          + \delta_4
          - \frac{1}{2}\,\delta_3
          - \frac{1}{8}\,\delta_2
\nonumber\\&&\hskip1cm
          + \frac{1}{32}\,\delta_1
          - 2\,\gamma_5
          - \gamma_4
          - \frac{1}{2}\,\gamma_3
          - \frac{1}{8}\,\gamma_2
          - \frac{3}{32}\,\gamma_1
          - 2\,n\,\delta_5
          - \frac{n}{2}\,\delta_3
          - \frac{n}{8}\,\delta_2
\nonumber\\&&\hskip1cm
          - \frac{n}{32}\,\delta_1
          + 2\,n\,\gamma_4
          + \frac{n}{8}\,\gamma_1
          + 4\,n^2\,\delta_5
          + n^2\,\delta_3
          + \frac{n^2}{4}\,\delta_2
          + \frac{n^2}{16}\,\delta_1
          \Bigg)
\nonumber\\&&
       + \pi_{16}\,x_2^3 \, \Bigg(
            \frac{7}{32 n^3}\,L
          + \frac{15}{32 n^2}\,L
          + \frac{3}{n^2}\,L^r_{8}
          - \frac{1}{n^2}\,L^r_{5}
          + \frac{3}{n^2}\,L^r_{3}
          + \frac{3}{n^2}\,L^r_{0}
          - \frac{29}{32 n}\,L
\nonumber\\&&\hskip1cm
          + \frac{6}{n}\,L^r_{8}
          - \frac{2}{n}\,L^r_{6}
          - \frac{2}{n}\,L^r_{5}
          + \frac{6}{n}\,L^r_{4}
          + \frac{6}{n}\,L^r_{3}
          - \frac{2}{n}\,L^r_{2}
          - \frac{2}{n}\,L^r_{1}
          + \frac{6}{n}\,L^r_{0}
\nonumber\\&&\hskip1cm
          - \frac{55}{32}\,L
          - 13\,L^r_{8}
          + 3\,L^r_{5}
          + 8\,L^r_{4}
          - 13\,L^r_{3}
          - 13\,L^r_{0}
          + \frac{21 n}{16}\,L
          - 16\,n\,L^r_{8}
\nonumber\\&&\hskip1cm
          + 6\,n\,L^r_{6}
          + 4\,n\,L^r_{5}
          - 26\,n\,L^r_{4}
          - 16\,n\,L^r_{3}
          + 6\,n\,L^r_{2}
          + 6\,n\,L^r_{1}
          - 16\,n\,L^r_{0}
\nonumber\\&&\hskip1cm
          + \frac{19 n^2}{8}\,L
          + 16\,n^2\,L^r_{8}
          - 12\,n^2\,L^r_{6}
          - 12\,n^2\,L^r_{4}
          + 16\,n^2\,L^r_{3}
          - 12\,n^2\,L^r_{2}
\nonumber\\&&\hskip1cm
          - 12\,n^2\,L^r_{1}
          + 16\,n^2\,L^r_{0}
          - \frac{5 n^3}{2}\,L
          + 16\,n^3\,L^r_{6}
          + 16\,n^3\,L^r_{4}
          + 16\,n^3\,L^r_{2}
\nonumber\\&&\hskip1cm
          + 16\,n^3\,L^r_{1}
          \Bigg)
\nonumber\\&&
       + \pi_{16}^2\,x_2^3 \, \Bigg(
           \frac{85}{48}
          - \frac{3}{8 n^3}
          - \frac{3}{4 n^2}
          + \frac{5}{4 n}
          - \frac{3 n}{8}
          - \frac{29 n^2}{48}
          + \frac{n^3}{12}
          \Bigg)
\nonumber\\&&
       + \pi^2\,\pi_{16}^2\,x_2^3 \, \Bigg(
          - \frac{89}{288}
          + \frac{1}{16 n^3}
          + \frac{1}{8 n^2}
          - \frac{5}{24 n}
          + \frac{n}{16}
          + \frac{49 n^2}{288}
          - \frac{5 n^3}{72}
          \Bigg)\,,
\\
\lefteqn{\pi a^A_0 =
       + x_2 \, \Bigg(
          - \frac{1}{32}
          - \frac{1}{16 n}
          + \frac{n}{16}
          \Bigg)
}&&
\nonumber\\&&
       + x_2^2 \, \Bigg(
          - \frac{2}{n}\,\alpha_4
          - \frac{1}{n}\,\alpha_3
          - \frac{1}{4 n}\,\alpha_2
          - \frac{3}{16 n}\,\alpha_1
          - \alpha_4
          - \frac{1}{2}\,\alpha_3
          - \frac{1}{8}\,\alpha_2
          - \frac{3}{32}\,\alpha_1
\nonumber\\&&\hskip1cm
          + \beta_4
          + \frac{1}{16}\,\beta_1
          + n\,\alpha_4
          + \frac{n}{16}\,\alpha_1
          \Bigg)
\nonumber\\&&
       + \pi_{16}\,x_2^2 \, \Bigg(
          - \frac{7}{32}
          + \frac{1}{8 n^2}
          + \frac{1}{8 n}
          - \frac{n}{8}
          + \frac{n^2}{8}
          \Bigg)
\nonumber\\&&
       + x_2^3 \, \Bigg(
          - \frac{4}{n}\,\gamma_5
          - \frac{2}{n}\,\gamma_4
          - \frac{1}{n}\,\gamma_3
          - \frac{1}{4 n}\,\gamma_2
          - \frac{3}{16 n}\,\gamma_1
          + \delta_4
          + \frac{1}{16}\,\delta_1
          - 2\,\gamma_5
          - \gamma_4
\nonumber\\&&\hskip1cm
          - \frac{1}{2}\,\gamma_3
          - \frac{1}{8}\,\gamma_2
          - \frac{3}{32}\,\gamma_1
          + n\,\gamma_4
          + \frac{n}{16}\,\gamma_1
          \Bigg)
\nonumber\\&&
       + \pi_{16}\,x_2^3 \, \Bigg(
            \frac{7}{8 n^3}\,L
          + \frac{19}{16 n^2}\,L
          + \frac{12}{n^2}\,L^r_{8}
          - \frac{4}{n^2}\,L^r_{5}
          + \frac{12}{n^2}\,L^r_{3}
          + \frac{12}{n^2}\,L^r_{0}
          - \frac{13}{16 n}\,L
\nonumber\\&&\hskip1cm
          + \frac{12}{n}\,L^r_{8}
          - \frac{8}{n}\,L^r_{6}
          - \frac{4}{n}\,L^r_{5}
          + \frac{8}{n}\,L^r_{4}
          + \frac{12}{n}\,L^r_{3}
          - \frac{8}{n}\,L^r_{2}
          - \frac{8}{n}\,L^r_{1}
          + \frac{12}{n}\,L^r_{0}
\nonumber\\&&\hskip1cm
          - \frac{41}{32}\,L
          - 13\,L^r_{8}
          - 4\,L^r_{6}
          + 3\,L^r_{5}
          + 4\,L^r_{4}
          - 13\,L^r_{3}
          - 4\,L^r_{2}
          - 4\,L^r_{1}
          - 13\,L^r_{0}
\nonumber\\&&\hskip1cm
          + \frac{3 n}{8}\,L
          - 8\,n\,L^r_{8}
          + 8\,n\,L^r_{6}
          + 2\,n\,L^r_{5}
          - 8\,n\,L^r_{4}
          - 8\,n\,L^r_{3}
          + 8\,n\,L^r_{2}
\nonumber\\&&\hskip1cm
          + 8\,n\,L^r_{1}
          - 8\,n\,L^r_{0}
          + \frac{n^2}{2}\,L
          + 4\,n^2\,L^r_{8}
          + 4\,n^2\,L^r_{3}
          + 4\,n^2\,L^r_{0}
          - \frac{n^3}{4}\,L
          \Bigg)
\nonumber\\&&
       + \pi_{16}^2\,x_2^3 \, \Bigg(
           \frac{37}{48}
          - \frac{5}{4 n^3}
          - \frac{13}{8 n^2}
          + \frac{3}{8 n}
          + \frac{31 n}{48}
          + \frac{3 n^2}{16}
          - \frac{7 n^3}{24}
          \Bigg)
\nonumber\\&&
       + \pi^2\,\pi_{16}^2\,x_2^3 \, \Bigg(
          - \frac{41}{288}
          + \frac{5}{24 n^3}
          + \frac{13}{48 n^2}
          - \frac{1}{16 n}
          - \frac{29 n}{288}
          - \frac{n^2}{96}
          + \frac{5 n^3}{144}
          \Bigg)\,,
\\
\lefteqn{\pi a^S_1 =
       + x_2 \, \Bigg(
          - \frac{1}{48}
          + \frac{n}{48}
          \Bigg)
}&&
\nonumber\\&&
       + x_2^2 \, \Bigg(
            \frac{1}{3}\,\alpha_4
          + \frac{1}{24}\,\alpha_2
          + \frac{1}{3}\,\beta_4
          + \frac{1}{24}\,\beta_2
          - \frac{n}{3}\,\alpha_4
          - \frac{n}{24}\,\alpha_2
          \Bigg)
\nonumber\\&&
       + \pi_{16}\,x_2^2 \, \Bigg(
          - \frac{11}{864}
          + \frac{1}{288 n^2}
          + \frac{1}{288 n}
          + \frac{7 n}{864}
          - \frac{n^2}{432}
          \Bigg)
\nonumber\\&&
       + x_2^3 \, \Bigg(
            \frac{2}{3}\,\delta_6
          + \frac{1}{3}\,\delta_4
          + \frac{1}{24}\,\delta_2
          + \frac{2}{3}\,\gamma_6
          + \frac{1}{3}\,\gamma_4
          + \frac{1}{24}\,\gamma_2
          - \frac{2 n}{3}\,\gamma_6
          - \frac{n}{3}\,\gamma_4
\nonumber\\&&\hskip1cm
          - \frac{n}{24}\,\gamma_2
          \Bigg)
\nonumber\\&&
       + \pi_{16}\,x_2^3 \, \Bigg(
            \frac{7}{288 n^3}\,L
          + \frac{1}{36 n^2}\,L
          + \frac{1}{3 n^2}\,L^r_{8}
          - \frac{1}{9 n^2}\,L^r_{5}
          + \frac{5}{27 n^2}\,L^r_{3}
          + \frac{5}{27 n^2}\,L^r_{0}
\nonumber\\&&\hskip1cm
          - \frac{5}{288 n}\,L
          + \frac{1}{3 n}\,L^r_{8}
          - \frac{2}{9 n}\,L^r_{6}
          - \frac{1}{9 n}\,L^r_{5}
          + \frac{2}{9 n}\,L^r_{4}
          + \frac{5}{27 n}\,L^r_{3}
          - \frac{2}{27 n}\,L^r_{2}
\nonumber\\&&\hskip1cm
          - \frac{2}{9 n}\,L^r_{1}
          + \frac{5}{27 n}\,L^r_{0}
          - \frac{67}{1296}\,L
          - \frac{4}{9}\,L^r_{8}
          + \frac{2}{9}\,L^r_{6}
          + \frac{1}{54}\,L^r_{5}
          + \frac{10}{27}\,L^r_{4}
\nonumber\\&&\hskip1cm
          - \frac{5}{27}\,L^r_{3}
          + \frac{2}{9}\,L^r_{2}
          - \frac{2}{27}\,L^r_{1}
          - \frac{10}{27}\,L^r_{0}
          + \frac{125 n}{2592}\,L
          - \frac{4 n}{9}\,L^r_{6}
          + \frac{7 n}{54}\,L^r_{5}
\nonumber\\&&\hskip1cm
          - \frac{4 n}{27}\,L^r_{4}
          - \frac{n}{27}\,L^r_{3}
          - \frac{8 n}{27}\,L^r_{2}
          - \frac{4 n}{27}\,L^r_{1}
          + \frac{2 n}{27}\,L^r_{0}
          - \frac{7 n^2}{864}\,L
          - \frac{n^2}{27}\,L^r_{5}
\nonumber\\&&\hskip1cm
          + \frac{n^3}{432}\,L
          \Bigg)
\nonumber\\&&
       + \pi_{16}^2\,x_2^3 \, \Bigg(
           \frac{13}{96}
          + \frac{1}{96 n^3}
          - \frac{7}{288 n^2}
          - \frac{281 n}{1728}
          + \frac{151 n^2}{2592}
          - \frac{25 n^3}{5184}
          \Bigg)
\nonumber\\&&
       + \pi^2\,\pi_{16}^2\,x_2^3 \, \Bigg(
          - \frac{173}{10368}
          + \frac{1}{432 n^3}
          + \frac{5}{864 n^2}
          - \frac{1}{576 n}
          + \frac{169 n}{10368}
          - \frac{n^2}{432}
\nonumber\\&&\hskip1cm
          - \frac{n^3}{1296}
          \Bigg)\,,
\\
\lefteqn{\pi a^{FA}_0 =
       + \frac{1}{16}x_2
}&&
\nonumber\\&&
       + x_2^2 \, \Bigg(
          - \alpha_3
          - \frac{1}{4}\,\alpha_2
          - \frac{1}{16}\,\alpha_1
          + \beta_4
          + \frac{1}{16}\,\beta_1
       + \frac{1}{8}\pi_{16}\Bigg)
\nonumber\\&&
       + x_2^3 \, \Bigg(
           \delta_4
          + \frac{1}{16}\,\delta_1
          - 4\,\gamma_5
          - \gamma_3
          - \frac{1}{4}\,\gamma_2
          - \frac{1}{16}\,\gamma_1
          \Bigg)
\nonumber\\&&
       + \pi_{16}\,x_2^3 \, \Bigg(
          - \frac{1}{8 n^2}\,L
          - \frac{1}{4 n}\,L
          - \frac{3}{8}\,L
          - 4\,L^r_{8}
          + 8\,L^r_{6}
          + 4\,L^r_{5}
          - 8\,L^r_{4}
          - 4\,L^r_{3}
\nonumber\\&&\hskip1cm
          + 8\,L^r_{2}
          + 8\,L^r_{1}
          - 4\,L^r_{0}
          \Bigg)
\nonumber\\&&
       + \pi_{16}^2\,x_2^3 \, \Bigg[
            \frac{17}{24}
          + \frac{1}{4 n^2}
          + \frac{1}{2 n}
          - \frac{n}{24}
       + \pi^2 \, \Bigg(
         \frac{7 n}{144}
          - \frac{13}{144}
          - \frac{1}{24 n^2}
          - \frac{1}{12 n}
          \Bigg)\Bigg]\,,
\\
\lefteqn{\pi a^{MA}_1 =
       + x_2^2 \, \Bigg[
            \frac{1}{3}\,\beta_4
          + \frac{1}{24}\,\beta_2
       + \pi_{16}\, \Bigg(
          - \frac{1}{288}
          + \frac{1}{288 n^2}
          \Bigg) \Bigg]
}&& \nonumber\\&&
       + x_2^3 \, \Bigg(
            \frac{2}{3}\,\delta_6
          + \frac{1}{3}\,\delta_4
          + \frac{1}{24}\,\delta_2
          \Bigg)
\nonumber\\&&
       + \pi_{16}\,x_2^3 \, \Bigg(
            \frac{7}{288 n^3}\,L
          + \frac{1}{72 n^2}\,L
          + \frac{1}{3 n^2}\,L^r_{8}
          - \frac{1}{9 n^2}\,L^r_{5}
          + \frac{5}{27 n^2}\,L^r_{3}
          + \frac{5}{27 n^2}\,L^r_{0}
\nonumber\\&&\hskip1cm
          - \frac{1}{72 n}\,L
          - \frac{2}{9 n}\,L^r_{6}
          + \frac{2}{9 n}\,L^r_{4}
          - \frac{2}{27 n}\,L^r_{2}
          - \frac{2}{9 n}\,L^r_{1}
          - \frac{11}{2592}\,L
          - \frac{1}{9}\,L^r_{8}
\nonumber\\&&\hskip1cm
          - \frac{2}{9}\,L^r_{6}
          + \frac{2}{9}\,L^r_{4}
          - \frac{1}{27}\,L^r_{3}
          - \frac{2}{27}\,L^r_{2}
          - \frac{2}{9}\,L^r_{1}
          - \frac{1}{9}\,L^r_{0}
          + \frac{7 n}{1296}\,L
          \Bigg)
\nonumber\\&&
       + \pi_{16}^2\,x_2^3 \, \Bigg(
          - \frac{17}{2592}
          + \frac{1}{96 n^3}
          + \frac{1}{144 n^2}
          + \frac{n}{648}
          \Bigg)
\nonumber\\&&
       + \pi^2\,\pi_{16}^2\,x_2^3 \, \Bigg(
            \frac{1}{1296}
          + \frac{1}{432 n^3}
          + \frac{1}{864 n^2}
          - \frac{1}{864 n}
          - \frac{n}{5184}
          \Bigg)\,,
\\
\lefteqn{\pi a^{MS}_0 =
        - \frac{1}{32} x_2
}&&
\nonumber\\&&
       + x_2^2 \, \Bigg(
            \frac{1}{2}\,\alpha_3
          + \frac{1}{8}\,\alpha_2
          + \frac{1}{32}\,\alpha_1
          + \beta_4
          + \frac{1}{16}\,\beta_1
          + \frac{1}{32}\pi_{16}
          \Bigg)
\nonumber\\&&
       + x_2^3 \, \Bigg(
          + \delta_4
          + \frac{1}{16}\,\delta_1
          + 2\,\gamma_5
          + \frac{1}{2}\,\gamma_3
          + \frac{1}{8}\,\gamma_2
          + \frac{1}{32}\,\gamma_1
          \Bigg)
\nonumber\\&&
       + \pi_{16}\,x_2^3 \, \Bigg(
          + \frac{1}{16 n^2}\,L
          - \frac{1}{16 n}\,L
          + \frac{3}{32}\,L
          - L^r_{8}
          - 4\,L^r_{6}
          + L^r_{5}
          + 4\,L^r_{4}
          - L^r_{3}
\nonumber\\&&\hskip1cm
          - 4\,L^r_{2}
          - 4\,L^r_{1}
          - L^r_{0}
          \Bigg)
\nonumber\\&&
       + \pi_{16}^2\,x_2^3 \, \Bigg[
            \frac{1}{8 n}
          - \frac{1}{96}
          - \frac{1}{8 n^2}
          - \frac{11n}{96}
       + \pi^2 \, \Bigg(
           \frac{5}{576}
          + \frac{1}{48 n^2}
          - \frac{1}{48 n}
          + \frac{7 n}{576}
          \Bigg)\Bigg]\,.
\ea

\section{Loop integrals}
\label{loopintegrals}

\subsection{One-loop integrals}
\label{oneloopint}

We use dimensional regularization here throughout in $d$ dimensions
with $d=4-2\epsilon$.
We need integrals with one, two and three propagators in principle.
These one propagator integral is
\be
A(m^2) = \frac{1}{i}\int \frac{d^d q}{(2\pi)^d}\frac{1}{q^2-m^2}\,.
\ee

The two propagator integrals we encountered are
\ba
B(m_1^2,m_2^2,p^2)&=& \frac{1}{i}\int\frac{d^dq}{(2\pi)^d}
\frac{1}{(q^2-m_1^2)((q-p)^2-m_2^2)}\,,
\nonumber\\
B_\mu(m_1^2,m_2^2,p)
&=& \frac{1}{i}\int\frac{d^dq}{(2\pi)^d}
\frac{q_\mu}{(q^2-m_1^2)((q-p)^2-m_2^2)}
\nonumber\\
&=& p_\mu B_1(m_1^2,m_2^2,p^2)\,,
\nonumber\\
B_{\mu\nu}(m_1^2,m_2^2,p)
&=& \frac{1}{i}\int\frac{d^dq}{(2\pi)^d}
\frac{q_\mu q_\nu}{(q^2-m_1^2)((q-p)^2-m_2^2)}
\nonumber\\
&=& p_\mu p_\nu B_{21}(m_1^2,m_2^2,p^2)
+ g_{\mu\nu} B_{22}(m_1^2,m_2^2,p^2)\,.
\ea

All the cases with three propagator integrals that show up can be absorbed
into the two-propagator ones by moving to the real masses rather than the
lowest order masses. This provides a consistency check on the
calculations.

The explicit expressions are well known
\ba
\label{resultintegrals}
A(m^2) & = &\frac{m^2}{16\pi^2}\Bigg\{\lambda_0-\ln(m^2)
+\epsilon\Big(\frac{C^2}{2}+\frac{1}{2}+\frac{\pi^2}{12}
+\frac{1}{2}\ln^2(m^2)
\nonumber \\& &
- C\ln(m^2)\Big)\Bigg\}
+{\cal O}(\epsilon^2)
\,,
\nonumber\\
B(m_1^2,m_2^2,p^2) &=&
\frac{1}{16\pi^2}\left(\lambda_0
-\frac{m_1^2\ln(m_1^2)-m_2^2\ln(m_2^2)}{m_1^2-m_2^2}\right)
+\bar{J}(m_1^2,m_2^2,p^2)+{\cal O}(\epsilon)\,,
\nonumber\\
\bar{J}(m_1^2,m_2^2,p^2) &=&
-\frac{1}{16\pi^2}\int_0^1 dx
\ln\left(\frac{m_1^2 x+m_2^2(1-x)-x(1-x)p^2}{m_1^2 x+m_2^2(1-x)}\right)\,,
\ea
$C=\ln(4\pi)+1-\gamma$ and $\lambda_0 = 1/\epsilon+C$.
The function $\bar{J}(m_1^2,m_2^2,p^2)$ develops an imaginary
part for $p^2\ge (m_1+m_2)^2$. Using $\Delta=m_1^2-m_2^2$, $\Sigma=m_1^2+m_2^2$
and $\nu^2 = p^4+m_1^4+m_2^4-2p^2m_1^2-2p^2m_2^2-2m_1^2m_2^2$
it is given by
\ba
(32\pi^2)\bar{J}(m_1^2,m_2^2,p^2) &=&
2+\left(-\frac{\Delta}{p^2}+\frac{\Sigma}{\Delta}\right)\ln\frac{m_1^2}{m_2^2}
-\frac{\nu}{p^2}\ln\frac{(p^2+\nu)^2-\Delta^2}{(p^2-\nu)^2-\Delta^2}\,.
\ea

The two-propagator integrals can all be reduced to $B$ and $A$ via
\ba
B_1(m_1^2,m_2^2,p^2)& = &-\frac{1}{2p^2}\left(
  A(m_1^2)-A(m_2^2)+(m_2^2-m_1^2-p^2) B(m_1^2,m_2^2,p^2)\right)\,,
\nonumber\\
B_{22}(m_1^2,m_2^2,p^2)& = & \frac{1}{2(d-1)}\Big(
 A(m_2^2)+2 m_1^2 B(m_1^2,m_2^2,p^2)
\nonumber\\&&
+(m_2^2-m_1^2-p^2) B_1(m_1^2,m_2^2,p^2)
\Big)\,,
\nonumber\\
B_{21}(m_1^2,m_2^2,p^2)& = &\frac{1}{p^2}\left(
  A(m_2^2)+m_1^2 B(m_1^2,m_2^2,p^2)-d B_{22}(m_1^2,m_2^2,p^2)\right)\,.
\ea
The basic method used here is the one from Passarino and Veltman \cite{PV}.

\subsection{Sunset integrals}

Sunset integrals have been treated in many places already,
in general and for various special cases. We use here a method that
is a hybrid of various other approaches. We only cite the literature
actually used.
We define
\be
\lla X \rra
= \frac{1}{i^2}\int \frac{d^d q}{(2\pi)^d} \frac{d^d r}{(2\pi)^d}
\frac{X}
{\left(q^2-m_1^2\right)\left(r^2-m_2^2\right)\left((q+r-p)^2-m_3^2\right)}\,,
\ee
\ba
\label{defsunset}
H(m_1^2,m_2^2,m_3^2;p^2) &=&\lla 1\rra\,,\nonumber\\
H_\mu(m_1^2,m_2^2,m_3^2;p^2) &=&\lla q_\mu\rra
= p_\mu H_1(m_1^2,m_2^2,m_3^2;p^2)\,,\nonumber\\
H_{\mu\nu}(m_1^2,m_2^2,m_3^2;p^2) &=&\lla q_\mu q_\nu\rra\nonumber\\
& = &
p_\mu p_\nu H_{21}(m_1^2,m_2^2,m_3^2;p^2)
+g_{\mu\nu} H_{22}(m_1^2,m_2^2,m_3^2;p^2)
\,.
\nonumber\\
\ea
By redefining momenta the others can
be simply related to the above three:
\ba
\lla r_\mu\rra &=& p_\mu H_1(m_2^2,m_1^2,m_3^2;p^2)\,,\nonumber\\
\lla r_\mu r_\nu\rra &=& p_\mu p_\nu H_{21}(m_2^2,m_1^2,m_3^2;p^2)
+ g_{\mu\nu} H_{22}(m_2^2,m_1^2,m_3^2;p^2)\,,\nonumber\\
\lla q_\mu r_\nu \rra &=& \lla r_\mu q_\nu\rra \,,\nonumber\\
\lla q_\mu r_\nu\rra &=&p_\mu p_\nu H_{23}(m_1^2,m_2^2,m_3^2;p^2)
+g_{\mu\nu} H_{24}(m_1^2,m_2^2,m_3^2;p^2)\,,
\ea
with
\ba
2H_{23}(m_1^2,m_2^2,m_3^2;p^2)&=&
-H_{21}(m_1^2,m_2^2,m_3^2;p^2)
-H_{21}(m_2^2,m_1^2,m_3^2;p^2)
\nonumber\\&&
+H_{21}(m_3^2,m_1^2,m_2^2;p^2)
+2 H_1(m_1^2,m_2^2,m_3^2;p^2)
\nonumber\\&&
+2 H_1(m_2^2,m_1^2,m_3^2;p^2)
-H(m_1^2,m_2^2,m_3^2;p^2)\,,
\nonumber\\
2 H_{24}(m_1^2,m_2^2,m_3^2;p^2)&=&
-H_{22}(m_1^2,m_2^2,m_3^2;p^2)
-H_{22}(m_2^2,m_1^2,m_3^2;p^2)
\nonumber\\&&
+H_{22}(m_3^2,m_1^2,m_2^2;p^2)\,.
\ea
The first two follow from interchanging $q$ and $r$ and the third
from the fact that it is proportional to $g_{\mu\nu}$ or $p_\mu p_\nu$, which
are both symmetric in $\mu$ and $\nu$. The last one is deribed using
\ba
\left(q_\mu r_\nu+r_\mu q_\nu\right) &=&
\left(q_\mu+r_\mu-p_\mu\right)\left(q_\nu+r_\nu-p_\nu\right)
-q_\mu q_\nu - r_\mu r_\nu -p_\mu p_\nu
\nonumber\\&&+2p_\mu\left(q_\nu+r_\nu\right)
+2 p_\nu\left(q_\mu+r_\mu\right)
\ea
and redefining momenta and masses on the r.h.s..
In addition we have the relation
\ba
\label{H22relation}
\lefteqn{
p^2 H_{21}(m_1^2,m_2^2,m_3^2;p^2)+d H_{22}(m_1^2,m_2^2,m_3^2;p^2)
= }&&
\nonumber\\&&
 m_1^2 H(m_1^2,m_2^2,m_3^2;p^2)+A(m_2^2) A(m_3^2)\,.
\ea
which allows to express $H_{22}$ in a simple way in terms
of $H_{21}$.
There is also a relation between $H_1$ and $H$
\ba
\label{relationH1}
 H_1(m_1^2,m_2^2,m_3^2;p^2)+ H_1(m_2^2,m_1^2,m_3^2;p^2)
 + H_1(m_3^2,m_1^2,m_2^2;p^2)=
\nonumber\\
H(m_1^2,m_2^2,m_3^2;p^2)\,,
\ea
which  allows to write $H_1(m^2,m^2,m^2;p^2)
= 1/3\;H(m^2,m^2,m^2;p^2)$ in the case of equal masses.
The function $H$ is fully symmetric in $m_1^2,m_2^2$ and $m_3^2$, while
$H_1$, $H_{21}$ and $H_{22}$ are symmetric under the interchange
of $m_2^2$ and $m_3^2$.

We only need the sunset integrals at $p^2=m_1^2=m_2^2=m_3^2$ and their
derivatives w.r.t. $p^2$.
These have been calculated using the methods of \cite{GS}.
With $H_{id}=\frac{\partial}{\partial p^2}H_{i}$ we obtain
\ba
H&=&\lambda_1 m^2\left(\frac{5  \pi_{16}^2}{4}-3 L \pi_{16}\right)
+\frac{3}{2} \lambda_2 m^2 \pi_{16}^2\nonumber\\
&&+m^2\left(3 L^2 -\frac{5}{2} L  \pi_{16}+\frac{1}{4}  \pi ^2 \pi_{16}^2
+\frac{15 \pi_{16}^2}{8}\right)
\\
H_{21}&=&
\lambda_1 m^2 \left(\frac{11  \pi_{16}^2}{72}-\frac{2}{3} L  \pi_{16}\right)
+\frac{1}{3} \lambda_2 m^2 \pi_{16}^2\nonumber\\
&&+m^2\left(-\frac{11}{36} L  \pi_{16}
+\frac{1}{18}  \pi ^2 \pi_{16}^2+\frac{493 \pi_{16}^2}{864}\right)
\\
H_{22}&=&
\lambda_1 m^4\left(\frac{157  \pi_{16}^2}{288}-\frac{13}{12} L  \pi_{16}\right)
+\frac{13}{24} \lambda_2 m^4 \pi_{16}^2\nonumber\\
&& +m^4\left(-\frac{157}{144} L  \pi_{16}+\frac{13 L^2 4}{12}
+\frac{13}{144} \pi ^2 \pi_{16}^2+\frac{2933 \pi_{16}^2}{3456}\right)
\\
H_d&=&
\frac{L \pi_{16}}{2}-\frac{\lambda_1 \pi_{16}^2}{4}+\frac{7 \pi_{16}^2}{8}\\
H_{21d}&=&
\frac{L \pi_{16}}{12}-\frac{\lambda_1 \pi_{16}^2}{24}+\frac{43 \pi_{16}^2}{288}
\\
H_{22d}&=&
\frac{5}{48} L m^2 \pi_{16}-\frac{5}{96} \lambda_1 m^2 \pi_{16}^2+\frac{179  \pi_{16}^2}{1152}m^2
\ea
 $H_1$ and $H_{1d}$ follow immmediately using (\ref{relationH1}).

\subsection{Vertex integrals}

The vertex diagram (16) in Fig.~\ref{pipi2loop} is the most difficult
two-loop diagram
in $\phi\phi$ scattering, and it can also appear in other process.
The two loop integral for the equal mass case can be written as
\ba
\lla X \rra&=&\frac{\mu^{4\epsilon}}{i^2}\int\int dr ds \frac{X}
{(r^2-m^2)\cdot[(r-q)^2-m^2]\cdot(s^2-m^2) \cdot [(s+r-p)^2-m^2]}
\nonumber\\
\ea
The Lorentz decompositions of the vertex integrals are \cite{BT1}
\begin{eqnarray}
\label{lorentz}
\langle\langle 1 \rangle\rangle &=& V\,,
\nonumber\\
 \lla r_\mu \rra
&=&p_\mu V_{11} + q_\mu V_{12}\,,
\nonumber\\
 \langle\langle s_\mu \rangle\rangle
&=&  p_\mu V_{13} + q_\mu V_{14}\,,
\nonumber\\
 \langle\langle r_\mu r_\nu \rangle\rangle
 &=& g_{\mu\nu}V_{21}
+p_\mu p_\nu V_{22} + q_\mu q_\nu V_{23} +(p_\mu q_\nu + q_\mu
p_\nu)V_{24}\,,
\nonumber\\
\langle\langle r_\mu s_\nu \rangle\rangle
 &=& g_{\mu\nu}V_{25}
+p_\mu p_\nu V_{26} + q_\mu q_\nu V_{27} +q_\mu p_\nu V_{28}+ p_\mu
q_\nu V_{29}\,,
\nonumber\\
\langle\langle s_\mu s_\nu \rangle\rangle
 &=& g_{\mu\nu}V_{210}
+p_\mu p_\nu V_{211} + q_\mu q_\nu V_{212} +(q_\mu p_\nu + p_\mu
q_\nu) V_{213}\,,
\nonumber\\
\langle\langle r_\mu r_\nu r_\alpha\rangle\rangle
 &=& (g_{\mu\nu}p_\alpha+g_{\mu\alpha}p_\nu+g_{\nu\alpha}p_\mu)V_{31}
+(g_{\mu\nu}q_\alpha+g_{\mu\alpha}q_\nu+g_{\nu\alpha}q_\mu)V_{32}
\nonumber\\&& +p_\mu p_\nu p_\alpha V_{33} + q_\mu q_\nu q_\alpha
V_{34} \nonumber\\&&+(p_\mu p_\nu q_\alpha+ p_\mu q_\nu p_\alpha +
q_\mu p_\nu p_\alpha)V_{35}  +(q_\mu q_\nu p_\alpha+ q_\mu p_\nu
q_\alpha + p_\mu q_\nu q_\alpha)V_{36}\,,
\nonumber\\
\langle\langle r_\mu r_\nu s_\alpha\rangle\rangle
 &=& g_{\mu\nu}p_\alpha V_{37}+g_{\mu\nu}q_\alpha V_{38}
+(g_{\mu\alpha}p_\nu+g_{\nu\alpha}p_\mu)V_{39}+(g_{\mu\alpha}q_\nu+g_{\nu\alpha}q_\mu)V_{310}
\nonumber\\&&
 +p_\mu p_\nu
p_\alpha V_{311} + q_\mu q_\nu q_\alpha V_{312} +p_\mu p_\nu
q_\alpha V_{313}+ q_\mu q_\nu p_\alpha V_{314} \nonumber\\&& +(p_\mu
q_\nu+q_\mu p_\nu)p_\alpha V_{315} +(p_\mu q_\nu+q_\mu
p_\nu)q_\alpha V_{316}\,,
\nonumber\\
\langle\langle r_\mu s_\nu s_\alpha\rangle\rangle
 &=& p_\mu g_{\nu\alpha} V_{317}+q_\mu g_{\nu\alpha} V_{318}
+(g_{\mu\nu}p_\alpha+g_{\mu\alpha}p_\nu)V_{319}+(g_{\mu\nu}q_\alpha+g_{\mu\alpha}q_\nu)V_{320}
\nonumber\\&&
 +p_\mu p_\nu p_\alpha V_{321} + q_\mu q_\nu q_\alpha V_{322}
+p_\mu q_\nu q_\alpha V_{323} + q_\mu p_\nu p_\alpha V_{324}
\nonumber\\&&
 +p_\mu(p_\nu q_\alpha+q_\nu p_\alpha) V_{325}
+q_\mu(p_\nu q_\alpha+q_\nu p_\alpha) V_{326}\,.
\nonumber\\
%
\end{eqnarray}
The $\langle\langle s_\mu s_\nu s_\alpha\rangle\rangle$ does not show up
in $\phi\phi$ scattering.
Most of those $V_i$ functions have been calculated analytically
in \cite{BCEGS2} except the
$\langle\langle s_\mu s_\nu \rangle\rangle$ and $\langle\langle r_\mu s_\nu
s_\alpha\rangle\rangle$. We have calculated the rest of them in this work,
which are $V_{210} - V_{213}$ and $V_{317} - V_{326}$.
Again, the methods of \cite{GS} were used here, somewhat extended to the cases
at hand.
We have compared our results with the numerical
evaluation for general masses described in \cite{BT1}.
The quantity $B^\epsilon$ is the next term in the expansion of $B$ in 
(\ref{resultintegrals}) but these terms always cancel in the final result.

\begin{eqnarray*}
V_0&=&\lambda_1 \left(\pi_{16} \bar B+\frac{\pi^2_{16}}{2}\right)
+\lambda_2\frac{ \pi^2_{16}}{2}+\pi_{16} B^\epsilon
\\&&+\bar J  (2 \pi_{16}-L)+\frac{k_1 }{2}-\frac{k_3 }{3}
+\frac{L^2}{2}-\pi^2_{16}
\\
V_{11}&=&\lambda_1\frac{ \pi^2_{16}}{4}
+\frac{1}{2} \pi_{16} \bar J
-\frac{1}{3}k_3 -2 k_4 -\frac{L \pi_{16}}{2}-\frac{7 \pi^2_{16}}{8}\\
V_{12}&=&\lambda_1 \left(\frac{1}{2} \pi_{16} \bar B+\frac{\pi^2_{16}}{8}\right)+\lambda_2\frac{ \pi^2_{16}}{4}
+\frac{1}{2} \pi_{16} B^\epsilon\\
&&+\bar J\left(\frac{3 \pi_{16}}{4}-\frac{L}{2}\right)
+\frac{k_1 }{4}+k_4
 +\frac{L^2}{4}+\frac{L \pi_{16}}{4}-\frac{\pi^2_{16}}{16}\\
V_{13}&=&\lambda_1 \left(\frac{1}{2} \pi_{16} \bar B+\frac{\pi^2_{16}}{8}\right)
+\lambda_2\frac{ \pi^2_{16}}{4}+\frac{1}{2} \pi_{16} B^\epsilon
\\&&+\bar J  \left(\frac{3 \pi_{16}}{4}
-\frac{L}{2}\right)+\frac{k_1 }{4}+k_4
+\frac{L^2}{4}+\frac{L \pi_{16}}{4}-\frac{\pi^2_{16}}{16}\\
V_{14}&=&-\lambda_1 \left(\frac{1}{4} \pi_{16} \bar B+\frac{\pi^2_{16}}{16}\right)
-\lambda_2 \frac{\pi^2_{16}}{8}-\frac{1}{4} \pi_{16} B^\epsilon
\\&&+\bar J  \left(\frac{L}{4}-\frac{3 \pi_{16}}{8}\right)-\frac{k_1 }{8}
-\frac{k_4 }{2}
-\frac{L^2}{8}-\frac{L \pi_{16}}{8}+\frac{\pi^2_{16}}{32}\\
V_{21}&=&\frac{1}{2}\lambda_1 \left\{\pi_{16}
\left[  s \bar B_{21}+  \bar B\left(m^2 -\frac{1}{2} s \right)-  L m^2  \right]
+\pi^2_{16}\left(\frac{11 m^2 }{6}-\frac{13 s}{72}\right)\right\}\\
&&+\frac{1}{2} \lambda_2  \pi^2_{16}\left( m^2 -\frac{ s}{12}\right)
+B^\epsilon \pi_{16} \left(\frac{m^2 }{3}-\frac{ s}{12}\right)\\
&&+\bar J \left(\frac{L s}{12}-\frac{L m^2}{3}+\frac{5 m^2 \pi_{16}}{12}
-\frac{5 \pi_{16} s}{24}\right)
-\frac{ m^2}{6} k_1 -\frac{s}{24} k_2 -\frac{m^2}{6}  k_3 -m^2 k_4\\
&&+\frac{5 L^2 m^2}{6}-\frac{L^2 s}{24}
+\pi_{16}L\left(\frac{  s}{24}-\frac{7 m^2 }{6}\right)
+\pi^2_{16} \left(\frac{m^2 \pi ^2}{18}+\frac{11 m^2}{24}-\frac{\pi ^2 s}{72}-\frac{5 s}{96}\right)
\\
V_{22}&=&\lambda_1\frac{ \pi^2_{16}}{12}
+\frac{1}{6} \pi_{16} \bar J -\frac{k_3 }{3}-6 k_4 -12 k_5
-\frac{L \pi_{16}}{6}-\frac{7 \pi^2_{16}}{24}\\
V_{23}&=&\lambda_1 \left(\pi_{16} \bar B_{21}+\frac{\pi^2_{16}}{36}\right)
+\lambda_2\frac{ \pi^2_{16}}{6}+B^\epsilon\pi_{16}  \left(\frac{1}{3}-\frac{m^2  }{3 s}\right)
\\&&+\bar J  \left(\frac{L m^2}{3 s}-\frac{L}{3}-\frac{2 m^2 \pi_{16}}{3 s}
+\frac{\pi_{16}}{2}\right)+\frac{3 k_1 }{16}-\frac{k_2 }{48}-\frac{k_4 }{4}-3 k_5
+\frac{k_9 }{12}\\
&&
+\frac{L^2 m^2}{6 s}+\frac{L^2}{6}
+\pi_{16}L\left(\frac{  m^2 }{3 s}+\frac{1 }{6}\right)+\pi^2_{16} \left(\frac{m^2 \pi ^2}{36 s}+\frac{m^2}{6 s}+\frac{5}{24}\right)
\\
V_{24}&=&\lambda_1\frac{ \pi^2_{16}}{12}
+\frac{1}{6} \pi_{16} \bar J +2 k_4 +6 k_5 -\frac{L \pi_{16}}{6}-\frac{7 \pi^2_{16}}{24}
\\
V_{25}&=&\frac{1}{4}\lambda_1 \left\{\pi_{16}
\left[-  s \bar B_{21}+\bar B\left(\frac{s}{2} -  m^2\right)+ L m^2 \right]
+\pi^2_{16}\left(\frac{13  s}{72}-\frac{11}{6} m^2 \right)\right\}\\
&&+\frac{1 }{4}\lambda_2 \pi^2_{16}\left(\frac{ s}{12}-m^2\right)
+B^\epsilon  \pi_{16}\left(\frac{ s}{24}-\frac{m^2 }{6}\right)\\
&&+\bar J  \left(\frac{L m^2}{6}-\frac{L s}{24}-\frac{5 m^2 \pi_{16}}{24}
+\frac{5 \pi_{16} s}{48}\right)+\frac{ m^2}{12} k_1 +\frac{s}{48}  k_2
+\frac{m^2}{12}  k_3 +\frac{m^2}{2}  k_4\\
&& -\frac{5 L^2 m^2}{12}+\frac{L^2 s}{48}
+L \pi_{16}\left(\frac{7  m^2 }{12} -\frac{ s}{48}\right)
 +\pi^2_{16} \left(\frac{\pi ^2 s}{144}+\frac{5 s}{192}-\frac{m^2 \pi ^2}{36}-\frac{11 m^2}{48}
 \right)
\\
V_{26}&=&\lambda_1\frac{ \pi^2_{16}}{12}
+\frac{1}{6} \pi_{16} \bar J +2 k_4 +6 k_5 -\frac{L \pi_{16}}{6}-\frac{7 \pi^2_{16}}{24}
\\
V_{27}&=&-\frac{1}{2}\lambda_1 \left( \pi_{16} \bar B_{21}+\frac{\pi^2_{16}}{36}\right)
-\lambda_2\frac{ \pi^2_{16}}{12}+B^\epsilon \pi_{16} \left(\frac{m^2 }{6 s}-\frac{1}{6}\right)
\\&&+\bar J  \left(\frac{L}{6}-\frac{L m^2}{6 s}+\frac{m^2 \pi_{16}}{3 s}
-\frac{\pi_{16}}{4}\right)-\frac{3 k_1 }{32}+\frac{k_2 }{96}+\frac{k_4 }{8}
+\frac{3 k_5 }{2}-\frac{k_9 }{24}\\
&&
-\frac{L^2 m^2}{12 s}-\frac{L^2}{12}-L \pi_{16}\left(\frac{ m^2 }{6 s}+\frac{1}{12}\right)
-\pi^2_{16} \left(\frac{m^2 \pi ^2}{72 s}+\frac{m^2}{12 s}+\frac{5}{48}\right)
\\
V_{28}&=&\frac{1}{4}\lambda_1 \left( \pi_{16} \bar B+\frac{\pi^2_{16}}{12}\right)
+\lambda_2\frac{ \pi^2_{16}}{8}+\frac{1}{4} \pi_{16} B^\epsilon
\\&&+\bar J  \left(\frac{7 \pi_{16}}{24}-\frac{L}{4}\right)
+\frac{k_1 }{8}-\frac{k_4 }{2}-3 k_5
+\frac{L^2}{8}+\frac{5 L \pi_{16}}{24}+\frac{11 \pi^2_{16}}{96}
\\
V_{29}&=&-\lambda_1 \frac{\pi^2_{16}}{24}
-\frac{1}{12} \pi_{16} \bar J -k_4 -3 k_5 +\frac{L \pi_{16}}{12}+\frac{7 \pi^2_{16}}{48}
\\
V_{210}&=&\lambda_1 \left\{\frac{1}{4}\pi_{16}\left[-\frac{s}{3} \bar B_{21}
+ \bar B\left(\frac{5 m^2}{3}  +\frac{s}{6}   \right)
- L m^2  \right]
+5\pi^2_{16}\left(\frac{ m^2 }{18}+\frac{  s}{864}\right)\right\}\\
&&+\lambda_2 \pi^2_{16}\left(\frac{m^2 }{3}+\frac{  s}{144}\right)
 + B^\epsilon   \pi_{16} \left(\frac{4 m^2  }{9}+\frac{s}{72}\right)
\\&&+\bar J  \left(\frac{499 m^2 \pi_{16}}{864}
+\frac{7 \pi_{16} s}{432}-\frac{4 L m^2}{9}-\frac{L s}{72}\right)\\
&&+k_1  \left(\frac{17 m^2}{288}+\frac{s}{96}\right)
-\frac{s}{288}  k_2 -\frac{m^2}{12}  k_3 -\frac{ m^2}{24} k_4
+k_5  \left(\frac{m^2}{2}-\frac{s}{8}\right)\\
&&+\frac{4 L^2 m^2}{9}+\frac{L^2 s}{144}
+\pi_{16}L\left(\frac{5  s}{432}-\frac{ m^2 }{6}\right)
+\pi^2_{16} \left(\frac{m^2 \pi ^2}{54}-\frac{47 m^2}{216}
-\frac{\pi ^2 s}{864}-\frac{11 s}{1296}\right)
\\
V_{211}&=&\lambda_1 \left(\frac{1}{3} \pi_{16} \bar B+\frac{\pi^2_{16}}{18}\right)
+\lambda_2\frac{ \pi^2_{16}}{6}+\frac{1}{3} \pi_{16} B^\epsilon
\\&&+\bar J  \left(\frac{4 \pi_{16}}{9}-\frac{L}{3}\right)
+\frac{k_1 }{6}-2 k_5  +\frac{L^2}{6}+\frac{2 L \pi_{16}}{9}+\frac{\pi^2_{16}}{27}
\\
V_{212}&=&\lambda_1 \left(\frac{1}{3} \pi_{16} \bar B_{21}+\frac{\pi^2_{16}}{54}\right)
+\lambda_2\frac{ \pi^2_{16}}{18}+B^\epsilon  \pi_{16}\left(\frac{1}{9}-\frac{m^2 }{9 s}\right)
\\&&+\bar J  \left(\frac{L m^2}{9 s}-\frac{L}{9}+\frac{m^2 \pi_{16}}{54 s}
+\frac{5 \pi_{16}}{27}\right)+\frac{k_1 }{24}+\frac{k_2 }{72}-\frac{k_5 }{2}+\frac{k_8 }{6}\\
&&+{L^2\over18}\left({m^2\over s}+1\right)+L\pi_{16} \left({m^2\over 9s}+{1\over27}\right)
+\pi^2_{16} \left(\frac{m^2 \pi ^2}{108 s}+\frac{m^2}{18 s}-\frac{1}{108}\right)
\\
V_{213}&=&-\lambda_1 \left(\frac{1}{6} \pi_{16} \bar B+\frac{\pi^2_{16}}{36}\right)
-\lambda_2\frac{ \pi^2_{16}}{12}-\frac{1}{6} \pi_{16} B^\epsilon
\\&&+\bar J  \left(\frac{L}{6}-\frac{2 \pi_{16}}{9}\right)-\frac{k_1 }{12}+k_5
 -\frac{L^2}{12}-\frac{L \pi_{16}}{9}-\frac{\pi^2_{16}}{54}
\\
V_{31}&=&\lambda_1 \left(\frac{19 m^2 \pi^2_{16}}{144}-\frac{1}{6} L m^2 \pi_{16}-\frac{\pi^2_{16} s}{96}\right)
+\frac{1}{12} \lambda_2 m^2 \pi^2_{16}
\\&&-\bar J  \left(\frac{2 m^2 \pi_{16}}{9}+\frac{\pi_{16} s}{48}\right)-\frac{1}{12} m^2 k_1
-\frac{1}{6} m^2 k_3 -3 m^2 k_4 -5 m^2 k_5\\
&& +\frac{L^2 m^2}{6}-\frac{19 L m^2 \pi_{16}}{72}+\frac{L \pi_{16} s}{48}+\pi^2_{16} \left(\frac{m^2 \pi ^2}{72}+\frac{47 m^2}{1728}-\frac{s}{384}\right)
\\
V_{32}&=&\frac{1}{2}\lambda_1 \left[ \pi_{16}
\left( s \bar B_{31}-  s \bar B_{21}+\frac{1}{2} m^2   \bar B -\frac{1}{3} L m^2  \right)
+ \pi^2_{16}\left(\frac{113 m^2}{144}-\frac{23 s}{288}\right)\right]\\
&&+\lambda_2 \pi^2_{16}\left(\frac{5 m^2 }{24}-\frac{ s}{48}\right)
+B^\epsilon  \pi_{16}\left(\frac{m^2 }{6}-\frac{s}{24}\right)
\\&&+\bar J  \left(\frac{L s}{24}-\frac{L m^2}{6}+\frac{23 m^2 \pi_{16}}{72}
-\frac{3 \pi_{16} s}{32}\right)-\frac{m^2}{24}  k_1 -\frac{s}{48} k_2 +m^2 k_4
 +\frac{5m^2}{2}  k_5\\
&& +\frac{L^2 m^2}{3}-\frac{L^2 s}{48}+L \pi_{16}\left(\frac{ s}{96}-\frac{65  m^2 }{144}\right)
 +\pi^2_{16} \left(\frac{m^2 \pi ^2}{48}+\frac{745 m^2}{3456}-\frac{\pi ^2 s}{144}-\frac{19 s}{768}\right)
\\
V_{33}&=&\lambda_1 \frac{\pi^2_{16}}{24}
+\frac{1}{12} \pi_{16} \bar J -\frac{k_3 }{3}-12 k_4 -60 k_5 -20 k_6 -\frac{L \pi_{16}}{12}-\frac{43 \pi^2_{16}}{288}
\\
V_{34}&=&\lambda_1 \left(\pi_{16} \bar B_{31}-\frac{\pi^2_{16}}{96}\right)
+\lambda_2\frac{ \pi^2_{16}}{8}+B^\epsilon \pi_{16} \left(\frac{1}{4}-\frac{m^2 }{2 s}\right)
\\&&+\bar J  \left(\frac{L m^2}{2 s}-\frac{L}{4}-\frac{5 m^2 \pi_{16}}{6 s}
+\frac{19 \pi_{16}}{48}\right)\\
&&+\frac{19 k_1 }{128}-\frac{3 k_2 }{128}-\frac{9 k_4 }{32}
+\frac{9 k_5 }{8}+\frac{5 k_6 }{2}+\frac{3 k_9 }{32}\\
&&
+\frac{L^2 m^2}{4 s}+\frac{L^2}{8}+ \pi_{16}L\left(\frac{ m^2 }{2 s}+\frac{5 }{48}\right)
+\pi^2_{16} \left(\frac{m^2 \pi ^2}{24 s}+\frac{m^2}{4 s}+\frac{323}{1152}\right)
\\
V_{35}&=&\lambda_1\frac{ \pi^2_{16}}{48}
+\frac{1}{24} \pi_{16} \bar J +3 k_4 +24 k_5 +10 k_6 -\frac{L \pi_{16}}{24}-\frac{41 \pi^2_{16}}{576}
\\
V_{36}&=&\lambda_1\frac{ \pi^2_{16}}{24}-\frac{L \pi_{16}}{12}-\frac{37 \pi^2_{16}}{288}
\\
&&+\bar J  \left(\frac{\pi_{16}}{12}-\frac{m^2 \pi_{16}}{9 s}\right)+\frac{k_1 }{192}
-\frac{k_2 }{192}-\frac{k_4 }{16}-\frac{31 k_5 }{4}-5 k_6 +\frac{k_9 }{48}
\\
V_{37}&=&\lambda_1 \left\{\frac{1}{4}\pi_{16}\left[  s \bar B_{21}
+\bar B\left( m^2  -\frac{1}{2} s \right)
-\frac{2}{3} L m^2  \right]
+\pi^2_{16}\left(\frac{113 m^2 }{288}-\frac{23  s}{576}\right)\right\}\\
&&+\lambda_2 \pi^2_{16}\left(\frac{5 m^2 }{24}-\frac{s}{48}\right)
+B^\epsilon \pi_{16} \left(\frac{m^2 }{6}-\frac{ s}{24}\right)
\\&&+\bar J  \left(\frac{L s}{24}-\frac{L m^2}{6}+\frac{23 m^2 \pi_{16}}{72}
-\frac{3 \pi_{16} s}{32}\right)-\frac{m^2 }{24} k_1 -\frac{s}{48}  k_2 +m^2 k_4
+\frac{5m^2}{2}  k_5\\
&& +\frac{L^2 m^2}{3}
 -\frac{L^2 s}{48}-\frac{65 L m^2 \pi_{16}}{144}+\frac{L \pi_{16} s}{96}
  +\pi^2_{16} \left(\frac{m^2 \pi ^2}{48}+\frac{745 m^2}{3456}-\frac{\pi ^2 s}{144}-\frac{19 s}{768}\right)
\\
V_{38}&=&\lambda_1 \left[\frac{1}{4}  \pi_{16}
\left(  s \bar B_{21}- s \bar B_{31}-\frac{m^2}{2}    \bar B+\frac{L m^2}{3}  \right)
+\pi^2_{16}\left(\frac{23s}{1152}-\frac{113}{576} m^2 \right)\right]\\
&&+\lambda_2 \pi^2_{16}\left(\frac{ s}{96}-\frac{5 m^2 }{48}\right)
+B^\epsilon  \pi_{16}\left(\frac{ s}{48}-\frac{m^2}{12}\right)\\&&
+\bar J  \left(\frac{L m^2}{12}-\frac{L s}{48}-\frac{23 m^2 \pi_{16}}{144}
+\frac{3 \pi_{16} s}{64}\right)+\frac{m^2}{48}  k_1 +\frac{s}{96} k_2 -\frac{m^2}{2}  k_4
-\frac{5 m^2}{4} k_5\\
&&-\frac{L^2 m^2}{6}+\frac{L^2 s}{96}+\frac{65 L m^2 \pi_{16}}{288}-\frac{L \pi_{16} s}{192}
+\pi^2_{16} \left(\frac{\pi ^2 s}{288}+\frac{19 s}{1536}-\frac{m^2 \pi ^2}{96}-\frac{745 m^2}{6912}\right)
\\
V_{39}&=&\lambda_1 \left(\frac{1}{12} L m^2 \pi_{16}-\frac{19}{288} m^2 \pi^2_{16}
+\frac{\pi^2_{16} s}{192}\right)-\frac{1}{24} \lambda_2 m^2 \pi^2_{16}
\\
&&+\bar J  \left(\frac{m^2 \pi_{16}}{9}+\frac{\pi_{16} s}{96}\right)+\frac{1}{24} m^2 k_1
+\frac{1}{12} m^2 k_3 +\frac{3}{2} m^2 k_4 +\frac{5}{2} m^2 k_5 \\
&&-\frac{L^2 m^2}{12}
+\frac{19 L m^2 \pi_{16}}{144}-\frac{L \pi_{16} s}{96}
+\pi^2_{16} \left(\frac{s}{768}-\frac{m^2 \pi ^2}{144}-\frac{47 m^2}{3456}\right)
\\
V_{310}&=&\lambda_1 \left[\frac{1}{4} \pi_{16}\left( s \bar B_{21}-  s \bar B_{31}
-\frac{1}{2} m^2  \bar B+\frac{1}{3} L m^2 \right)+
\pi^2_{16}\left(\frac{23 s}{1152}-\frac{113}{576} m^2\right)\right]\\
&&+\lambda_2 \left(\frac{\pi^2_{16} s}{96}-\frac{5 m^2 \pi^2_{16}}{48}\right)
+B^\epsilon \pi_{16} \left(\frac{ s}{48}-\frac{m^2  }{12}\right)
\\&&+\bar J  \left(\frac{L m^2}{12}-\frac{L s}{48}-\frac{23 m^2 \pi_{16}}{144}
+\frac{3 \pi_{16} s}{64}\right)+\frac{m^2}{48}  k_1 +\frac{s}{96}  k_2 -\frac{m^2}{2}  k_4
 -\frac{ m^2}{4} k_5 \\
&&
 -\frac{L^2 m^2}{6}+\frac{L^2 s}{96}+\frac{65 L m^2 \pi_{16}}{288}-\frac{L \pi_{16} s}{192}
+\pi^2_{16} \left(\frac{\pi ^2 s}{288}+\frac{19 s}{1536}-\frac{m^2 \pi ^2}{96}-\frac{745 m^2}{6912}\right)
\\
V_{311}&=&\lambda_1\frac{ \pi^2_{16}}{48}
+\frac{1}{24} \pi_{16} \bar J +3 k_4 +24 k_5 +10 k_6 -\frac{L \pi_{16}}{24}-\frac{41 \pi^2_{16}}{576}
\\
V_{312}&=&\lambda_1 \left(\frac{\pi^2_{16}}{192}-\frac{1}{2} \pi_{16} \bar B_{31}\right)
-\lambda_2\frac{ \pi^2_{16}}{16}+B^\epsilon  \pi_{16} \left(\frac{m^2}{4 s}-\frac{1}{8}\right)
\\&&+\bar J  \left(\frac{L}{8}-\frac{L m^2}{4 s}+\frac{5 m^2 \pi_{16}}{12 s}
-\frac{19 \pi_{16}}{96}\right)-\frac{19 k_1 }{256}+\frac{3 k_2 }{256}+\frac{9 k_4 }{64}
-\frac{9 k_5 }{16}-\frac{5 k_6 }{4}-\frac{3 k_9 }{64}\\
&&-\frac{L^2 m^2}{8 s}-\frac{L^2}{16}-\frac{L m^2 \pi_{16}}{4 s}-\frac{5 L \pi_{16}}{96}
-\pi^2_{16} \left(\frac{m^2 \pi ^2}{48 s}+\frac{m^2}{8 s}+\frac{323}{2304}\right)
\\
V_{313}&=&-\lambda_1\frac{ \pi^2_{16}}{96}
-\frac{1}{48} \pi_{16} \bar J -\frac{3 k_4 }{2}-12 k_5 -5 k_6 +\frac{L \pi_{16}}{48}+\frac{41 \pi^2_{16}}{1152}
\\
V_{314}&=&\lambda_1 \left(\frac{1}{2} \pi_{16} \bar B_{21}-\frac{\pi^2_{16}}{144}\right)
+\lambda_2\frac{ \pi^2_{16}}{12}+B^\epsilon \pi_{16} \left(\frac{1}{6}-\frac{m^2 }{6 s}\right)
\\&&+\bar J  \left(\frac{L m^2}{6 s}-\frac{L}{6}-\frac{5 m^2 \pi_{16}}{18 s}
+\frac{5 \pi_{16}}{24}\right)+\frac{35 k_1 }{384}-\frac{k_2 }{128}-\frac{3 k_4 }{32}
+\frac{19 k_5 }{8}+\frac{5 k_6 }{2}+\frac{k_9 }{32}\\
&&
+\frac{L^2 m^2}{12 s}+\frac{L^2}{12}+\frac{L m^2 \pi_{16}}{6 s}+\frac{L \pi_{16}}{8}
+\pi^2_{16} \left(\frac{m^2 \pi ^2}{72 s}+\frac{m^2}{12 s}+\frac{97}{576}\right)
\\
V_{315}&=&\lambda_1 \frac{\pi^2_{16}}{32}
+\frac{1}{16} \pi_{16} \bar J -\frac{k_4 }{2}-9 k_5 -5 k_6 -\frac{L \pi_{16}}{16}-\frac{127 \pi^2_{16}}{1152}
\\
V_{316}&=&-\lambda_1\frac{ \pi^2_{16}}{48}
+\bar J  \left(\frac{m^2 \pi_{16}}{18 s}-\frac{\pi_{16}}{24}\right)\\
&&-\frac{k_1 }{384}+\frac{k_2 }{384}+\frac{k_4 }{32}+\frac{31 k_5 }{8}+\frac{5 k_6 }{2}-\frac{k_9 }{96}+\frac{L \pi_{16}}{24}+\frac{37 \pi^2_{16}}{576}
\\
V_{317}&=&\lambda_1
\left[ \frac{1}{12}\pi_{16} \left(s \bar B_{21}
+  m^2   \bar B -\frac{1}{2}   s \bar B -  L m^2 \right)
+ \pi^2_{16}\left(\frac{31 m^2}{144}-\frac{7  s}{432}\right)\right]\\
&&+\lambda_2\pi^2_{16} \left(\frac{m^2 }{12}-\frac{ s}{144}\right)
+B^\epsilon  \pi_{16}\left(\frac{m^2 }{18}-\frac{  s}{72}\right)
\\&&+\bar J  \left(\frac{L s}{72}-\frac{L m^2}{18}+\frac{101 m^2 \pi_{16}}{864}
-\frac{\pi_{16} s}{27}\right)-\frac{11}{288} m^2 k_1 -\frac{1}{144} s k_2 -\frac{1}{12} m^2 k_3\\
&&
-\frac{11}{24} m^2 k_4 +k_5  \left(\frac{7 m^2}{4}-\frac{3 s}{8}\right)
+k_6  \left(m^2-\frac{s}{4}\right)\\
&&
+\frac{5 L^2 m^2}{36}-\frac{L^2 s}{144}-\frac{23 L m^2 \pi_{16}}{72}+\frac{L \pi_{16} s}{108}
+\pi^2_{16} \left(\frac{m^2 \pi ^2}{108}-\frac{m^2}{1728}-\frac{\pi ^2 s}{432}-\frac{173 s}{10368}\right)
\\
V_{318}&=&\lambda_1 \left[ \frac{1}{6}\pi_{16}\left( s \bar B_{21}- s \bar B_{31}
+ m^2   \bar B-\frac{1}{2} L m^2  \right)
+ \pi^2_{16}\left(\frac{m^2}{32}+\frac{19   s}{1728}\right)\right]\\
&&+\lambda_2 \pi^2_{16}\left(\frac{m^2 }{8}+\frac{s}{144}\right)
+B^\epsilon \pi_{16} \left(\frac{7 m^2 }{36}+\frac{ s}{72}\right)
\\&&+\bar J  \left(-\frac{7 L m^2}{36}-\frac{L s}{72}+\frac{199 m^2 \pi_{16}}{864}
+\frac{23 \pi_{16} s}{864}\right)\\
&&+k_1  \left(\frac{7 m^2}{144}+\frac{s}{192}\right)
+\frac{s}{576}  k_2 +\frac{5m^2}{24}  k_4 +k_5  \left(\frac{s}{8}-\frac{5 m^2}{8}\right)
+k_6  \left(\frac{s}{8}-\frac{m^2}{2}\right)\\
&&+\frac{11 L^2 m^2}{72}+\frac{L^2 s}{144}+\frac{11 L m^2 \pi_{16}}{144}+\frac{L \pi_{16} s}{864}
+\pi^2_{16} \left(\frac{m^2 \pi ^2}{216}-\frac{125 m^2}{1152}+\frac{\pi ^2 s}{1728}+\frac{85 s}{20736}\right)
\\
V_{319}&=&\lambda_1 \left\{\frac{1}{6} \pi_{16}
\left[- s \bar B_{21}+\bar B\left(\frac{1}{2}  s -  m^2 \right) +  L m^2  \right]+
\pi^2_{16}\left(\frac{25 s}{864}-\frac{35}{144} m^2 \right)\right\}\\
&&+\lambda_2 \pi^2_{16}\left(\frac{ s}{72}-\frac{m^2  }{6}\right)
+B^\epsilon  \pi_{16}\left(\frac{ s}{36}-\frac{m^2 }{9}\right)
\\&&+\bar J  \left(\frac{L m^2}{9}-\frac{L s}{36}-\frac{2 m^2 \pi_{16}}{27}
+\frac{29 \pi_{16} s}{432}\right)+\frac{1}{18} m^2 k_1 +\frac{1}{72} s k_2
 -\frac{1}{3} m^2 k_4 -m^2 k_5\\
&&
 -\frac{5 L^2 m^2}{18}+\frac{L^2 s}{72}+\frac{19 L m^2 \pi_{16}}{72}-\frac{5 L \pi_{16} s}{432}
 +\pi^2_{16} \left(\frac{\pi ^2 s}{216}+\frac{169 s}{10368}-\frac{m^2 \pi ^2}{54}-\frac{397 m^2}{1728}\right)
\\
V_{320}&=&\lambda_1 \left[\frac{1}{6} \pi_{16}\left(- s \bar B_{21}+  s \bar B_{31}
+\frac{1}{2} m^2   \bar B-\frac{1}{2} L m^2 \right)+\pi^2_{16}\left(\frac{35 m^2 }{288}
-\frac{25   s}{1728}\right)\right]\\
&&+\lambda_2 \pi^2_{16}\left(\frac{m^2 }{12}-\frac{  s}{144}\right)
+B^\epsilon  \pi_{16}\left(\frac{m^2 }{18}-\frac{  s}{72}\right)
\\
&&+\bar J  \left(\frac{L s}{72}-\frac{L m^2}{18}+\frac{m^2 \pi_{16}}{27}
-\frac{29 \pi_{16} s}{864}\right)-\frac{m^2}{36}  k_1 -\frac{s}{144}  k_2 +\frac{m^2}{6}  k_4
 +\frac{m^2}{2}  k_5 \\
&&+\frac{5 L^2 m^2}{36}-\frac{L^2 s}{144}-\frac{19 L m^2 \pi_{16}}{144}+\frac{5 L \pi_{16} s}{864}
+\pi^2_{16} \left(\frac{m^2 \pi ^2}{108}+\frac{397 m^2}{3456}-\frac{\pi ^2 s}{432}-\frac{169 s}{20736}\right)
\\
V_{321}&=&\lambda_1\frac{ \pi^2_{16}}{24}
+\frac{1}{12} \pi_{16} \bar J -6 k_5 -4 k_6 -\frac{L \pi_{16}}{12}-\frac{43 \pi^2_{16}}{288}
\\
V_{322}&=&\lambda_1 \left(\frac{1}{3} \pi_{16} \bar B_{31}+\frac{\pi^2_{16}}{288}\right)
+\lambda_2\frac{ \pi^2_{16}}{24}+B^\epsilon  \pi_{16}\left(\frac{1}{12}-\frac{m^2}{6 s}\right)
\\&&+\bar J  \left(\frac{L m^2}{6 s}-\frac{L}{12}+\frac{m^2 \pi_{16}}{36 s}
+\frac{7 \pi_{16}}{48}\right)+\frac{k_1 }{48}+\frac{k_2 }{48}+\frac{k_5 }{2}+\frac{k_6 }{2}
-\frac{3 k_7 }{2}+\frac{k_8 }{4}\\
&&+\frac{L^2 m^2}{12 s}+\frac{L^2}{24}+\frac{L m^2 \pi_{16}}{6 s}+\frac{L \pi_{16}}{48}
+\pi^2_{16} \left(\frac{m^2 \pi ^2}{72 s}+\frac{m^2}{12 s}+\frac{49}{3456}\right)
\\
V_{323}&=&\lambda_1 \frac{\pi^2_{16}}{72}
+\frac{1}{36} \pi_{16} \bar J -\frac{3 k_5 }{2}-k_6 +k_7 -\frac{L \pi_{16}}{36}-\frac{43 \pi^2_{16}}{864}
\\
V_{324}&=&\lambda_1 \left(\frac{1}{6} \pi_{16} \bar B+\frac{\pi^2_{16}}{144}\right)
+\lambda_2 \frac{\pi^2_{16}}{12}+\frac{1}{6} \pi_{16} B^\epsilon
\\&&+\bar J  \left(\frac{13 \pi_{16}}{72}-\frac{L}{6}\right)+\frac{k_1 }{12}+2 k_5 +2 k_6
 +\frac{L^2}{12}+\frac{11 L \pi_{16}}{72}+\frac{161 \pi^2_{16}}{1728}
\\
V_{325}&=&-\lambda_1\frac{ \pi^2_{16}}{48}
-\frac{1}{24} \pi_{16} \bar J +3 k_5 +2 k_6 +\frac{L \pi_{16}}{24}+\frac{43 \pi^2_{16}}{576}
\\
V_{326}&=&-\lambda_1 \left(\frac{1}{3} \pi_{16} \bar B_{21}+\frac{\pi^2_{16}}{216}\right)
-\lambda_2\frac{ \pi^2_{16}}{18}+B^\epsilon  \pi_{16}\left(\frac{m^2 }{9 s}-\frac{1}{9}\right)
\\&&+\bar J  \left(\frac{L}{9}-\frac{L m^2}{9 s}-\frac{m^2 \pi_{16}}{54 s}
-\frac{17 \pi_{16}}{108}\right)-\frac{k_1 }{24}-\frac{k_2 }{72}-k_5 -k_6 +k_7 -\frac{k_8 }{6}\\
&&-\frac{L^2 m^2}{18 s}
-\frac{L^2}{18}-\frac{L m^2 \pi_{16}}{9 s}
-\pi^2_{16} \left(\frac{m^2 \pi ^2}{108 s}+\frac{m^2}{18 s}+\frac{35}{864}\right)
\end{eqnarray*}
Where the $\bar J$  and $k_i$ function are defined as
\begin{eqnarray*}
\label{defki}
\sigma &=&\sqrt{1-{4\over s}},\\
h &=& \frac{1}{\sigma}\ln\frac{\sigma-1}{\sigma+1}\\
\bar J&=& \pi_{16}(\sigma^2 h +2)\\
k_1&=& \pi_{16}^2\sigma^2h^2\\
k_2&=& \pi_{16}^2(\sigma^4h^2-4)\\
k_3&=&{1\over(16\pi^2)^2}\left[{\sigma^2\over s}h^3 +\pi^2{1\over s}h -{\pi^2\over2} \right]\\
k_4&=&\frac{1}{s\sigma^2}\left[\frac{1}{2}k_1+\frac{1}{3}k_3+
\pi_{16}\bar{J} +\frac{\pi_{16}^2}{12}(\pi^2-6)s\right]\; .\nonumber\\
k_5&=&{1\over s \sigma^2}\left[k_4 - {1\over12} k_1 - {\pi_{16}\over12} \bar J
+ \pi^2_{16} \left({5\over6} - {\pi^2\over9} \right) \right]
+ {\pi^2_{16}\over12} \left({5\over2} - {1\over3} pi^2\right)\\
k_6&=&{1\over s \sigma^2}\left[5k_5 + {1\over12} k_1+ {\pi_{16}\over18} \bar J
+ {\pi^2_{16}\over6 }\left(\pi^2-{49\over6} \right)\right]
       +{1\over24}\pi^2_{16} \left(\pi^2-{49\over6 } \right)\\
k_7&=& {1\over s}  \left(  k_5 + {1\over2} k_4 + {1\over24} k_1
+ {5\over24} \pi_{16} \bar J \right) + {1\over72} \pi^2_{16}\\
k_8&=&{1\over s} \left( k_4 + {7\over12} k_1 + {25\over36} \pi_{16}\bar J \right)
+ \pi^2_{16}\left( {47\over216}  + {1\over36} \pi^2\right)\\
k_9&=&  {1\over s}  \left( k_3 - {5\over2} k_1 \right) - \pi^2_{16}\left(2 + {\pi^2\over12} \right)
\end{eqnarray*}
The $\bar J$ and $k_i$ vanish at $s=0$
and are well behaved for $s\to \infty$.
They have discontinuities in the derivative at threshold
but there no poles there. The functions $k_i$ are constructed using the
arguments and methods of \cite{knecht}.

All the $k_i$ above show up at intermediate stages of the
calulations but in the final result $k_5(s),\ldots,k_9(s)$ always appear
multiplied by powers of $s$ and can thus be removed.

Finally, to get the scattering lengths we need to expand these functions
around $t,u=0$ and $s=4$.
The expansion using $s = 4(1+q^2)$ around $s=4$ reads up to order $q^4$.
\ba
\bar J(s) &=& \pi_{16} \, \left( 2-2q^2+\frac{4}{3}q^4 \right)
\nonumber\\
k_1(s) &=& \pi_{16}^2\,\left(-\pi^2+4q^2-\frac{4}{3}q^4 \right)
\nonumber\\
k_2(s) &=&
        \pi_{16}^2 \,  \left( - 4 -\pi^2 q^2 +(4+\pi^2)q^4\right)
\nonumber\\
k_3(s) &=&
       \pi_{16}^2 \, \left(\frac{1}{2} \pi^2
     - \left(2+\frac{2}{3}\pi^2\right)q^2
     +\left( 2+\frac{8}{15}\pi^2\right)q^4\right)
\nonumber\\
k_4(s) &=&
        \pi_{16}^2 \, \left(   - \frac{2}{3} + \frac{1}{36}\pi^2
       +  \left(\frac{1}{3} +\frac{2}{45} \pi^2 \right)q^2
       - \left(\frac{1}{3}+\frac{4}{105} \pi^2\right) q^4\right)\,.
\ea
The expansion around $t=0$ up to order $t^2$ are
\ba
\bar J(t) &=& \pi_{16}\left(\frac{1}{6}t +\frac{1}{60}t^2\right)\,,
\nonumber\\
k_1(t) &=&
       \pi_{16}^2 \, \left(  - t -\frac{1}{12} t^2 \right)
\nonumber\\
k_2(t) &=&
        \pi_{16}^2 \,\left(  -\frac{2}{3}t -\frac{7}{180}t^2\right)
\nonumber\\
k_3(t) &=&
       \pi_{16}^2 \, \left(\left(  -\frac{1}{2} + \frac{1}{12}\pi^2 \right)t
       + \left(  - \frac{1}{8} + \frac{1}{60}\pi^2 \right)t^2\right)
\nonumber\\
k_4(t) &=&
       \pi_{16}^2 \,\left(\left(\frac{1}{4}- \frac{1}{36}\pi^2\right)t
       + \left(\frac{19}{240} - \frac{1}{120}\pi^2 \right)t^2\right)\,.
\ea

\end{document}